\documentclass[3p,times,11pt]{elsarticle}
\usepackage{algorithm} 
\usepackage{algpseudocode}
\usepackage[titletoc,title]{appendix}
\usepackage{mdframed}
\usepackage{float}
\usepackage{wrapfig,blindtext}
\usepackage{graphicx} 
\usepackage{epstopdf} 
\usepackage{pslatex} 
\usepackage{amsmath,amssymb}
\usepackage{caption}
\usepackage{subcaption}
\usepackage{amsfonts,amsthm} 
\usepackage[english]{babel} 
\usepackage[dvipsnames]{xcolor}
\usepackage{enumitem}
\usepackage{booktabs}
\usepackage{wasysym}
\usepackage{pdfpages}
\usepackage{float}
\usepackage{comment}
\usepackage{soul}

\usepackage{nomencl}
\usepackage{multirow}
\usepackage{soul}
\usepackage{cancel}
\usepackage{ulem}

\usepackage[font=small,labelfont=bf,
   justification=justified,format=plain]{caption}

\usepackage{comment}

\usepackage{nomencl}
  \makenomenclature

\newcommand\norm[1]{\left\lVert#1\right\rVert_2}

\newcommand{\mb}[1]{{\mathbf{#1}}}

\newcommand{\MC}{\mathcal}
\DeclareMathOperator*{\argmin}{arg\,min}

\newcommand{\itermaxLinMS}{\ensuremath{l}}      
\newcommand{\itermaxRK}{\ensuremath{s}}         
\newcommand{\sumIdx}{\ensuremath{j}}
\newcommand{\sumIdxTwo}{\ensuremath{m}}
\newcommand{\iterIdx}{\ensuremath{n}}           
\newcommand{\subIdx}{\ensuremath{k}}            
\newcommand{\normVarIdx}{\ensuremath{v}}        
\newcommand{\varIdx}{\ensuremath{i}}            
\newcommand{\genIdx}{\ensuremath{i}}            

\newcommand{\rkSolVar}{\ensuremath{s}}
\newcommand{\rkSol}{\ensuremath{\mathbf{\rkSolVar}}}

\newcommand{\stateVar}{\ensuremath{q}}
\newcommand{\solConsFOM}{\ensuremath{\mathbf{\stateVar}}}
\newcommand{\solPrimFOM}{\ensuremath{\mathbf{\stateVar}_p}}
\newcommand{\solConsFOMFunc}[1]{\ensuremath{\solConsFOM \left( #1 \right)}}

\newcommand{\solConsFOMRef}{\ensuremath{\mathbf{\stateVar}_{\text{ref}}}}
\newcommand{\solPrimFOMRef}{\ensuremath{\mathbf{\stateVar}_{p,\text{ref}}}}
\newcommand{\solConsFOMUnst}{\ensuremath{\mathbf{\stateVar}^\prime}}
\newcommand{\solPrimFOMUnst}{\ensuremath{\mathbf{\stateVar}_p^\prime}}
\newcommand{\dataMatCons}{\ensuremath{\mathbf{\MakeUppercase{\stateVar}}}}

\newcommand{\solConsROMFull}{\ensuremath{\widetilde{\mathbf{\stateVar}}}}
\newcommand{\solConsROMRed}{\textcolor{black}{\ensuremath{\widehat{\mathbf{\stateVar}}}}}
\newcommand{\solPrimROMFull}{\ensuremath{\widetilde{\mathbf{\stateVar}}_p}}
\newcommand{\solPrimROMRed}{\ensuremath{\textcolor{black}{\widehat{\mathbf{\stateVar}}}_{\textcolor{black}{p}}}}
\newcommand{\solPrimFOMVar}{\ensuremath{\mathbf{\stateVar}_{p,\varIdx}}}
\newcommand{\solPrimROMFullVar}{\ensuremath{\mathbf{\widetilde{\stateVar}}_{p,\varIdx}}}
\newcommand{\solPrimROMProj}{\ensuremath{\mathbf{\widetilde{\stateVar}}_{p,\text{pod}}}}
\newcommand{\solPrimROMProjVar}{\ensuremath{\mathbf{\widetilde{\stateVar}}_{p,\text{pod},\varIdx}}}

\newcommand{\rhsVar}{\ensuremath{f}}
\newcommand{\rhs}{\ensuremath{\mathbf{\rhsVar}}}
\newcommand{\rhsFunc}[1]{\ensuremath{\rhs \left( #1 \right)}}
\newcommand{\rhsPrim}{\ensuremath{\mathbf{\rhsVar}_p}}
\newcommand{\rhsFuncPrim}[1]{\ensuremath{\rhsPrim \left( #1 \right)}}

\newcommand{\jacobVar}{\ensuremath{j}}
\newcommand{\jacobConsFOM}{\ensuremath{\mathbf{\MakeUppercase{\jacobVar}}}}
\newcommand{\jacobConsROM}{\ensuremath{\mathbf{\widetilde{\MakeUppercase{\jacobVar}}}}}
\newcommand{\jacobPrimROM}{\ensuremath{\textcolor{black}{\mathbf{\widetilde{\MakeUppercase{\jacobVar}}}}_{\textcolor{black}{p}}}}
\newcommand{\gm}{\ensuremath{\boldsymbol{\Gamma}}}
\newcommand{\gmInv}{\ensuremath{\boldsymbol{\Gamma}^{-1}}}
\newcommand{\gmROM}{\ensuremath{\textcolor{black}{\widetilde{\boldsymbol{\Gamma}}}}}

\newcommand{\resVar}{\ensuremath{r}}
\newcommand{\res}{\ensuremath{\mathbf{\resVar}}}
\newcommand{\resFunc}[1]{\ensuremath{\res \left( #1 \right)}}
\newcommand{\resPrim}{\ensuremath{\mathbf{\resVar}_p}}
\newcommand{\resPrimFunc}[1]{\ensuremath{\resPrim \left( #1 \right)}}

\newcommand{\scaleVarCons}{\ensuremath{p}}
\newcommand{\scaleMatCons}{\ensuremath{\mathbf{\MakeUppercase{\scaleVarCons}}}}
\newcommand{\scaleVarPrim}{\ensuremath{h}}
\newcommand{\scaleMatPrim}{\ensuremath{\mathbf{\MakeUppercase{\scaleVarPrim}}}}

\newcommand{\trialBasisVar}{\ensuremath{v}}
\newcommand{\trialBasisCons}{\ensuremath{\mathbf{\MakeUppercase{\trialBasisVar}}}}
\newcommand{\trialBasisPrim}{\textcolor{black}{\ensuremath{\mathbf{\MakeUppercase{\trialBasisVar}}_p}}}
\newcommand{\testBasisVar}{\ensuremath{w}}
\newcommand{\testBasisCons}{\ensuremath{\mathbf{\MakeUppercase{\testBasisVar}}}}
\newcommand{\testBasisPrim}{\ensuremath{\textcolor{black}{\mathbf{\MakeUppercase{\testBasisVar}}}_{\textcolor{black}{p}}}}
\newcommand{\testBasisPrimPrim}{\ensuremath{\textcolor{black}{\mathbf{\widetilde{\MakeUppercase{\testBasisVar}}}}_{\textcolor{black}{p}}}}
\newcommand{\testBasisPrimGPOD}{\ensuremath{\textcolor{black}{\mathbf{\overline{\MakeUppercase{\testBasisVar}}}}_{\textcolor{black}{p}}}}

\newcommand{\sigmaPOD}{\ensuremath{\widetilde{\sigma}}}
\newcommand{\linStabDiacritic}[1]{\widehat{#1}}
\newcommand{\sigmaLinStab}{\textcolor{black}{\ensuremath{\linStabDiacritic{\sigma}}}}
\newcommand{\kappaLinStab}{\textcolor{black}{\ensuremath{\linStabDiacritic{\kappa}}}}
\newcommand{\lambdaLinStab}{\textcolor{black}{\ensuremath{\linStabDiacritic{\lambda}}}}

\newcommand{\resApprox}{\ensuremath{\mathbf{\overline{\resVar}}}}
\newcommand{\resBasisVar}{\ensuremath{u}}
\newcommand{\resBasis}{\ensuremath{\mathbf{\MakeUppercase{\resBasisVar}}}}
\newcommand{\sampVar}{\ensuremath{s}}
\newcommand{\sampMat}{\ensuremath{\mathbf{\MakeUppercase{\sampVar}}}}

\newcommand{\trialSpaceCons}{\ensuremath{\MC{\MakeUppercase{\trialBasisVar}}}}
\newcommand{\trialSpacePrim}{\ensuremath{\MC{\MakeUppercase{\trialBasisVar}}_p}}

\newcommand{\numDOF}{\ensuremath{N}}            
\newcommand{\numElements}{\ensuremath{N_{elem}}}    
\newcommand{\numVars}{\ensuremath{N_{var}}}     
\newcommand{\numSolModes}{\ensuremath{n_p}}     
\newcommand{\numSolModesTotal}{\ensuremath{n_{p,total}}}
\newcommand{\numResModes}{\ensuremath{n_d}}     
\newcommand{\numSamps}{\ensuremath{n_s}}
\newcommand{\numSnaps}{\ensuremath{n_t}}
\newcommand{\numSpec}{\ensuremath{n_{\text{spec}}}}

\newcommand{\timeVar}{\text{t}}                 
\newcommand{\dt}{\ensuremath{\Delta \text{t}}}  
\newcommand{\dTimeVar}{\text{dt}}               
\newcommand{\dPTimeVar}{\ensuremath{\text{d}\tau}} 
\newcommand{\dtau}{\ensuremath{\Delta \tau}}    

\newcommand{\errROMVsFOM}{\ensuremath{\epsilon}}
\newcommand{\errROMVsProj}{\ensuremath{\widetilde{\epsilon}}}
\newcommand{\errProjVsFOM}{\ensuremath{\widehat{\epsilon}}}

\newcommand{\ltiMatA}{\ensuremath{\mathbf{J}}}
\newcommand{\ltiMatB}{\ensuremath{\mathbf{B}}}
\newcommand{\linStabMat}{\ensuremath{\mathbf{C}}}

\newcommand{\water}{\ensuremath{\text{H}_2\text{O}}}
\newcommand{\oxygen}{\ensuremath{\text{O}_2}}
\newcommand{\methane}{\ensuremath{\text{CH}_4}}
\newcommand{\carbonDiox}{\ensuremath{\text{CO}_2}}

\newcommand{\lp}{\left(}
\newcommand{\rp}{\right)}
\newcommand{\yFunc}[1]{\ensuremath{\mathbf{y} \left( #1 \right)}}
\newcommand{\zeroVec}{\ensuremath{\mathbf{0}}}

\newcommand{\pde}[2]{\ensuremath{\frac{\partial #1}{\partial #2}}}

\journal{Journal of Computational Physics, May 2020}

\begin{document}
\topmargin -1.5cm
\textheight 23cm

\begin{frontmatter}

\title{Model Reduction for Multi-Scale Transport Problems using {\color{black} Model-form Preserving} Least-Squares Projections with Variable Transformation}

\author{Cheng Huang}
\ead{huangche@umich.edu}
\author{Christopher R. Wentland}
\ead{chriswen@umich.edu}
\author{Karthik Duraisamy}
\ead{kdur@umich.edu}
\address{University of Michigan, Ann Arbor, MI}
\author{Charles Merkle}
\ead{merkle@purdue.edu}
\address{Purdue University, West Lafayette, IN}

\begin{abstract}
 A projection-based formulation is presented for non-linear model reduction of problems with extreme scale disparity. The approach allows for the selection of an arbitrary, but complete, set of solution variables while preserving the structure of the governing equations. Least-squares-based minimization is leveraged to guarantee symmetrization and discrete consistency with the full-order model (FOM). Two levels of scaling are used to achieve the conditioning required to effectively handle problems with extremely disparate physical phenomena, characterized by extreme stiffness in the system of equations.  The formulation -- referred to as {\color{black}model-form preserving} least-squares with variable transformation ({\color{black} MP-LSVT}) -- provides global stabilization for both implicit and explicit time integration schemes. To achieve computational efficiency, a pivoted QR decomposition is used with oversampling, and adapted to the {\color{black} MP-LSVT} method. The framework is demonstrated in  representative two- and three-dimensional reacting flow problems, and the {\color{black} MP-LSVT} is shown to exhibit improved stability and accuracy over standard projection-based ROM techniques. Physical realizability and local stability are promoted by enforcing limiters in both temperature and species mass fractions. These limiters are demonstrated to be important in eliminating regions of spurious burning, thus enabling the ROMs to provide accurate representations of the heat release rate and flame propagation speed. In the 3D application, it is shown that more than two orders of magnitude acceleration in computational efficiency can be achieved, while also providing reasonable future-state predictions. A key contribution of this work is the development and demonstration of a comprehensive ROM formulation that targets highly challenging multi-scale transport-dominated problems. 
\end{abstract}

\end{frontmatter}

\section{Introduction}
\label{intro}
With  advances in computing architectures and computational algorithms, high-fidelity multi-physics simulations are becoming integral to the analysis and design of complex systems. A pertinent example is that of reacting flow simulations~\cite{adityaDNS, OefeleinSupercritical}, which play an important role in the investigation and understanding of combustion dynamics in propulsion systems operating at high pressures and temperatures. Such simulations provide details which cannot be quantitatively accessed through experiments. Accurate and efficient modeling of combustion dynamics in a practical environment has the potential to improve engine performance and reduce failures. However, large eddy simulations (LES) of combustion dynamics -- even for small-scale engines  ~\cite{UrbanoLES2016} -- typically require $O(10^7)$ CPU-hours per simulation, and are thus impractical in many-query applications such as engineering design and uncertainty quantification. In the present work, we develop model order reduction techniques to achieve efficient and accurate simulations of complex multi-physics problems, and use rocket combustion examples to motivate and evaluate our algorithms.

Projection-based reduced-order models~\cite{Lumley1997, Graham1997, Lucia2003} have proven to be effective in reducing partial differential equations (PDE)-based dynamical systems to low-dimensional manifolds and have been successfully applied in problems such as flow control~\cite{Barbagallo2012, Barbagallo2009, Barbagallo2011} and aeroelasticity~\cite{Lieu2007, Lucia2004}. When applied to complex multi-scale problems that involve transport phenomena such as convection, it is well-recognized that projection-based ROMs suffer from accuracy and stability issues. These issues may arise from the inherent lack of numerical stability of the projection method itself (e.g. Galerkin projection~\cite{Rempfer2000}), mode truncation (e.g. removal of low-energy spatial modes~\cite{Bergmann2009}), or simplifications of model equations~\cite{Noack2005}. 

Several remedies for these stability issues have been proposed in the literature. For linear dynamical systems, balanced truncation~\cite{Moore1981, Pernebo1982} develops stable ROMs by eliminating system modes which are largely uncontrollable and unobservable, but is not tractable for large-scale systems. The balanced proper orthogonal decomposition technique~\cite{Willcox2002, Rowley2005} approximates balanced truncation, eliminating stability guarantees but making it tractable for large-scale systems. The use of adaptive bases~\cite{San2015, peherstorfer2015online, PeherstorferADEIM, CarlbergAdaptiveBasis2015, EtterAdaptiveBasis2020,ZimmermannAdaptiveBasis2018} and projection onto non-linear manifolds via deep convolutional autoencoders~\cite{LeeNonlinearManifold2020} have also been investigated to mitigate ROM accuracy and stability issues. Others have approached stabilization from a control perspective via eigenvalue reassignment~\cite{Kalashnikova2014}, convex optimization~\cite{Amsallem2012}, or a combination of the two~\cite{rezaian2020hybrid}.

Another group of studies address stability issues from the perspective of closure modeling, accounting for the effect of truncated ROM dynamics based on the resolved ROM dynamics (analogous to the closure problem in LES). Bergmann et al.~\cite{Bergmann2009} proposed to use residuals of the full-order models to account for the absence of low-energy dissipative spatial modes. Lucia et al.~\cite{Lucia2003} demonstrated the effectiveness of constructing stable ROMs by including additional artificial dissipation terms. Variational multi-scale closure models~\cite{SanVMS2015,IliescuVMS2014,StabileVMS2019} for projection-based ROMs have been demonstrated to offer stabilization for multi-scale problems. Parish et al.~\cite{parish2017non,parish2017dynamic} leveraged the Mori–Zwanzig formalism to develop a closed representation of the unresolved scales. Extended to projection-based ROMs~\cite{Parish2019}, the method is referred to as the Adjoint Petrov–Galerkin (APG) method. The Markovian first order approximation to the memory kernel yields a technique that is analogous to adjoint-stabilization~\cite{hughes1995multiscale} in the finite element community.

Researchers have also attempted to leverage the underlying numerical discretization to improve ROM stability. Rowley et al.~\cite{Rowley2004} pointed out that defining a physically-meaningful inner product to obtain low-dimensional manifolds (e.g. proper orthogonal decomposition bases) can yield  a quadratic reduced system that is more stable and much simpler to implement for model reduction of compressible flows. Barone et al.~\cite{Barone2008,Barone2009JCP} proposed to stabilize the reduced system by symmetrizing the higher-order PDE with a preconditioning matrix, and also highlighted the importance of formulating a proper inner product in preserving stability in the reduced system. For aeroelastic applications, Amsallem and Farhat~\cite{Amsallem2014} have shown the advantages of using the descriptor form over the non-descriptor form of the governing equations. While these methods can enable stable ROMs, they can compromise the conservative properties of the governing equations during the model reduction procedure. Afkham et al.~\cite{Afkham2020} highlighted the importance of preserving the conservative form of the governing equations in ROM development. 

Following developments in the finite element community (e.g. the Galerkin Least-Squares technique of stabilization~\cite{LSFEM}), Carlberg et al.~\cite{carlberg2013gnat,Carlberg2017}  demonstrated that minimizing the least-squares residual of the projected solution yields stabilized non-linear ROMs. This technique is referred to as least-squares Petrov-Galerkin (LSPG) projection. The GNAT method~\cite{carlberg2013gnat} extends hyper-reduction to LSPG projection.  Strategies have also been developed to explicitly enforce desirable properties such as discrete conservation~\cite{CarlbergConsvLSPG}. Grimberg et al.~\cite{GrimbergPROM2020} demonstrated improved stability, accuracy and efficiency of hyper-reduced LSPG ROMs (HPROMs) in the context of convection-dominated turbulent flows.  In practice, least-squares-based techniques have been restricted to implicit time integration schemes, as applying residual minimization to an explicit time integrator results in a test basis that is  identical to the trial basis. 

The above investigations have been demonstrated to be effective in the applications pursued in the respective publications. Many multi-scale, multi-physics contexts, such as reacting flows, further exacerbate accuracy and robustness issues  with high numerical stiffness arising from the chemical kinetics. This stiffness can lead to  Jacobian matrices with high condition numbers $O(10^{12})$, and may produce many issues even in full-order models~\cite{StoneKinetics}. For instance, a recent investigation~\cite{HuangAIAAJ2019} has identified the appearance of localized spurious oscillations near sharp, dispersed flame fronts as a major contributor to the stability issues in reacting flow ROMs. These oscillations are a consequence of under-resolution, or the inability of the ROM to resolve resolve sharp flame fronts, and can often lead to features such as negative temperatures (i.e. T $\leq$ 0 K) that terminate the calculations. A temperature limiter was required to improve ROM robustness even in relatively simple reacting flow configurations. Similar ideas of enforcing physical realizability were also employed by Blonigan et al.~\cite{BloniganHypersonics} for hypersonic flow applications in which strong temperature gradients are present.

In the present work, we develop a comprehensive framework for projection-based reduced-order model development for complex multi-scale applications to achieve improved robustness and computational efficiency. Variable transformations have long been used to improve the stability of high-fidelity CFD models~\cite{Tadmor1984, Hughes1986, Nam2011, Hassler2020} and have recently proven useful in constructing low-cost, accurate ROMs~\cite{Kramer2019LiftAndLearn, Swischuk2020AIAAJ_LL, qian2020lift}.  Inspired by these developments, we introduce a {\color{black}model-form preserving} variable transformation and leverage least-squares minimization~\cite{Carlberg2017} to achieve global stabilization. To promote local stability, we enforce physical realizability on both temperature and species mass fraction fields. This method achieves discrete consistency and symmetrization, results in well-conditioned ROMs, and allows for the use of implicit and explicit time integrators. We refer to our new formulation as the {\color{black}model-form preserving} least-squares with variable transformation ({\color{black}MP-LSVT}) technique. Detailed evaluations of the {\color{black}MP-LSVT} technique and comparisons with Galerkin and LSPG projection are presented in challenging reacting flow applications.

The remainder of the paper is organized as follows. Section~\ref{fom} presents  the full-order model (FOM) and time discretization. Section~\ref{StandardROM} reviews the procedure for standard model reduction via Galerkin and LSPG projection. Identifying a gap in the literature, Section~\ref{proof} presents a proof of linear stability of least-squares-based ROMs. Section~\ref{SPLSPG} introduces the selection of transformed solution variables, and the procedure for {\color{black}MP-LSVT} projection to achieve symmetrization and discrete consistency. Section~\ref{limiters} discusses a limiter-based strategy to improve local robustness. Section~\ref{results} presents numerical results  and analysis for ROMs of benchmark reacting flow problems and assesses their accuracy, robustness, and efficiency. In Section~\ref{sec:conclusion}, we provide concluding remarks and perspectives.

\section{Full-Order Model  and Time Discretization}
\label{fom}

We represent the governing equations of the full-order model as a generic dynamical system
\begin{equation}
    \frac{\text{d} \solConsFOM}{\dTimeVar} = \rhsFunc{\solConsFOM, \timeVar}, \ \ \solConsFOM(0) = \solConsFOM_0,
    \label{cfd:discretized}
\end{equation}
where $\timeVar \in [0,T]$ is the solution time, $\solConsFOM: [0, T] \rightarrow {\mathbb{R}^{\numDOF}}$ is the vector of state variables, and $\rhs: \mathbb{R}^{\numDOF} \times [0,T] \rightarrow \mathbb{R}^{\numDOF}$ is a (potentially non-linear) function. While our formulation is general,  demonstrations are performed on dynamical systems that are derived from a spatial discretization of partial differential equations. For a set of discretized partial differential equations (PDE), where $\numElements$ is the total number of elements  and $\numVars$ is the number of state variables in each element. The function $\rhs$ would thus represent surface fluxes, source terms, and body forces arising from the spatial discretization of the governing equations. Equation~\ref{cfd:discretized} (or its fully-discrete counterpart) is referred to as the full-order model (FOM), and for reasonably complex systems with well-refined spatial discretizations, the dimension $\numDOF$ may be $O(10^5 \; \text{-} \; 10^8)$.

Two classes of time-discretization methods to solve Eq.~\ref{cfd:discretized} are introduced: linear multi-step methods and Runge--Kutta methods.

\subsection{Linear Multi-step Methods}
\label{LinearMultistep}
The solution to the governing equations (Eq.~\ref{cfd:discretized}) can be computed using linear multi-step methods, the $\itermaxLinMS$-step version of which can be expressed as
\begin{equation}
    \solConsFOM^{\iterIdx} + \sum^{\itermaxLinMS}_{j=1} \alpha_j \solConsFOM^{\iterIdx - j} = \dt \beta_0 \rhsFunc{\solConsFOM^\iterIdx, \timeVar^\iterIdx} + \dt \sum^{\itermaxLinMS}_{j=1} \beta_j \rhsFunc{\solConsFOM^{\iterIdx-j}, \timeVar^{\iterIdx-j}},
    \label{cfd:LinearMultiStep}
\end{equation}
where $\dt \in \mathbb{R}^{+}$ is the physical time step for the numerical solution, and the coefficients $\alpha_j$, $\beta_j \in \mathbb{R}$ are determined  based on $\itermaxLinMS$. If $\beta_0 = 0$ , the method is explicit; otherwise, the method is implicit. The FOM equation residual $\res: \mathbb{R}^{\numDOF} \rightarrow \mathbb{R}^{\numDOF}$ is  defined as
\begin{equation}
    \resFunc{\solConsFOM^\iterIdx} \triangleq \solConsFOM^{\iterIdx} + \sum^{\itermaxLinMS}_{j=1} \alpha_j \solConsFOM^{\iterIdx-j} - \dt \beta_0 \rhsFunc{\solConsFOM^\iterIdx, \timeVar^\iterIdx} - \dt \sum^{\itermaxLinMS}_{j=1} \beta_j \rhsFunc{\solConsFOM^{\iterIdx-j}, \timeVar^{\iterIdx-j}}.
    \label{EqRes:LinearMultiStep}
\end{equation}
The state variables, $\solConsFOM^\iterIdx$, are solved for at each time step so that $\resFunc{\solConsFOM^\iterIdx} = \zeroVec$.

\subsection{Runge--Kutta Methods}
\label{RK}
Alternatively, the governing equations can be solved numerically using the $\itermaxRK$-stage Runge--Kutta method
\begin{equation}
    \solConsFOM^{\iterIdx} = \solConsFOM^{\iterIdx-1} + \dt \sum^{\itermaxRK}_{j=1} b_j \mathbf{s}_j,
    \label{cfd:RK}
\end{equation}
where $\mathbf{s}_1 = \rhsFunc{\solConsFOM^{\iterIdx-1}, \timeVar^{\iterIdx-1}}$ and

\begin{equation}
    \mathbf{s}_j = \rhsFunc{\solConsFOM^{\iterIdx-1} + \dt \sum^{s}_{m=1} a_{jm}{\mathbf{s}_m}, \timeVar^{\iterIdx-1} + c_j \dt},
    \label{cfd:RK_stage}
\end{equation}
and the coefficients, $a_{jm}$, $b_j$, $c_j \in \mathbb{R}$ are determined for different Runge--Kutta methods. The methods are explicit if $a_{jm} = 0$, $\forall \; m \geq j$, and are diagonally implicit if $a_{jm} = 0$, $\forall \; m > j$. Otherwise, the methods are implicit.  The FOM equation residual is defined as
\begin{equation}
    \resFunc{\solConsFOM^\iterIdx} \triangleq \solConsFOM^{\iterIdx} - \solConsFOM^{\iterIdx-1} - \dt \sum^{\itermaxRK}_{j=1} b_j \mathbf{s}_j.
    \label{EqRes:RK}
\end{equation}
where, again, the solution variables, $\solConsFOM^\iterIdx$, are solved at each time step such that $\resFunc{\solConsFOM^\iterIdx} = \zeroVec$.
\section{Standard Model Reduction}
\label{StandardROM}
In this section, we introduce the standard Galerkin and least-squares Petrov-Galerkin (LSPG) projection methods for developing ROMs of the governing equations in Eq.~\ref{cfd:discretized}. 

\subsection{Construction of Proper Orthogonal Decomposition Bases} 
\label{pod:consv}
In both approaches, the state $\solConsFOM$ is expressed in a trial space $\trialSpaceCons \triangleq \text{Range}(\trialBasisCons)$, where $\trialBasisCons \in \mathbb{R}^{\numDOF \times \numSolModes}$ is the trial basis matrix. We define $\solConsFOMRef(\timeVar) \triangleq \solConsFOM(\timeVar) - \solConsFOMRef$, where $\solConsFOMRef$ is a reference state. Possible reference states include the initial FOM solution, $\solConsFOMRef = \solConsFOM(\timeVar = \timeVar_0)$, or the time-averaged FOM solution, $\solConsFOMRef = \frac{1}{\Delta \text{T}} \int_{\timeVar_0}^{\timeVar_0 + \Delta \text{T}} \solConsFOM(\timeVar) \text{dt}$. 

We then seek a representation  $\solConsROMFull: [0,T] \rightarrow \trialSpaceCons$ such that
\begin{equation}
   \scaleMatCons \lp \solConsROMFull(\timeVar) - \solConsFOMRef \rp = \trialBasisCons \solConsROMRed(\timeVar),
    \label{pod:expansion_consv}
\end{equation}
where $\solConsROMRed: [0,T] \rightarrow \mathbb{R}^{\numSolModes}$ is the reduced state with $\numSolModes$ representing the number of trial basis modes. In this work, $\trialBasisCons$ is computed via the proper orthogonal decomposition (POD)~\cite{Lumley1997} from the singular value decomposition (SVD), which is a solution to
\begin{equation}
  \min_{\trialBasisCons \in \mathbb{R}^{\numDOF \times \numSolModes}} ||\dataMatCons - \trialBasisCons \trialBasisCons^T \dataMatCons ||_F\ \ s.t. \ \ \trialBasisCons^T \trialBasisCons = \mathbf{I},
    \label{pod:orthogonality_consv}
\end{equation}
Here, $\dataMatCons$ is a data matrix in which each column is a snapshot of the solution $\solConsFOMUnst$ at different time instances. A scaling matrix, $\scaleMatCons \in{\mathbb{R}^{\numDOF \times \numDOF}}$, must be applied to $\solConsFOMUnst$ such that the variables corresponding to different physical quantities in the data matrix $\dataMatCons$ have similar orders of magnitude. Otherwise, $\dataMatCons$ may be biased by physical quantities of higher magnitudes (e.g. total energy). In this work, we normalize all quantities by their $L^2$-norm, as proposed by Lumley and Poje~\cite{Lumley1997}
\begin{equation}
    \scaleMatCons = diag \lp \scaleMatCons_1, \ldots, \scaleMatCons_i, \ldots, \scaleMatCons_{\numElements} \rp,
    \label{pod:normalization_consv}
\end{equation} 
where $\scaleMatCons_i = diag\left( \varphi^{-1}_{1,norm}, \ldots , \varphi^{-1}_{\numVars,norm} \right)$. Here, $\varphi_{\normVarIdx,norm}$ represents the $\normVarIdx^{th}$ state  variable and
\begin{equation}
    \varphi_{\normVarIdx,norm} = {\frac{1}{\Delta \text{T}}\int^{t_0 + \Delta T}_{\timeVar_0} \frac{1}{\Omega} \int_{\Omega} \varphi'^2_\normVarIdx(\mathbf{x}, \timeVar) \; \text{d}\mathbf{x} \; \text{dt}}.
\end{equation}

\subsection{Galerkin Projection}
\label{galerkin_rom}
Model reduction via Galerkin projection is formulated for the continuous-time representation of the FOM (Eq.~\ref{cfd:discretized}). This is done by first scaling Eq.~\ref{cfd:discretized} using the scaling matrix $\scaleMatCons$, then projecting onto the test space $\trialSpaceCons$
\begin{equation}
    \trialBasisCons^T \scaleMatCons \frac{\text{d} \solConsFOM}{\dTimeVar} = \trialBasisCons^T \scaleMatCons \rhsFunc{\solConsFOM, \timeVar}.
    \label{rom:galerkin}
\end{equation}
This specification that the trial and test spaces are identical defines Galerkin projection. The scaling of Eq.~\ref{cfd:discretized} is necessary to ensure that each equation makes similar contributions to the reduced system after projection. Otherwise, the reduced system may be biased by equations for quantities of higher magnitudes (e.g. the energy equation), exacerbating floating-point errors. With the low-rank representation in Eq.~\ref{pod:expansion_consv}, $\solConsROMFull(\timeVar) \triangleq \solConsFOMRef+\scaleMatCons^{-1}\trialBasisCons \solConsROMRed(\timeVar)$, substituted into Eq.~\ref{rom:galerkin}, a reduced-order ODE system can be obtained 
\begin{equation}
    \frac{\text{d}{\solConsROMRed}}{\dTimeVar} = \trialBasisCons^T \scaleMatCons \rhsFunc{\solConsROMFull, \timeVar}, \ \ \solConsROMRed(0) = \trialBasisCons^T\solConsFOM_0.
    \label{rom:galerkin_ode}
\end{equation}
 The dimension of the ROM ODE is $\numSolModes$, which can be orders of magnitudes smaller than  $\numDOF$  in Eq.~\ref{cfd:discretized}.

\subsection{Least-squares Petrov-Galerkin Projection}
\label{slspg_rom}
Least-squares Petrov-Galerkin (LSPG) projection~\cite{Carlberg2017} is formulated from the discrete-time representation of Eq.~\ref{cfd:discretized}. The objective of LSPG is to minimize the fully-discrete FOM equation residual, $\res$, defined in Eqs.~\ref{EqRes:LinearMultiStep} and~\ref{EqRes:RK}, with respect to the state, $\solConsFOM$, as approximated in the trial space $\trialSpaceCons$,  $\solConsROMFull = \solConsFOMRef + \scaleMatCons^{-1} \trialBasisCons \solConsROMRed$.
The problem statement of LSPG is to seek a solution to the minimization problem
\begin{equation}
    \solConsROMFull^\iterIdx \triangleq  \argmin_{ \solConsROMFull^\iterIdx \in \textrm{Range}(\trialBasisCons)} \norm{\scaleMatCons \resFunc{\solConsROMFull^\iterIdx}}^2,
    \label{rom:slspg_def}
\end{equation}
with the equation residual, $\res$, scaled by $\scaleMatCons$ such that each equation in $\res$ has similar contributions to the minimization. The norm used is the the Euclidean vector-induced matrix norm $L_{2,2}$, which we will represent by $\norm{\cdot}.$

The solution of this minimization problem seeks to satisfy
\begin{equation}
    \lp \testBasisCons^\iterIdx \rp^T \scaleMatCons \resFunc{\solConsROMFull^\iterIdx} = \zeroVec,
    \label{rom:slspg_proj}
\end{equation}
where $\testBasisCons^{\iterIdx}$ is the Petrov-Galerkin projection test basis given by
\begin{equation}
    \testBasisCons^\iterIdx = \frac{\partial \scaleMatCons \resFunc{\solConsROMFull^\iterIdx}}{\partial \solConsROMRed^\iterIdx}.
    \label{rom:slspg_w}
\end{equation}

For linear multi-step methods, the equation residual in Eq.~\ref{EqRes:LinearMultiStep} yields the test basis
\begin{equation}
    \testBasisCons^\iterIdx = \scaleMatCons \lp \mathbf{I} - \dt \beta_0 \jacobConsROM^\iterIdx \rp \scaleMatCons^{-1} \trialBasisCons,
    \label{rom:slspg_w_LinearMultistep}
\end{equation}
where $\jacobConsROM^\iterIdx = \left[\partial \rhs / \partial \solConsFOM \right]^\iterIdx_{\solConsFOM = \solConsROMFull}$. For explicit schemes ($\beta_0 = 0$), $\testBasisCons^\iterIdx = \trialBasisCons$, thus reverting to Galerkin projection. Similarly, for Runge--Kutta methods, the equation residual in Eq.~\ref{EqRes:RK} yields the test basis
\begin{equation}
    \testBasisCons^\iterIdx = \trialBasisCons,
    \label{rom:slspg_w_RK}
\end{equation}
for explicit and diagonally implicit schemes, which is identical to the trial basis as in Galerkin projection. For implicit Runge--Kutta methods,
\begin{equation}
    \testBasisCons^\iterIdx = \scaleMatCons \lp \mathbf{I} - \dt b_s \jacobConsROM^\iterIdx \rp \scaleMatCons^{-1} \trialBasisCons.
    \label{rom:slspg_w_RK_imp}
\end{equation}
\section{Stability
of Least-Squares-based ROMs for Linear Problems}
\label{proof}
In this section, we provide a complete proof of least-squares-based ROMs for a linear time-invariant (LTI) system. While the finite element community is rich with literature on the analysis of least-squares-based techniques (e.g.~\cite{LSFEM}), stability analysis in the ROM literature is not commonly found. Pursuing a viewpoint of contractivity, stability preservation for projection-based model order reduction in continuous time is presented in ~\cite{Selga_stabilityPreservation2012}. In contrast, we pursue a linear algebra standpoint in discrete time, which will be useful for the analysis of numerical results in Section~\ref{2d:LinearStability}.

For the full order model, we assume an autonomous LTI system given by
\begin{equation}
    \frac{\text{d} \solConsFOM}{\dTimeVar} = \ltiMatA \solConsFOM, \ \ \solConsFOM(0) = \solConsFOM_0,
    \label{LTI}
\end{equation}
where $\solConsFOM \in {\mathbb{R}^{\numDOF}}$, and $\ltiMatA \in \mathbb{R}^{\numDOF \times \numDOF}$. 
For clarity of presentation, we will consider a backward Euler discretization for the FOM:

\begin{align}
  \solConsFOM^\iterIdx  &= (\mb{I} - \dt \ltiMatA)^{-1} \solConsFOM^{\iterIdx-1}, \ \  \solConsFOM^0 = \solConsFOM_0.
  \label{LTI:discrete}
\end{align}

Consider a low-rank representation $\solConsROMFull(\timeVar) \triangleq \trialBasisCons \solConsROMRed(\timeVar)$, where $\trialBasisCons \in \mathbb{R}^{\numDOF \times \numSolModes}$ and $\solConsROMRed \in \mathbb{R}^{\numSolModes}$. Note that the main conclusions also hold when $\solConsROMFull (\timeVar) \triangleq \solConsFOMRef + \scaleMatCons^{-1} \trialBasisCons \solConsROMRed (\timeVar)$.

\subsection{Least-Squares Petrov Galerkin ROM}
Following the development in Section~\ref{slspg_rom}, we get
\begin{equation}
\testBasisCons^T \frac{ \solConsROMFull^\iterIdx - \solConsROMFull^{\iterIdx-1} }{\dt} =  \testBasisCons^T \ltiMatA \solConsROMFull^\iterIdx,
\label{LTI:ROMT}
\end{equation}
where $\testBasisCons \triangleq (\mb{I} - \dt \ltiMatA) \trialBasisCons.$ Manipulating further, we get the governing equations for the modal coefficients in the form
\begin{equation}
(\testBasisCons^T \testBasisCons) \solConsROMRed^\iterIdx =  \testBasisCons^T \trialBasisCons \solConsROMRed^{\iterIdx-1}, \ \ \solConsROMRed^0 = \trialBasisCons^T \solConsFOM_0.
\label{LTI:ROM}
\end{equation}
 We will now prove the stability of the ROM (Eq.~\ref{LTI:ROM}) under the condition that the FOM (Eq.~\ref{LTI:discrete}) is linearly stable. 

We will pursue a notion of stability via the Euclidean vector-induced matrix norm $\norm{\cdot}.$ We will use $\sigma_\genIdx(\mb{A})$ to represent the $\genIdx^{th}$ singular value of a matrix $\mb{A}$. For a real symmetric matrix $\mb{S}$, we will use $\lambda_\genIdx(\mb{S})$ to represent the $\genIdx^{th}$ eigenvalue of $\mb{S}$, with the  eigenvalues arranged in descending order.

\vspace{0.75cm}

\noindent{\bf Theorem 1} : If the FOM given by Equation~\ref{LTI:discrete} is asymptotically stable in the sense of $\norm{(\mb{I} - \dt \ltiMatA)^{-1}} \leq 1$, then the ROM given by Equation~\ref{LTI:ROM} is also asymptotically stable with no further assumptions required.
\vspace{0.25cm}
\noindent {\em Proof}:
The problem statement reduces to showing that
\begin{equation}
\norm{(\testBasisCons^T \testBasisCons)^{-1} \testBasisCons^T \trialBasisCons}  \leq 1,
\end{equation}
with $\testBasisCons\triangleq \ltiMatB\trialBasisCons$, where $\ltiMatB \triangleq \mb{I} - \dt \ltiMatA$, given $\norm{\ltiMatB^{-1}} \leq 1.$

Since $\ltiMatB$ and $\trialBasisCons$ are both full-rank matrices, $\testBasisCons$ is also a full-rank matrix, and the left pseudo-inverse $\testBasisCons^+ \triangleq (\testBasisCons^T \testBasisCons)^{-1} \testBasisCons^T$ also exists. Therefore, given $\norm{\testBasisCons^+}= \frac{1}{\sigma_{\numSolModes}(\testBasisCons)}$ we have 
\begin{equation}
 \norm{(\testBasisCons^T \testBasisCons)^{-1} \testBasisCons^T \trialBasisCons} = \norm{\testBasisCons^+   \trialBasisCons} \leq  \norm{\testBasisCons^+} \norm{\trialBasisCons}
 = \frac{1}{\sigma_{\numSolModes}(\testBasisCons)}
 \label{eq:proof}
\end{equation}
and thus our problem reduces to showing 

\begin{equation}
    \sigma_{\numSolModes}(\testBasisCons) = \sigma_{\numSolModes}(\ltiMatB \trialBasisCons) \geq 1, \ \ \textrm{given} \ \ \sigma_{\numDOF}(\ltiMatB) \geq 1.
    \label{eq:neweq}
\end{equation}

We write $\testBasisCons^T \testBasisCons = \trialBasisCons^T \ltiMatB^T \ltiMatB \trialBasisCons$
and use the Poincar\'{e} separation theorem~(\ref{appendix:poincare})  which yields
\begin{equation}
\lambda_{\numSolModes}(\testBasisCons^T \testBasisCons) \geq \lambda_{\numDOF} (\ltiMatB^T \ltiMatB).
\end{equation}
Since $\lambda_{\numSolModes}(\testBasisCons^T \testBasisCons) = \sigma_{\numSolModes}(\testBasisCons)$ and  $\lambda_{\numDOF}(\ltiMatB^T \ltiMatB) = \sigma_{\numDOF}(\ltiMatB)$, we have
\begin{equation}
\sigma_{\numSolModes}(\testBasisCons) \geq \sigma_{\numDOF} (\ltiMatB).
\end{equation}
Since $\sigma_{\numDOF} (\ltiMatB) \geq 1$, we have $\sigma_{\numSolModes}(\testBasisCons) \geq 1$. This completes the proof following Equation~\ref{eq:proof}. 

\vspace{0.75cm}

\subsection{Galerkin ROMs}

Galerkin projection  on Eq.~\ref{LTI:discrete} yields
\begin{equation}
(\mb{I} - \dt \trialBasisCons^T \ltiMatA \trialBasisCons)\solConsROMRed^\iterIdx = \solConsROMRed^{\iterIdx-1}.
\label{LTI:ROMG}
\end{equation}
Stability demands $\norm{(\mb{I} - \dt \trialBasisCons^T \ltiMatA \trialBasisCons)^{-1}} \leq 1$. This can be written as $\norm{(\trialBasisCons^T \ltiMatB \trialBasisCons)^{-1}} \leq 1$, with $\ltiMatB \triangleq \mb{I} - \dt \ltiMatA$. 

$\bullet$ If $\ltiMatA$ is a symmetric matrix, $\norm{(\trialBasisCons^T \ltiMatB \trialBasisCons)^{-1}} = \frac{1}{\lambda_{\numSolModes}(\trialBasisCons^T \ltiMatB \trialBasisCons)}$. Also note that $\ltiMatA$ is negative definite, $\ltiMatB$ will be positive definite. Thus, we have
\begin{equation}
\lambda_{\numSolModes}(\trialBasisCons^T \ltiMatB \trialBasisCons) \geq \lambda_{\numDOF}(\ltiMatB) \geq 1,
\end{equation}
where the first inequality is due to the Poincar\'{e} separation theorem, and the second one is due to the fact that $\ltiMatA$ is negative-definite. Thus, given that the FOM is asymptotically stable, the Galerkin ROM is asymptotically stable with no further assumptions required.

$\bullet$ If $\ltiMatA$ is not a symmetric matrix, then $\norm{(\trialBasisCons^T \ltiMatB \trialBasisCons)^{-1}} = \frac{1}{\sigma_{\numSolModes}(\trialBasisCons^T \ltiMatB \trialBasisCons)}$ and thus we have to show that $\sigma_{\numSolModes}(\trialBasisCons^T \ltiMatB \trialBasisCons) \ge 1$. While a convenient upper bound exists (refer~\ref{appendix:singValIdent}) $\sigma_{\numSolModes}(\trialBasisCons^T \ltiMatB \trialBasisCons) \leq \sigma_{\numSolModes}(\ltiMatB)$,  tight lower bounds do not exist for general non-symmetric $\ltiMatB$. Thus unconditional stability (or instability) cannot be proven, and a case-by-case approach is required, depending on the structure of $\ltiMatA$ and the time-step $\dt$.  It should, however, be noted that, even when the FOM represents a discretization of Hyperbolic PDEs, the underlying numerical dissipation in the FOM, coupled with a small time-step $\dt$ can stabilize Galerkin ROMs depending on the level of mode truncation.

\vspace{0.75cm}

\noindent{\bf Theorem 2} : If the FOM given by Equation~\ref{LTI:discrete} is asymptotically stable in the sense of $\norm{(\mb{I} - \dt \ltiMatA)^{-1}} \leq 1$, then the Galerkin ROM given by Equation~\ref{LTI:ROMG} is also asymptotically stable if $\lambda_{\numDOF}\left( \mb{I} - \frac{1}{2} \dt (\ltiMatA + \ltiMatA^T)\right) \geq 1.$

\vspace{0.25cm}
\noindent {\em Proof:} Define $\ltiMatB \triangleq \mb{I} - \dt \ltiMatA$. The goal is to show $\sigma_{\numSolModes}(\trialBasisCons^T \ltiMatB \trialBasisCons) \geq 1.$

From~\ref{appendix:evalsHermitian}, 
\begin{equation}
\sigma_\genIdx(\trialBasisCons^T \ltiMatB \trialBasisCons) \geq \lambda_\genIdx \left(\trialBasisCons^T \frac{\ltiMatB + \ltiMatB^T}{2} \trialBasisCons \right)
\end{equation}

From~\ref{appendix:poincare}, 
\begin{equation}
\lambda_\genIdx \left(\trialBasisCons^T \frac{\ltiMatB + \ltiMatB^T}{2} \trialBasisCons  \right) \geq \lambda_{\numDOF - \numSolModes +\genIdx} \left(\frac{\ltiMatB + \ltiMatB^T}{2}\right).
\end{equation}

Therefore
\begin{equation}
\sigma_{\numSolModes}(\trialBasisCons^T \ltiMatB \trialBasisCons) \geq \lambda_{\numDOF} \left( \frac{\ltiMatB + \ltiMatB^T}{2}\right)
\end{equation}
and stability demands
\begin{equation}
\lambda_{\numDOF} \left(\frac{\ltiMatB + \ltiMatB^T}{2}\right) \geq 1.
\end{equation}
\vspace{0.75cm}

\noindent Note: It can be easily verified, for instance, that a Galerkin ROM of a linear upwind spatial discretization of the linear advection equation with Euler implicit time integration is unconditionally stable. 

\section{{\color{black}Model-form Preserving} Model Reduction for Transformed Solution Variables}
\label{SPLSPG}
In this section, we introduce the {\color{black}model-form preserving} least-squares with variable transformation ({\color{black}MP-LSVT}) ROM formulation. Since the targeted applications are complex multi-scale problems, this formulation involves a number of steps which are detailed below. 

\subsection{Semi-discrete Formulation  for Transformed Solution Variables}
\label{fom_qp}
It is typical to develop projection-based model reduction methods (Section~\ref{StandardROM}) for the state variables used  in the full-order model. In many circumstances, motivated either by physical or numerical considerations, it can be beneficial to solve the governing equations based on an alternate set of variables, which we denote generally by \solPrimFOM. 

The following are representative examples of variable transformations that serve as inspirations for the present work:
\begin{enumerate}
   
\item Kramer and Willcox ~\cite{Kramer2019LiftAndLearn,KramerBalancedROM}, Swischuk et al.~\cite{Swischuk2020AIAAJ_LL} and Qian et al.~\cite{qian2020lift} define a coordinate transformation to rewrite the governing equations in quadratic form. This simplified structure is leveraged to develop non-intrusive - yet interpretable - ROM operators. 

\item Similarly, Pettersson et al.~\cite{pettersson2014stochastic} use the Roe variables~\cite{Roe1981} to simplify the non-linearity of the compressible Euler equations. The simplified equations are used to develop intrusive stochastic Galerkin techniques for uncertainty quantification.

\item In contrast to hand-crafted variable transformations above, Pan et al.~\cite{pan2020physics} and Champion et al.~\cite{champion2019data} use neural network-based autoencoders to discover a latent space in which the dynamics is linear.

\item In computational fluid dynamic solvers, actions such as solution reconstruction and slope-limiting~\cite{cfd1-1,cfd1-2} and time marching updates~\cite{cfd2,Pulliam} are performed in terms of non-conservative variables, rather than the  conserved variables. For compressible flows, especially in reacting flow applications, a primitive variable formulation~\cite{HarvazinskiPoF} allows for easier computation of thermodynamic and transport properties and arbitrary equations of state~\cite{MerkleAIAAJ}. It also provides more flexibility when extending applications to complex fluid problems like liquid and supercritical fluids~\cite{Oefelein_supercrtical2005,Oefelein_supercrtical2006}. In addition, directly updating quantities such as pressure and temperature at the time-step (or sub-iteration) level can be beneficial from the viewpoint of enforcing physical realizability (e.g. positivity) and robustness.

\end{enumerate}

Influenced by the above examples, we describe a transformation to an arbitrary, but complete, set of state variables $\solPrimFOM$. In contrast to lifting-based approaches (items 1, 2, and 3 above), however, we do so in a manner so as to preserve the equation structure of the full-order model.

We begin by re-writing Eq.~\ref{cfd:discretized} as
\begin{equation}
    \frac{\text{d} \solConsFOM(\solPrimFOM)}{\dTimeVar} = \rhsFunc{\solPrimFOM, \timeVar},
    \label{cfd:discretized_qp}
\end{equation}
where $\solConsFOM$ $: \mathbb{R}^{\numDOF} \rightarrow \mathbb{R}^{\numDOF}$  and $\rhs$ $: \mathbb{R}^{\numDOF} \times [0,T] \rightarrow \mathbb{R}^{\numDOF}$ are functions of the solution variables, $\solPrimFOM$ $: [0, T] \rightarrow {\mathbb{R}^{\numDOF}}$. Note that, in general, computing $\rhs$ often requires a combination of the conserved and solution variables, and the notation here is simply one of convenience.  Defining the Jacobian of the transformation $\gm \triangleq {\partial \solConsFOM} / {\partial \solPrimFOM}$, applying the chain rule, and applying the left inverse of $\gm$ results in the following equation
\begin{equation}
    \frac{\text{d}\solPrimFOM}{\dTimeVar} = \gmInv \rhsFunc{\solPrimFOM, \timeVar} = \rhsFuncPrim{\solPrimFOM,\timeVar}.
    \label{cfd:discretized_qp_ge}
\end{equation}
Although this is formally equivalent to Eq.~\ref{cfd:discretized_qp} (assuming differentiability) in the semi-discrete setting, it will not be conservative in the fully-discrete setting. Thus, a more careful approach is required to formulate the discrete representation.

With arbitrary solution variables, the fully-discrete FOM equation residual for linear multi-step method in Section~\ref{fom} becomes
\begin{equation}
    \resFunc{\solPrimFOM^\iterIdx} = \solConsFOMFunc{\solPrimFOM^{\iterIdx}} + \sum^{\itermaxLinMS}_{j=1} \alpha_j \solConsFOMFunc{\solPrimFOM^{\iterIdx-j}} - \dt \beta_0 \rhsFunc{\solPrimFOM^\iterIdx, \timeVar^\iterIdx} - \dt \sum^{\itermaxLinMS}_{j=1} \beta_j \rhsFunc{\solPrimFOM^{\iterIdx-j}, \timeVar^{\iterIdx-j}}.
    \label{EqRes:LinearMultiStep_qp}
\end{equation}
The fully discrete FOM equation residual for Runge--Kutta method with arbitrary solution variables becomes
\begin{equation}
    \resFunc{\solPrimFOM^\iterIdx} = \solConsFOMFunc{\solPrimFOM^{\iterIdx}} - \solConsFOMFunc{\solPrimFOM^{\iterIdx-1}} - \dt \sum^{\itermaxRK}_{j=1} b_j \mathbf{s}_j,
    \label{EqRes:RK_qp}
\end{equation}
where $\mathbf{s}_1 \triangleq \rhsFunc{\solPrimFOM^{\iterIdx-1}, \timeVar^{\iterIdx-1}}$ and
\begin{equation}
    \mathbf{s}_j = \rhsFunc{\solPrimFOM^{\iterIdx-1} + \dt \sum^{j-1}_{m=1} a_{jm} \gmInv_m \mathbf{s}_m, \timeVar^{\iterIdx-1} + c_j \dt},
    \label{cfd:RK_stage_qp}
\end{equation}
with $\gm_1 \triangleq \gm \lp \solPrimFOM^{\iterIdx-1}, \timeVar^{\iterIdx-1} \rp$ and
\begin{equation}
    \gm_j = \gm \lp \solPrimFOM^{\iterIdx-1} + \dt \sum^{j-1}_{m=1} a_{jm} \gmInv_m \mathbf{s}_m, \timeVar^{\iterIdx-1} + c_j \dt \rp.
    \label{cfd:GammaInv_qp}
\end{equation}

\subsection{Construction of POD Bases for Arbitrary Solution Variables}

Similar to Section~\ref{pod:consv}, the state $\solPrimFOM$ can be expressed in a trial space $\trialSpacePrim \triangleq \text{Range}(\trialBasisPrim)$, where $\trialBasisPrim \in \mathbb{R}^{\numDOF \times \numSolModes}$ is the trial basis matrix. Define $\solPrimFOMUnst(\timeVar) \triangleq \solPrimFOM(\timeVar) - \solPrimFOMRef$, where \solPrimFOMRef\ is a reference state. Possible reference states include the initial FOM solution, $\solPrimFOMRef = \solPrimFOM(\timeVar = \timeVar_0)$, or the time-averaged FOM solution, $\solPrimFOMRef = \frac{1}{\Delta \text{T}} \int_{\timeVar_0}^{\timeVar_0 + \Delta \text{T}} \solPrimFOM(\timeVar) \; \dTimeVar$. 

We then seek a representation $\solPrimROMFull: [0,T] \rightarrow \trialSpacePrim$ such that
\begin{equation}
    \scaleMatPrim \lp \solPrimROMFull - \solPrimFOMRef \rp = \trialBasisPrim \solPrimROMRed.
    \label{pod:expansion_qp}
\end{equation}
Here, $\solPrimROMRed: [0,T] \rightarrow \mathbb{R}^{\numSolModes}$ is the reduced state, and $\numSolModes$ represents the number of trial basis modes. $\trialBasisPrim$ is the POD basis derived by the SVD from the FOM snapshots of $\solPrimFOMUnst$, similar to the solution of Eq.~\ref{pod:orthogonality_consv}.

Similar to Eq.~\ref{pod:normalization_consv}, a scaling matrix, $\scaleMatPrim \in{\mathbb{R}^{\numDOF \times \numDOF}}$, must be applied to scale $\solPrimFOMRef$ so that the numerical values of physical quantities (e.g. pressure, velocities, temperature and species mass fraction) have similar orders of magnitude in generating the POD basis. Again, in this work we choose to normalize all quantities by their $L^2$-norm
\begin{equation}
    \scaleMatPrim = diag \lp \scaleMatPrim_1, \ldots, \scaleMatPrim_i, \ldots, \scaleMatPrim_{\numElements} \rp,
    \label{pod:normalization_qp}
\end{equation} 
where $\scaleMatPrim_i = diag \left(\phi^{-1}_{1,norm}, \ldots , \phi^{-1}_{\numVars,norm} \right)$. Here, $\phi_{\normVarIdx,norm}$ represents the $\normVarIdx^{th}$ solution variable and
\begin{equation}
    \phi_{\normVarIdx,norm} = {\frac{1}{\Delta \text{T}} \int^{\timeVar_0 + \Delta \text{T}}_{\timeVar_0} \frac{1}{\Omega} \int_{\Omega} \phi'^2_\normVarIdx (\mathbf{x}, \timeVar) \; \text{d}\mathbf{x} \; \dTimeVar}.
\end{equation}

\subsection{Least-squares with Variable Transformation}
\label{lspg-vt}
As mentioned in Section~\ref{slspg_rom}, standard LSPG projection~\cite{Carlberg2017} requires an implicit, discrete-time representation of the governing equations. Otherwise, LSPG projection is equivalent to Galerkin projection. In this work, we seek to develop a {\color{black}model-form preserving} least-squares formulation for an arbitrary selection of solution variables as in Eq.~\ref{cfd:discretized_qp}, and for both explicit and implicit time discretization schemes.

Our objective is to minimize the fully-discrete FOM equation residual $\res$, defined in Eqs.~\ref{EqRes:LinearMultiStep_qp} and~\ref{EqRes:RK_qp} with respect to the approximate solution variables, $\solPrimROMFull = \solPrimFOMRef + \scaleMatPrim^{-1} \trialBasisPrim \solPrimROMRed$. It should be noted that for a fully-converged FOM $\resFunc{\solPrimFOM^n} = 0$, but $\resFunc{\solPrimROMFull^n}$ is not necessarily equal to zero, as the ROM may be under-resolved. As in the LSPG technique, we seek to minimize the residual
\begin{equation}
    \solPrimROMFull^n \triangleq  \argmin_{\solPrimROMFull^n \in \textrm{Range}(\trialBasisPrim)} \norm{\scaleMatCons \resFunc{\solPrimROMFull^n}}^2,
    \label{rom:lspg_def}
\end{equation}
with the equation residual, $\res$, scaled by $\scaleMatCons$ such that each equation in $\res$ has similar contributions to the minimization problem in Eq.~\ref{rom:lspg_def}. Similar to Eq.~\ref{rom:slspg_proj},  a reduced non-linear equation system of dimension $\numSolModes$ can be obtained and viewed as the result of a Petrov-Galerkin projection
\begin{equation}
    \lp \testBasisPrim^\iterIdx \rp^T \scaleMatCons \resFunc{\solPrimROMFull^n} = \zeroVec,
    \label{rom:lspg_proj}
\end{equation}
where $\testBasisPrim$ is the test basis
\begin{equation}
    \testBasisPrim^\iterIdx =  \frac{\partial \scaleMatCons \resFunc{\solPrimROMFull^n}}{\partial \solPrimROMRed^n}.
    \label{rom:lspg_w}
\end{equation}

We point out that two distinct scaling matrices, $\scaleMatCons$ and $\scaleMatPrim$, are introduced, so that each equation in FOM equation residual contributes equally to the minimization problem in Eq.~\ref{rom:lspg_def} and the POD bases are appropriately generated. In standard model reduction methods only one scaling matrix, $\scaleMatCons$, is used (as seen in Section~\ref{StandardROM}).

We now develop ROMs for the governing equations with transformed solution variables solved via linear multi-step and Runge--Kutta methods.  More detailed derivations of the results below can be found in~\ref{appendix:SP-LSVTTestFunctionPhysicalTime}. 
For linear multi-step methods, the residual in Eq.~\ref{EqRes:LinearMultiStep_qp} yields the test function
\begin{equation}
    \testBasisPrim^\iterIdx = \scaleMatCons \lp \gmROM^\iterIdx - \dt \beta_0 \jacobConsROM^\iterIdx \gmROM^\iterIdx \rp \scaleMatPrim^{-1} \trialBasisPrim,
    \label{rom:lspg_w_LinearMultistep}
\end{equation}
where $\jacobConsROM^\iterIdx = \left[\partial \rhs / \partial \solConsFOM \right]^\iterIdx_{\solPrimFOM = \solPrimROMFull}$ and $\gmROM^\iterIdx = \left[\partial \solConsFOM / \partial \solPrimFOM \right]^\iterIdx_{\solPrimFOM=\solPrimROMFull}$. It is worthwhile to point out that for an explicit scheme ($\beta_0 = 0$),
$\testBasisPrim^\iterIdx = \scaleMatCons \gmROM^\iterIdx \scaleMatPrim^{-1} \trialBasisPrim$, which is not equivalent to Galerkin projection. Rather, the two projection methods are only equivalent for an explicit integrator when the solution variables are chosen to be the same as those in the semi-discrete form of the equations (i.e. $\solConsFOM$ for Eq.~\ref{cfd:discretized} and $\solPrimFOM$ for Eq.~\ref{cfd:discretized_qp_ge} with $\gmROM^\iterIdx = \mathbf{I}$ and $\scaleMatPrim = \scaleMatCons$). 

Similarly for Runge--Kutta methods, the equation residual in Eq.~\ref{EqRes:RK_qp} yields the test function
\begin{equation}
    \testBasisPrim^\iterIdx = \scaleMatCons \gmROM^\iterIdx \scaleMatPrim^{-1} \trialBasisPrim,
    \label{rom:lspg_w_RK}
\end{equation}
for explicit and diagonally implicit schemes, which is identical to the test function for an explicit linear multi-step time integrator. For implicit Runge--Kutta methods, the test function becomes
\begin{equation}
    \testBasisPrim^\iterIdx = \scaleMatCons \lp \gmROM^\iterIdx - \dt b_s \jacobConsROM^\iterIdx \gmROM^\iterIdx \rp \scaleMatPrim^{-1} \trialBasisPrim,
    \label{rom:lspg_w_RK_imp}
\end{equation}

\subsection{Solution of Eq.~\ref{rom:lspg_proj} by Newton’s method}
For the numerical results presented in the current work, Newton's method is used to solve Eq.~\ref{rom:lspg_proj}. We derive the iterative procedure for solving {\color{black}MP-LSVT} ROMs to highlight one important property that the method provides -- that is, the symmetrization of the resulting ROM. We introduce a sub-iteration variable $\subIdx$, and define $\yFunc{\solPrimROMRed^{\subIdx-1}} \triangleq(\testBasisPrim^{\subIdx-1})^T \scaleMatCons \resFunc{\solPrimROMFull^{\subIdx-1}}$ for the sake of compactness. Applying Newton's method to Eq.~\ref{rom:lspg_proj} gives

\begin{equation}
    \frac{\partial \yFunc{\solPrimROMRed^{\subIdx-1}}}{\partial \solPrimROMRed^{\subIdx-1}} \lp \solPrimROMRed^k - \solPrimROMRed^{\subIdx-1} \rp + \yFunc{\solPrimROMRed^{\subIdx-1}} = \zeroVec,
    \label{rom:lspg_proj_newton}
\end{equation}
where 
\begin{equation}
    \frac{\partial \yFunc{\solPrimROMRed^{\subIdx-1}}}{\partial \solPrimROMRed^{\subIdx-1}} = \lp \testBasisPrim^{\subIdx-1} \rp^T \frac{\partial \scaleMatCons \resFunc{\solPrimROMFull^{\subIdx-1}}}{\partial \solPrimROMRed^{\subIdx-1}} = \lp \testBasisPrim^{\subIdx-1} \rp^T \testBasisPrim^{\subIdx-1}.
    \label{rom:lspg_yJacob}
\end{equation}
Therefore, Eq.~\ref{rom:lspg_proj_newton} can be readily seen as a symmetrized  system. Inserting Eq.~\ref{rom:lspg_yJacob} into Eq.~\ref{rom:lspg_proj_newton}, we arrive at the simplified form of Newton's method applied to {\color{black}MP-LSVT} ROMs
\begin{equation}
    \lp \testBasisPrim^{\subIdx-1} \rp^T \testBasisPrim^{\subIdx-1} \lp \solPrimROMRed^k - \solPrimROMRed^{\subIdx-1} \rp + \lp \testBasisPrim^{\subIdx-1} \rp^T \scaleMatCons \resFunc{ \solPrimROMFull^{\subIdx-1}} = \zeroVec.
    \label{rom:lspg_proj_newton_symm}
\end{equation}

\subsection{Solution of Eq.~\ref{rom:lspg_proj} by Pseudo-time Stepping}
\label{lspg-vt-pseudo}
Instead of using Newton's method, pseudo-time stepping~\cite{sankaranDualTime1995,Pulliam} can be directly employed to solve Eq.~\ref{rom:lspg_proj} to achieve further robustness in simulating highly stiff multi-scale problems
\begin{equation}
    \dt \frac{\text{d} \yFunc{\solPrimROMRed}}{\dPTimeVar} + \yFunc{\solPrimROMRed} = 0,
    \label{rom:lspg_proj_dualTime}
\end{equation}
where $\yFunc{\solPrimROMRed} = \testBasisPrim^T \scaleMatCons \resFunc{\solPrimROMFull}$ and the pseudo-time ($\tau$) derivative is represented using the implicit Euler method in the following form
 \begin{equation}
     \left. \frac{\text{d} \yFunc{\solPrimROMRed}}{\dPTimeVar} \right\rvert_{\tau = \tau_k} = \left. \lp \frac{\partial \yFunc{\solPrimROMRed}}{\partial \solPrimROMRed} \frac{\text{d} \solPrimROMRed}{\dPTimeVar} \rp \right\rvert_{\tau = \tau_k} = \lp \testBasisPrim^{\subIdx-1} \rp^T \testBasisPrim^{\subIdx-1} \left. \frac{\text{d}\solPrimROMRed}{\dPTimeVar} \right\rvert_{\tau = \tau_k}  \approx \lp \testBasisPrim^{\subIdx-1} \rp^T \testBasisPrim^{\subIdx-1} \frac{\lp \solPrimROMRed^{\subIdx} - \solPrimROMRed^{\subIdx-1} \rp}{\dtau}.
 \end{equation}
Then Eq.~\ref{rom:lspg_proj_dualTime} becomes
\begin{equation}
    \lp \testBasisPrim^{\subIdx-1} \rp^T \testBasisPrim^{\subIdx-1} \frac{\dt}{\dtau} \lp \solPrimROMRed^{\subIdx} - \solPrimROMRed^{\subIdx-1} \rp + \yFunc{\solPrimROMRed^{\subIdx}} = \zeroVec,
    \label{rom:lspg_proj_dualTime_second}
\end{equation}
Linearizing $\yFunc{\solPrimROMRed^{\subIdx}}$
\begin{equation}
    \yFunc{\solPrimROMRed^{\subIdx}} \approx \yFunc{\solPrimROMRed^{\subIdx - 1}} + \frac{\partial \yFunc{\solPrimROMRed}}{\partial \solPrimROMRed} \lp \solPrimROMRed^{\subIdx} - \solPrimROMRed^{\subIdx - 1} \rp,
\end{equation}
and collecting terms, the final form of Eq.~\ref{rom:lspg_proj_dualTime_second} becomes
\begin{equation}
    \lp \frac{\dt}{\dtau} + 1 \rp \lp \testBasisPrim^{\subIdx-1} \rp^T \testBasisPrim^{\subIdx-1} \lp \solPrimROMRed^{\subIdx} - \solPrimROMRed^{\subIdx-1} \rp + \lp \testBasisPrim^{\subIdx-1} \rp^T \scaleMatCons \resFunc{\solPrimROMFull^{\subIdx-1}} = \zeroVec,
    \label{rom:lspg_proj_dualTime_final}
\end{equation}
which leads to the same symmetrized system as Newton’s method except that the left-hand-side operator is multiplied by the term, $\lp \frac{\dt}{\dtau} + 1 \rp > 1$.  Thus, for finite ${\dtau}$, the solution change is reduced in the same manner as when line search is incorporated in Newton’s method. In highly nonlinear equations, optimum convergence is obtained with finite ${\dtau}$. Also, it shall be pointed out that Eqs.~\ref{rom:lspg_proj_newton_symm} and~\ref{rom:lspg_proj_dualTime_final} are different iterative methods to solve for Eq.~\ref{EqRes:LinearMultiStep_qp_dualtime} and lead to the same solutions when convergence is achieved (i.e., $(\solPrimROMRed^{\subIdx} - \solPrimROMRed^{\subIdx-1}) \rightarrow \zeroVec$).

\subsection{Solution of Eq.~\ref{cfd:discretized_qp} by Pseudo-time Stepping}
Instead of using the pseudo-time stepping to the fully-discrete ROM equation (Eq.~\ref{rom:lspg_proj}), it can be directly employed to the fully-discrete FOM equation (Eq.~\ref{EqRes:LinearMultiStep_qp} or Eq.~\ref{EqRes:RK_qp}). Similarly, the pseudo-time ($\tau$) derivative is represented using the implicit Euler method in the following form 
 \begin{align}
     \frac{\text{d} \solConsFOM}{\dPTimeVar}|_{\tau = \tau_k} &= \gm \frac{\text{d} \solPrimFOM}{\dPTimeVar}|_{\tau = \tau_k} \approx \gm^{\subIdx-1} \frac{\solPrimFOM^{\subIdx} - \solPrimFOM^{\subIdx-1}}{\dtau},
 \end{align}
 where $\subIdx$ represents the pseudo-time iteration number. During the dual time-stepping procedure, this term is driven to zero. For linear multi-step time integrators, the fully-discrete FOM equation residual with pseudo-time, $\resPrim: \mathbb{R}^{\numDOF} \rightarrow \mathbb{R}^{\numDOF}$, takes the form
\begin{align}
\resPrimFunc{\solPrimFOM^\subIdx} &\triangleq \gm^{\subIdx-1} \frac{\dt}{\dtau} \lp \solPrimFOM^{\subIdx} - \solPrimFOM^{\subIdx-1} \rp + \resFunc{\solPrimFOM^{\subIdx}} \\
&= \gm^{\subIdx-1} \frac{\dt}{\dtau} \lp \solPrimFOM^{\subIdx} - \solPrimFOM^{\subIdx-1} \rp + \solConsFOMFunc{\solPrimFOM^{\subIdx}} + \sum^{\itermaxLinMS}_{j=1} \alpha_j \solConsFOMFunc{\solPrimFOM^{\iterIdx-j}} - \dt \beta_0 \rhsFunc{\solPrimFOM^\subIdx, \timeVar^\subIdx} - \dt \sum^{\itermaxLinMS}_{j=1} \beta_j \rhsFunc{\solPrimFOM^{\iterIdx-j}, \timeVar^{\iterIdx-j}}
\label{EqRes:LinearMultiStep_qp_dualtime}    
\end{align}
Linearizing the second and fourth terms  on the right-hand side
\begin{align}
   \solConsFOMFunc{\solPrimFOM^{\subIdx}} &\approx  \solConsFOMFunc{\solPrimFOM^{\subIdx-1}} +  \lp \frac{\partial \solConsFOM}{\partial \solPrimFOM} \rp^{\subIdx-1} \lp \solPrimFOM^{\subIdx} - \solPrimFOM^{\subIdx-1} \rp = \solConsFOMFunc{\solPrimFOM^{\subIdx-1}} +  \gm^{\subIdx-1} \lp \solPrimFOM^{\subIdx} - \solPrimFOM^{\subIdx-1} \rp, \\
   - \dt \beta_0 \rhsFunc{\solPrimFOM^\subIdx, \timeVar^\subIdx} &\approx - \dt \beta_0 \rhsFunc{\solPrimFOM^{\subIdx-1}, \timeVar^{\subIdx-1}} - \dt \beta_0 \jacobConsFOM^{\subIdx-1} \gm^{\subIdx-1} \lp \solPrimFOM^{\subIdx} - \solPrimFOM^{\subIdx-1} \rp,
\end{align}
and collecting terms, the final form of Eq.~\ref{EqRes:LinearMultiStep_qp_dualtime} becomes
\begin{equation}
    \resPrimFunc{\solPrimFOM^\subIdx} = \left[ \lp \frac{\dt}{\dtau} + 1 \rp \gm^{\subIdx-1} - \dt \beta_0 \jacobConsFOM^{\subIdx-1} \gm^{\subIdx-1}  \right] \lp \solPrimFOM^{\subIdx} - \solPrimFOM^{\subIdx-1} \rp + \resFunc{\solPrimFOM^{\subIdx-1}}.
    \label{EqRes:LinearMultiStep_qp_DualTime}
\end{equation}
If $\beta_0 = 0$, the time integration formulation is explicit; otherwise, it is implicit. 

Following the same procedure, the fully-consistent discrete FOM equation residual for Runge--Kutta methods becomes
\begin{equation}
    \resPrimFunc{\solPrimFOM^\subIdx} = \left[ \lp \frac{\dt}{\dtau} + 1 \rp \gm^{\subIdx-1} \right] \lp \solPrimFOM^{\subIdx} - \solPrimFOM^{\subIdx-1} \rp + \resFunc{\solPrimFOM^{\subIdx-1}}.
    \label{EqRes:RK_qp_dualtime}
\end{equation}

With sufficient convergence, $\resPrimFunc{\solPrimFOM^{\subIdx}} \rightarrow \mathbf{0}$ and $(\solPrimFOM^{\subIdx} - \solPrimFOM^{\subIdx-1}) \rightarrow \mathbf{0}$. Thus, the fully-discrete FOM residual is also driven to zero (i.e. $\resFunc{\solPrimFOM^{\subIdx}} \rightarrow \mathbf{0}$), achieving consistency with the conservative formulation. At this point, the solution is advanced in physical time, $\solPrimFOM^{\iterIdx} = \solPrimFOM^{\subIdx-1}$.

Therefore, if a dual-time algorithm is used  to solve Eq.~\ref{cfd:discretized_qp}, the {\color{black}MP-LSVT} ROM formulation applied to the physical time residual becomes
\begin{equation} 
\lp \testBasisPrim^{\subIdx-1} \rp^T \lp \frac{\dt}{\dtau}\gmROM^{\subIdx-1} + \testBasisPrim^{\subIdx-1} \rp \lp \solPrimROMRed^k - \solPrimROMRed^{\subIdx-1} \rp + \lp \testBasisPrim^{\subIdx-1} \rp^T \resFunc{\solPrimROMFull^{\subIdx-1}} = \zeroVec.
\label{rom:lspg_proj_dualtime_nonsymmetric}
\end{equation}
Similar as Eq.~\ref{rom:lspg_proj_dualTime_final}, it can be easily recognized that when $\dtau \rightarrow \infty$, we recover the Newton's method form (Eq.~\ref{rom:lspg_proj_newton_symm}) and optimum convergence is obtained. For finite $\dtau$, however, the above equation is not symmetrized. This issue is addressed in the following discussion.

\subsection{Discrete Consistency and Symmetrization}
\label{lspg-vt-dualtime}

Alternatively, we can reformulate the objective to minimize the fully-discrete FOM equation residual at the sub-iteration (or pseudo-time step) level. In this context, Eqs.~\ref{EqRes:LinearMultiStep_qp_DualTime} and~\ref{EqRes:RK_qp_dualtime} are minimized with respect to $\solPrimROMFull^{\subIdx} = \solPrimFOMRef + \scaleMatPrim^{-1} \trialBasisPrim \solPrimROMRed^{\subIdx}$. Therefore, we define
\begin{equation}
     \solPrimROMFull^{\subIdx}  \triangleq  \argmin_{ \solPrimROMFull^{\subIdx}  \in \textrm{Range}(\trialBasisPrim)} \norm{\scaleMatCons \resPrimFunc{\solPrimROMFull^{\subIdx}}}^2,
    \label{rom:lspg_def_dualtime}
\end{equation}
with the equation residual, $\res$, scaled by matrix, $\scaleMatCons$. Again, the solution to Eq.~\ref{rom:lspg_def_dualtime} has an equivalent semi-discrete Petrov-Galerkin projection form given by
\begin{equation}
    \lp \testBasisPrimPrim^\subIdx \rp^T \scaleMatCons \resPrimFunc{\solPrimROMFull^{\subIdx}} = \zeroVec,
    \label{rom:lspg_proj_dualtime}
\end{equation}
where $\testBasisPrimPrim$ is the test basis
\begin{equation}
    \testBasisPrimPrim^\subIdx = \frac{\partial \scaleMatCons \resPrimFunc{\solPrimROMFull^{\subIdx}}}{\partial \solPrimROMRed^{\subIdx}}.
\end{equation}

For linear multi-step methods, the FOM equation residual $\resPrim$ is defined in Eq.~\ref{EqRes:LinearMultiStep_qp_DualTime}. Substituting the approximate solutions, $\solPrimROMFull^{\subIdx}$ and $\solPrimROMFull^{\subIdx-1}$, we have 
\begin{equation}
    \frac{\partial \scaleMatCons \resPrimFunc{\solPrimROMFull^{\subIdx}}}{\partial \solPrimROMRed^{\subIdx}} = \frac{\partial \scaleMatCons \resPrimFunc{\solPrimROMFull^{\subIdx}}}{\partial \solPrimROMFull^{\subIdx}} \frac{\partial \solPrimROMFull^{\subIdx}}{\partial \solPrimROMRed^{\subIdx}} = \scaleMatCons \left[ \lp \frac{\dt}{\dtau} + 1 \rp \gmROM^{\subIdx-1} - \dt \beta_0 \jacobConsROM^{\subIdx-1} \gmROM^{\subIdx-1} \right] \scaleMatPrim^{-1} \trialBasisPrim.
\end{equation}
Hence, the test basis $\testBasisPrimPrim^\subIdx$ becomes
\begin{equation}
    \testBasisPrimPrim^\subIdx = \scaleMatCons \left[ \lp \frac{\dt}{\dtau} + 1 \rp \gmROM^{\subIdx-1} - \dt \beta_0 \jacobConsROM^{\subIdx-1} \gmROM^{\subIdx-1} \right] \scaleMatPrim^{-1} \trialBasisPrim.
    \label{rom:lspg_w_qp_DualTime}
\end{equation}

Similarly, we can derive the discretely-consistent {\color{black}MP-LSVT} ROM formulation for Runge--Kutta methods. The FOM equation residual is defined in Eq.~\ref{EqRes:RK_qp_dualtime}. Again substituting the approximate solutions, $\solPrimROMFull^{\subIdx}$ and $\solPrimROMFull^{\subIdx-1}$, results in the ROM residual

\begin{equation}
    \resPrimFunc{\solPrimROMFull^\subIdx} = \left[ \lp \frac{\dt}{\dtau} + 1 \rp \gmROM^{\subIdx-1} \right] \lp \solPrimROMFull^{\subIdx} - \solPrimROMFull^{\subIdx-1} \rp + \resFunc{\solPrimROMFull^{\subIdx-1}}.
\end{equation}
Therefore
\begin{equation}
    \frac{\partial\scaleMatCons \resPrimFunc{\solPrimROMFull^{\subIdx}}}{\partial \solPrimROMRed^{\subIdx}} = \frac{\partial \scaleMatCons \resPrimFunc{\solPrimROMFull^{\subIdx}}}{\partial \solPrimROMFull^{\subIdx}} \frac{\partial \solPrimROMFull^{\subIdx}}{\partial \solPrimROMRed^{\subIdx}} = \scaleMatCons \left[ \lp \frac{\dt}{\dtau} + 1 \rp \gmROM^{\subIdx-1} \right] \scaleMatPrim^{-1} \trialBasisPrim.
\end{equation}
Hence, the test basis $\testBasisPrimPrim^\subIdx$ becomes
\begin{equation}
    \testBasisPrimPrim^\subIdx = \scaleMatCons\left[ \lp \frac{\dt}{\dtau} + 1 \rp \gmROM^{\subIdx-1} \right] \scaleMatPrim^{-1} \trialBasisPrim.
    \label{rom:lspg_w_qp_DualTime_rk}
\end{equation}

It can be readily seen that {\color{black}MP-LSVT} ROMs using either linear multi-step and Runge--Kutta methods lead to the symmetrized formulation
\begin{equation}
\begin{split}
 \lp \testBasisPrimPrim^{\subIdx} \rp^T \testBasisPrimPrim^{\subIdx} \lp \solPrimROMRed^{\subIdx} - \solPrimROMRed^{\subIdx-1} \rp + \lp \testBasisPrimPrim^{\subIdx} \rp^T \resFunc{\solPrimROMFull^{\subIdx-1}} = \zeroVec.
\end{split}
\label{rom:lspg_proj_dualtime_symmetric_rk}
\end{equation}

\subsection{A Note on Symmetrization}
\label{sec:symmetry}
The benefits of symmetrization have been examined in  many contexts in the literature. For instance, Ref.~\cite{SymmetrizedODE} discusses applications in ODE systems. Barone et al.~\cite{Barone2009JCP} have shown that symmetrization provides enhanced stabilization and convergence of the discrete ROM system. In our context, symmetrization guarantees linear stability of the ROMs, subject to linear stability restrictions on the FOM, as shown in Section~\ref{proof}.

As shown in Eqs.~\ref{rom:lspg_proj_newton_symm} and~\ref{rom:lspg_proj_dualTime_final}, a symmetrized discrete system is obtained if the {\color{black}MP-LSVT} ROM formulation is applied to fully-discrete equation residual in physical time (Eq.~\ref{rom:lspg_proj}) and the reduced system is solved by either Newton’s method or dual time. In contrast, if a dual-time algorithm is introduced in the FOM equation, the {\color{black}MP-LSVT} ROM formulation applied to the physical time residual results in an unsymmetrized system (Eq.~\ref{rom:lspg_proj_dualtime_nonsymmetric}) except in the limit as the pseudo time step goes to infinity. It is noteworthy that for traditional finite-difference or finite-volume algorithms, the presence of a finite pseudo-time step corresponds to adding a sink term to the (already non-symmetric) operator in Eq.~\ref{EqRes:LinearMultiStep_qp_DualTime} so that finite pseudo-time steps provide faster convergence than infinite ones.  Nevertheless, upon projection the effect of this sink term is eliminated and an infinite pseudo-time step becomes optimum.

In summary, we have shown that with a more general selection of solution variables, symmetrized least-squares-based Petrov-Galerkin projections can be derived at the fully discrete sub-iteration level. These formulations are also applicable to systems discretized with explicit time integrators. In Section~\ref{results}, we will assess the robustness and accuracy of ROMs using Galerkin projection, LSPG projection, and the {\color{black}MP-LSVT} ROM formulation for challenging reacting flow problems.
\subsection{Hyper-reduction}
\label{hyper_reduction}

Even though projection-based ROMs lead to discrete systems of much lower dimension ($\numSolModes \ll \numDOF$), the evaluations of the non-linear terms remain a bottleneck as they involve $O(\numDOF)$ operations. A popular method to circumvent this problem is the discrete empirical interpolation method (DEIM)~\cite{chaturantabut2010nonlinear}, or its least-squares regression analogue, gappy POD~\cite{gappyPOD}. These methods use sparse samples and data-driven approximation to develop a full reconstruction of the non-linear function at non-sampled elements. This approximation of the non-linear function by gappy POD is given by the formulation

\begin{equation}
\resApprox \approx \resBasis \lp \sampMat^T \resBasis \rp^{+}  \sampMat^T \res,
\label{rom:deim_approx}
\end{equation}
where $\sampMat \in \mathbb{R}^{\numDOF \times \numSamps}$ is a selection operator that samples $\numSamps$ columns (i.e. sampling points) of the identity matrix, $\mathbf{I} \in \mathbb{R}^{\numDOF \times \numDOF}$, and $\resBasis \in \mathbb{R}^{\numDOF \times \numResModes}$ is a basis set used to approximate the non-linear term, $\res$. Typically, $\resBasis$ is constructed via POD from snapshots of $\res$. However, the authors have found that setting $\resBasis$ to the trial POD basis $\trialBasisCons$ in Eq.~\ref{pod:expansion_consv} or $\trialBasisPrim$ in Eq.~\ref{pod:expansion_qp} also produces excellent approximations; this method is used for all results presented in this paper. Further, although it is not strictly required that $\numResModes = \numSolModes$, this is true for all cases presented in this paper.

By applying sparse sampling, the cost of evaluating the non-linear term $\res$ now scales with $\numSamps$, where $\numSamps \ll \numDOF$. It has to be recognized that for coupled dynamical systems -- such as those generated via the spatial discretization of a PDE -- $\sampMat^T \resFunc{\solPrimFOM} \neq \resFunc{\sampMat^T \solPrimFOM}$; additional elements of $\res$ must be evaluated to compute the sub-sampled non-linear term. The scalable implementation of sparse sampling, especially critical for complex multi-scale problems, is discussed in greater detail in~\ref{appendix:hyperreductionDetails}. In the current work, the first $\numResModes$ sampling points are chosen using the rank-revealing QR algorithm suggested in Ref.~\cite{drmac_qdeim}. The remaining $\numSamps - \numResModes$ oversampling points are selected via a uniform random distribution, following evaluations performed by Peherstorfer et al.~\cite{PeherstorferODEIM}, which demonstrated that randomized oversampling can stabilize DEIM and is necessary especially when noise (e.g. from turbulence or numerical inaccuracies) is present in the problem.

We now extend the gappy POD formulation to the {\color{black}MP-LSVT} ROMs in a fashion similar to that of Carlberg et. al~\cite{carlberg2013gnat}. We begin by applying sparse sampling and reconstruction to the non-linear equation residual $\res$ in Eq.~\ref{rom:lspg_def} 
\begin{equation}
    \solPrimROMFull^{\iterIdx} \triangleq \argmin_{\solPrimROMFull^{\iterIdx} \in \textrm{Range}(\trialBasisPrim)} \norm{\resBasis \lp \sampMat^T \resBasis \rp^{+} \sampMat^T \scaleMatCons \resFunc{\solPrimROMFull^{\iterIdx}}}^2.
    \label{rom:lspg_def_deim}
\end{equation}
The resulting test basis $\testBasisPrimGPOD^\iterIdx$  is then given by
\begin{equation}
    \testBasisPrimGPOD^\iterIdx = \frac{ \partial \resBasis \lp \sampMat^T \resBasis \rp^{+} \sampMat^T \scaleMatCons \resFunc{\solPrimROMFull^{\iterIdx}}}{\partial \solPrimROMRed^{\iterIdx}} = \resBasis \lp \sampMat^T \resBasis \rp^{+} \sampMat^T \frac{ \partial \scaleMatCons \resFunc{\solPrimROMFull^{\iterIdx}}}{\partial \solPrimROMRed^{\iterIdx}} = \resBasis \lp \sampMat^T \resBasis \rp^{+} \sampMat^T \testBasisPrim^\iterIdx.
    \label{rom:lspg_w_deim}
\end{equation}
Thus, only $\numSamps$ rows of the test basis $\testBasisPrim^{\iterIdx}$ must be evaluated. This is yet another major step in reducing the number of necessary computations. 

With the approximated test basis in hand, the {\color{black}MP-LSVT} ROM in physical time (Eq.~\ref{rom:lspg_proj}) becomes
\begin{equation}
 \lp \testBasisPrimGPOD^\iterIdx \rp^T \resBasis \lp \sampMat^T \resBasis \rp^{+} \sampMat^T \scaleMatCons \resFunc{\solPrimROMFull^\iterIdx} = \zeroVec,
 \label{rom:lspg_proj_hyper}
\end{equation}
and, noting that $\resBasis^T \resBasis = \mathbf{I}$ by orthonormality, we arrive at the final form of the ROM
\begin{equation}
    \lp \sampMat^T \testBasisPrim^\iterIdx \rp^T \left[ \lp \sampMat^T \resBasis \rp^{+} \right]^T \lp \sampMat^T \resBasis \rp^{+} \sampMat^T \scaleMatCons \resFunc{\solPrimROMFull^{\iterIdx}} = \zeroVec.
    \label{rom:lspg_proj_deim}
\end{equation}
Although $[(\sampMat^T \resBasis)^{+}]^T (\sampMat^T \resBasis)^{+} \in \mathbb{R}^{\numSamps \times \numSamps}$ can be precomputed offline, this matrix may become quite large and therefore infeasible to store in memory. In reality, only $(\sampMat^T \resBasis)^{+} \in \mathbb{R}^{\numResModes \times \numSamps}$ is precomputed offline and loaded into memory at runtime.

Similarly, applying sparse sampling and reconstruction to $\resPrim$ in Eq.~\ref{rom:lspg_def_dualtime}, the {\color{black}MP-LSVT} ROM in pseudo-time (Eq.~\ref{rom:lspg_proj_dualtime}) is given by
\begin{equation}
    \lp \sampMat^T \testBasisPrimPrim^{\iterIdx} \rp^T \left[ \lp \sampMat^T \resBasis \rp^{+} \right]^T \lp \sampMat^T \resBasis \rp^{+} \sampMat^T \scaleMatCons \resPrimFunc{\solPrimROMFull^{\subIdx}} = \zeroVec.
    \label{rom:lspg_proj_deim_pseudo}
\end{equation}
\section{Enhancing Local Stability}
\label{limiters}
While symmetrization improves the prospects for global stability, ROMs of complex problems can fail because of spurious local behavior.  The current authors~\cite{HuangAIAAJ2019,Huang2020_SpeciesLimiters} have demonstrated that imposing physical realizability during ROM calculations can be critical to ROM stability. This is especially important when the trial basis is not rich enough to resolve sharp gradients in the flow field, resulting in spurious oscillations. In reacting flows, this often leads to unrealistic values of physical quantities (e.g. $T \leq 0$ K), terminating the calculations. 
The {\color{black}MP-LSVT} ROM formulation cannot guarantee positivity of quantities such as density or temperature. Therefore, to mitigate the production of such spurious oscillations near sharp gradients, local limiters can be deployed.

In the combustion ROMs presented in this work, two types of simple limiters are proposed: one for temperature and the other for species mass fractions. The temperature limiter follows the method proposed by the current authors~\cite{HuangAIAAJ2019} to restrict the temperature $\widetilde{T}$ in ROM calculations (based on Eq.~\ref{pod:expansion_qp}) to an interval bounded by $T_\text{min}$,  $T_\text{max}$. This range can be determined based on the underlying physics. In non-premixed flames, for example, physics dictates that the minimum and maximum temperatures in the simulation are bounded by the cold reactant temperature, $T_\text{c}$, and the adiabatic flame temperature, $T_\text{ad}$, respectively. This range can be can be determined \textit{a priori} based on the propellants and flow configuration  before running the FOM simulations. For example, $T_\text{min} = T_\text{c} - \delta T_\text{c}$ and $T_\text{min} = T_\text{ad} + \delta T_\text{ad}$, where values of $\delta T_\text{c}$ and $\delta T_\text{ad}$ can be estimated to account for temperature variations due to the effects of acoustics and thermodynamics. 

In contrast to  spurious oscillations in temperature which can terminate the ROM calculations immediately, the generation of spurious oscillations in the species mass fraction fields exhibits a more gradual  impact on the ROM results, as diagnosed in previous work~\cite{Huang2020_SpeciesLimiters}. This can be particularly problematic within the high temperature reaction regions. As a demonstrative example, a 1D premixed flame is considered in~\ref{appendix:species_oscillations}. To suppress these spurious oscillations in reaction regions, limiters on species mass fraction fields (denoted as the ``species limiter'' in the rest of the paper for simplicity) are proposed for premixed and non-premixed flames. 

For a premixed flame
\begin{equation}
    \text{if} \; \widetilde{T} > T_\text{th}, \; \widetilde{Y}_\text{R} = min \{ \widetilde{Y}_\text{R}, \delta \},
    \label{rom:species_limiter_premixed}
\end{equation}
The limiter is only activated in the reacting regions when  $\widetilde{T}$ exceeds a certain threshold value $T_\text{th}$, which can be predetermined based on the adiabatic flame temperature, $T_\text{ad}$ (e.g. $T_\text{th} = 0.8 T_\text{ad}$). $\widetilde{Y}_\text{R}$ represents the mass fraction of reactants in ROM calculations and $\delta$ is a small value (e.g. $1 \times 10^{-5}$).

Similarly, for a non-premixed flame, the limiter is only activated in the reacting regions
\begin{equation}
    \text{if} \; \widetilde{T} > T_\text{th}, \left\{\begin{array}{lll}
        \text{if} \; \widetilde{Z} > Z_\text{st}, & \widetilde{Y}_\text{ox} & = min \{ \widetilde{Y}_\text{ox}, \delta \} \\
        \text{if} \; \widetilde{Z} = Z_\text{st}, & \widetilde{Y}_\text{f} & = min \{ \widetilde{Y}_\text{f}, \delta \} \;\; \text{and} \;\; \widetilde{Y}_\text{ox} = min \{ \widetilde{Y}_\text{ox}, \delta \} \\
        \text{if} \; \widetilde{Z} < Z_\text{st}, & \widetilde{Y}_\text{f} & = min \{ \widetilde{Y}_\text{f}, \delta \}
    \end{array} \right .
    \label{rom:species_limiter_nonpremixed}
\end{equation}
where $\widetilde{Z}$ is the mixture fraction
\begin{equation}
\widetilde{Z} = \frac{ \nu_\text{st} \widetilde{Y}_\text{f} - \widetilde{Y}_\text{ox} + Y^0_\text{ox} }{ \nu_\text{st} Y^0_\text{f} + Y^0_\text{ox} }.
\label{mixture_fraction}
\end{equation}
Here, $\nu_\text{st}$ is the stoichiometric oxidizer-to-fuel mass fraction ratio, $\widetilde{Y}_\text{f}$ and $\widetilde{Y}_\text{ox}$ are the fuel and oxidizer mass fractions respectively, and $Y^0_\text{f}$ and $Y^0_\text{ox}$ are the mass fraction of fuel and oxidizer in the fuel and oxidizer inlet streams, respectively. The constant parameters $\nu_\text{st}$ , $Y^0_\text{f}$ and $Y^0_\text{ox}$ can be predetermined based on the reacting flow propellants and conditions before running the FOM simulations. $Z_{st}$ is the stoichiometric mixture fraction, $Z_\text{st} = Y^0_\text{ox} / ( \nu_\text{st} Y^0_\text{f} + Y^0_\text{ox} )$. As indicated in Eq.~\ref{rom:species_limiter_nonpremixed}, only the mass fraction of oxidizer, $\widetilde{Y}_\text{ox}$ is limited in fuel-rich reaction regions ($\widetilde{Z} > Z_\text{st}$). In fuel-lean reaction regions ($\widetilde{Z} < Z_\text{st}$), only the mass fraction of fuel, $\widetilde{Y}_\text{f}$, is limited. In stoichiometric reaction regions ($\widetilde{Z} = Z_\text{st}$), both are limited. 

Note that the species limiter in Eqs.~\ref{rom:species_limiter_premixed} and~\ref{rom:species_limiter_nonpremixed} is designated for chemical reactions modeled by multi-species transport equations. For a flamelet-type model~\cite{Pierce2004_FPV} where chemical reactions are modeled by representative transported scalars (e.g. mixture fraction $Z$ and progress variable $C$), the species limiter for a premixed flame becomes
\begin{equation}
    \text{if} \; \widetilde{T} > T_\text{th}, \; \widetilde{C} = min \{max \{ \widetilde{C}, C_\text{ref} - \delta \}, C_\text{ref} + \delta \},
    \label{rom:species_limiter_flpremixed}
\end{equation}
where $\widetilde{C}$ is the progress variable in ROM calculations that represents the progress of the chemical reaction and $C_\text{ref}$ is the highest progress variable value in the problem of interest (e.g. for a stoichiometric reaction, $C_\text{ref}=1$, while for non-stoichiometric reaction, $C_\text{ref}<1$). 

Similarly, for a non-premixed flame, the limiter becomes
\begin{equation}
    \text{if} \; \widetilde{T} > T_\text{th}, \left\{\begin{array}{ll}
        \widetilde{C} = min\{max \{ \widetilde{C}, C_\text{ref} - \delta \}, C_\text{ref} + \delta \} \\
        \widetilde{Z} = min \{ max \{ \widetilde{Z}, Z_\text{st} - \delta\}, Z_\text{st} + \delta \}
    \end{array} \right .
    \label{rom:species_limiter_flnonpremixed}
\end{equation}
where both the progress variables $\widetilde{C}$ and mixture fraction $\widetilde{Z}$ in ROM calculations need to be limited. The benefits of imposing the species limiter is further demonstrated using a 2D non-premixed reacting flow problem in Sec.~\ref{sec:2d_limiters}.

The above discussion emphasizes the importance of local robustness considerations in ROMs of complex problems.

\section{Numerical Results and Analysis}
\label{results}

To assess the capabilities of projection-based ROMs in predicting complex multi-scale multi-physics problems, such as reacting flows, two combustor configurations are established based on a generic laboratory-scale single injector combustor~\cite{YuJPP}. The first configuration is a simplified two-dimensional representation of the injector. This case is used to assess the accuracy and robustness of different ROM formulations, including Galerkin projection, LSPG projection, and the {\color{black}MP-LSVT} formulation. The second configuration is a full three-dimensional representation of the injector used in physical experiments, and is used to evaluate the computational efficiency and predictive capabilities of the {\color{black}MP-LSVT} ROM formulation. It should be pointed out that numerical results for the {\color{black}MP-LSVT} ROM reported here were obtained using the physical time formulation (Section~\ref{lspg-vt}). The dual-time formulation (Section~\ref{lspg-vt-dualtime}) was also found to produce very similar results, and are not reported here for brevity.

The computational infrastructure used for the full- and reduced-order models solves conservation equations for mass, momentum, energy and species  transport in a fully coupled way using the  in-house CFD code, the General Mesh and Equation Solver (GEMS). GEMS  has been used to model a variety of complex, practical reacting flow problems~\cite{HarvazinskiPoF,HuangLDI}. More details of the FOM equations can be found in~\ref{appendix:fom_eq}. The FOM employs a cell-centered second-order accurate finite volume method for spatial discretization. The Roe scheme~\cite{Roe1981} is used to evaluate the inviscid fluxes and a Green-Gauss gradient reconstruction procedure~\cite{MitchellReconScheme} is used to compute the face gradients and viscous fluxes. A gradient limiter by Barth and Jespersen~\cite{barth1989} is used to preserve monotonicity for flow fields with strong gradients. A ghost cell formulation is used for treatment of boundary conditions. Time integration for all FOM simulations uses the implicit second-order accurate backwards differentiation formula with dual time-stepping.

It should be recognized that in the case of reacting flow modeling, the chemical source terms in the species transport equations often lead to extreme numerical stiffness due to the kinetic model~\cite{WestbrookDryer}. The example in~\ref{appendix:species_oscillations} illustrates this problem. This stiffness is the result of rapid production or destruction of species at the microseconds timescale, which requires a very small time step for accurate resolution. More importantly, it can also produce stiff Jacobians with a high condition number, which can be as high as $O(10^{12})$ for high-pressure and high-temperature conditions. Solving the resulting linear system is very challenging, even at the FOM level.

\subsection{2D Reacting Injector}
\label{2d:injector}
\begin{figure}
	\centering
	\includegraphics[width=1.0\textwidth]{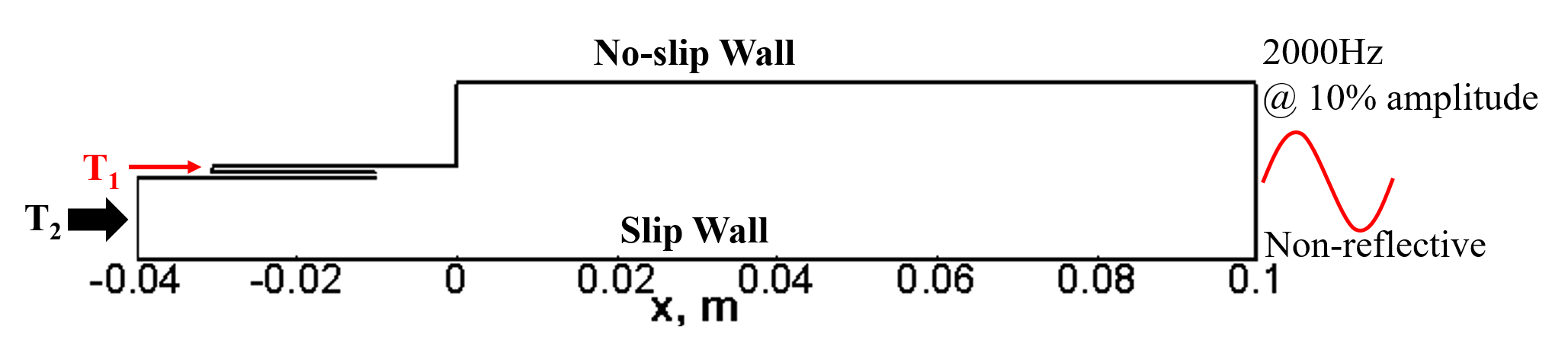}
	\caption{Computational configuration of 2D planar reacting  injector simulation.}\label{2d:geometry} 
\end{figure}

 We begin with a 2D-planar representation of the generic laboratory-scale rocket combustor~\cite{YuJPP}. This simplified model allows ROM capabilities to be evaluated in great detail without incurring exorbitant computational cost, while maintaining the essential physics of interest. The configuration is shown in Fig.~\ref{2d:geometry} and consists of a shear coaxial injector with an outer passage, $T_1$, that introduces fuel near the downstream end of the coaxial inner passage, $T_2$, that feeds oxidizer to the combustion chamber. The $T_1$ stream contains gaseous methane (100\% \methane) at 300 K. The $T_2$ stream is 42\% gaseous \oxygen\ by mass and 58\% gaseous \water\ by mass at 700 K. 

Operating conditions in the combustion chamber are maintained similar to conditions in the laboratory combustor~\cite{YuJPP,YuPhD}, with an adiabatic flame temperature of approximately 2,700 K and a mean chamber pressure of 1.1 MPa. Both the $T_1$ and $T_2$ streams are fed with constant mass flow rates, 0.37 kg/s and 5.0 kg/s, respectively. A non-reflective boundary condition is imposed at the downstream end to allow acoustic waves to properly exit the domain and control acoustic effects on the combustion dynamics. A sinusoidal pressure perturbation at 2,000 Hz, with amplitude 10\% of the mean pressure, is imposed at the downstream boundary. Transport of four chemical species (\methane, \oxygen,  \water, \carbonDiox) is modeled. The chemical reaction is modeled by the global single-step methane-oxygen reaction recommended by Westbrook and Dryer~\cite{WestbrookDryer}:
$
   \methane + 2 \oxygen \rightarrow \carbonDiox + 2 \water.
$
The chemical species are treated as thermally perfect gases.

As reported by the current authors~\cite{Huang2018Jan}, stable and accurate reconstruction of FOM solutions with stiff chemistry via projection-based ROMs is highly challenging. To mitigate these difficulties, studies have been performed by the current authors to evaluate ROM performance based on simulations using a reduced Arrhenius pre-exponential factor~\cite{HuangAIAAJ2019} ten times smaller than the value recommended by Westbrook and Dryer~\cite{WestbrookDryer}, producing a discrete system which is far less stiff and more amenable for ROM development and testing. The same dataset has also been used to investigate the use of operator inference learning in constructing ROMs for reacting flow simulations~\cite{Swischuk2020AIAAJ_LL}. It has been reported by the current authors~\cite{HuangAIAAJ2019} that even with this reduced reaction rate, construction of robust and accurate ROMs remains highly challenging and provides a clear example of the additional difficulties engendered when reactions are present. However, the 2D configuration (Fig.~\ref{2d:geometry}) in the current work is simulated with the stiff chemical kinetic model with the original pre-exponential factor to fully assess the capabilities of the proposed ROM framework. Therefore, more challenges are anticipated with the stiff chemical model in the current study.

\begin{figure}
	\centering
	\includegraphics[width=1.0\textwidth]{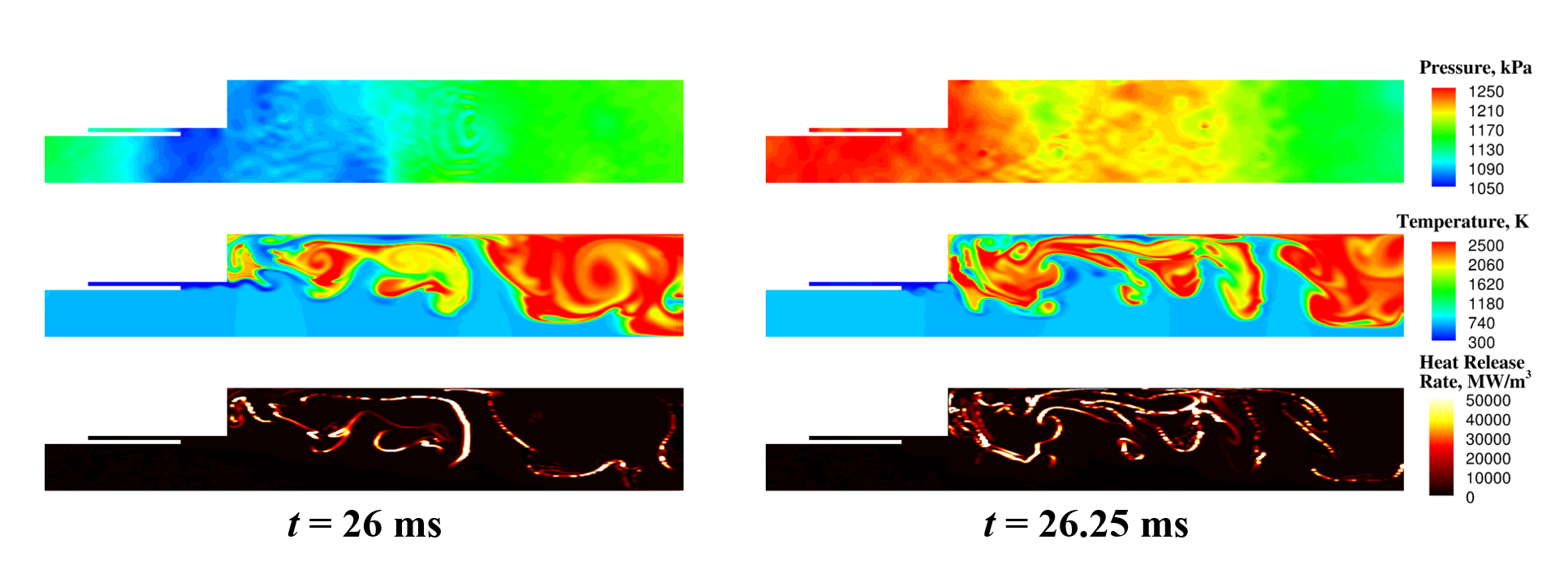}
	\caption{Representative instantaneous snapshots of pressure (top), temperature (middle) and heat release rate (bottom) from FOM simulation of the 2D reacting  injector.}\label{2d:fom_snapshots} 
\end{figure}

Two representative instantaneous snapshots of the 2D reacting single injector FOM solutions are shown in Fig.~\ref{2d:fom_snapshots} to demonstrate the overall characteristics of the flow field and highlight the dominant physics in the problem. The pressure exhibits global dynamics over the entire domain, while the combustion dynamics are characterized by local features like disperse pockets of intense heat release rate that are intermittently distributed in both space and time. {\color{black}The heat release rate quantifies the rate at which heat is generated by chemical reactions in the process of combustion.} The temperature and heat release rate contours span a wide range of scales, from the small eddies in the shear layers to the large-scale recirculation zone immediately downstream of the the dump plane at $x = 0$ m. More importantly, strong interactions can be identified between pressure and combustion dynamics. When the pressure  is low near the dump plane, high temperature pockets are distributed downstream of the combustor. Alternately, when the pressure is high at the dump plane, high temperature pockets are concentrated closer to the dump plane. These unique features and interactions introduce varying levels of difficulty in constructing a robust ROM.

The 2D FOM is computed using the second-order accurate  backwards difference formula and dual time-stepping, with a constant physical time step of 0.1 $\mu$s. The entire mesh contains a total of 38,523 finite volume cells with 7 solution variables ($p$, $u$, $v$, $T$, $Y_\text{CH4}$, $Y_\text{O2}$ and , $Y_\text{H2O}$), resulting in a total of 269,661 degrees of freedom. $Y_\text{CO2}$ is computed from the simple relation $\sum_i Y_i = 1$. Snapshots are stored at every physical time step over a duration of 1.0 ms, corresponding to a total of 10,000 snapshots, all of which are used to generate POD modes for ROM construction.

\subsubsection{POD Characteristics}
\label{2d:results-pod_res_energy}

\begin{figure}
	\centering
	\includegraphics[width=0.6\textwidth]{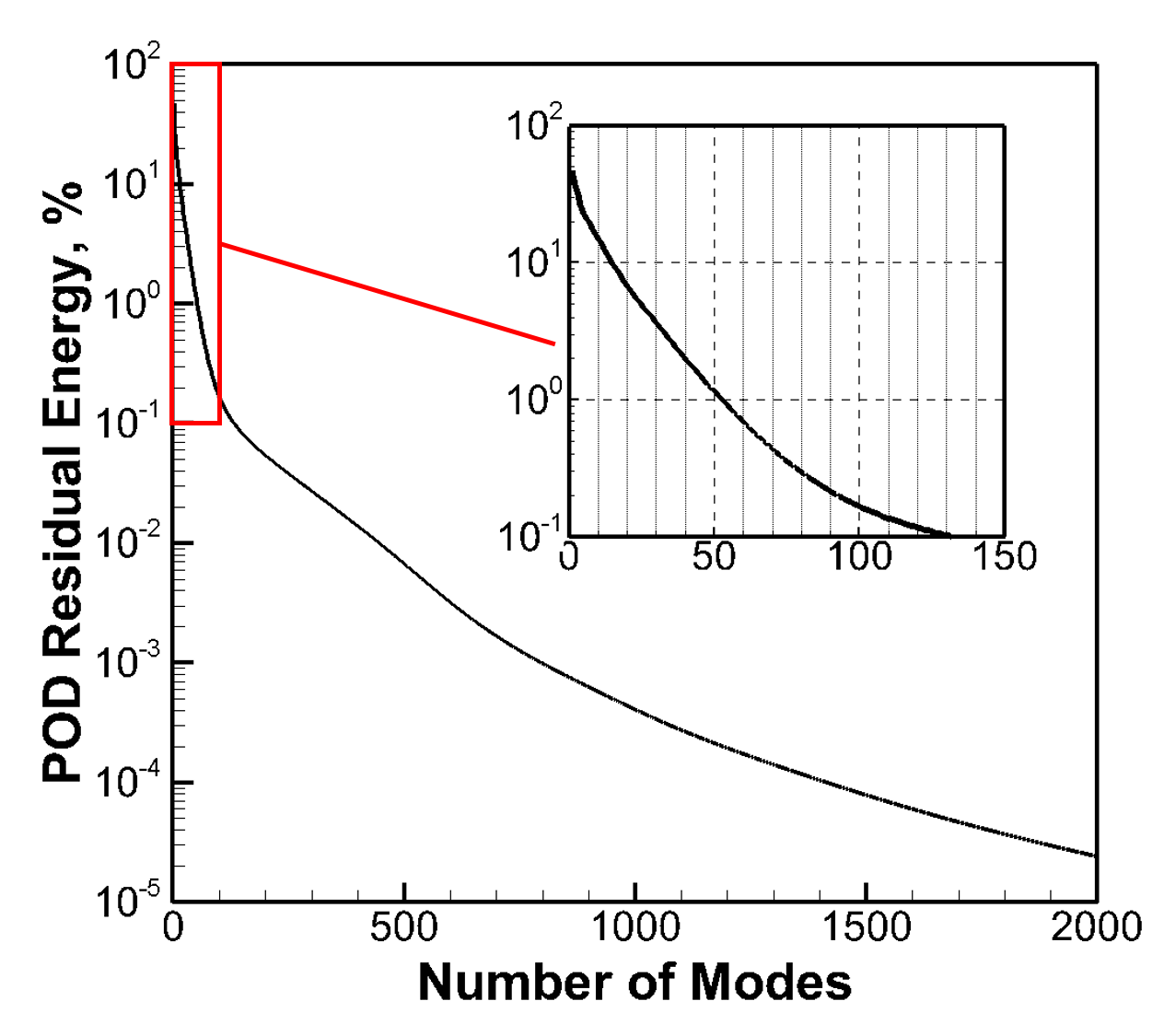}
	\caption{POD residual energy distribution for the 2D reacting  injector simulation.}\label{2d:pod_res_energy} 
\end{figure}

The POD characteristics of the reacting flow problem are first investigated to understand how well the POD trial basis represents the FOM dataset. The representation is evaluated using the POD residual energy: 
\begin{equation}
    \text{POD Residual Energy}(\numSolModes), \;\% = \lp 1 - \frac{\sum^{\numSolModes}_{\genIdx=1} \sigmaPOD^2_\genIdx}{\sum^{\numSolModesTotal}_{\genIdx=1} \sigmaPOD^2_\genIdx} \rp \times 100,
    \label{pod:res_energy}
\end{equation}
where $\sigmaPOD_\genIdx$ is the $\genIdx^\text{th}$ singular value of the SVD used to compute the trial basis \trialBasisPrim. The singular values are arranged in descending order. Again, $\numSolModes$ is the number of vectors retained in the POD trial basis, and $\numSolModesTotal$ (= 10,000) is the total number of snapshots in the dataset. The residual energy as a function of $\numSolModes$, as shown in Fig.~\ref{2d:pod_res_energy}, reveals the information excluded by the POD representation for a given  number of modes. The results show that the first 15 modes must be included to capture approximately 90\% of the total energy, 52 modes recover approximately 99\%, and at least 130 modes are needed to retrieve 99.9\% of the total energy. This slow energy decay is indicative of the significant complexity of the system dynamics. Many fundamental projection-based ROM methodologies are tested on relatively simple problems requiring only $\sim 10$ trial basis modes to achieve 99.9\% POD energy~\cite{LeeNonlinearManifold2020, Barone2009JCP, sanANNClosure}. ROMs for more practical engineering systems, however, generally require $\sim 100$ trial basis modes~\cite{StabileVMS2019, Carlberg2017, GrimbergPROM2020}.

\subsubsection{ROM Performance: Galerkin vs. LSPG vs. {\color{black}MP-LSVT}}
\label{2d:results-ROM}
The ability of a ROM to accurately model the evolution of the unsteady flow field is measured here in terms of the normalized, time-averaged, and variable-averaged $L^2$ error over the POD trial basis training window

\begin{equation}
    \errROMVsFOM = \frac{1}{\numVars} \sum^{\numVars}_{\varIdx=1} \frac{1}{\numSnaps} \sum^{\numSnaps}_{\iterIdx=1} \frac{\norm{\solPrimROMFullVar^\iterIdx - \solPrimFOMVar^\iterIdx}}{\norm{\solPrimFOMVar^\iterIdx}},
    \label{rom:reconstr_err}
\end{equation}
where $\solPrimROMFullVar^\iterIdx$ represents the $\varIdx^{th}$ solution variable of the state vector, $\solPrimROMFull^\iterIdx$,  at time step $\iterIdx$ from ROM simulations following Eq.~\ref{pod:expansion_qp}, and $\solPrimFOMVar^\iterIdx$ is obtained from FOM solutions. The entire error measure, $\errROMVsFOM$, is referred to as the reconstruction error, referring to the accuracy at which the ROM reconstructs the training data.

\begin{figure}
	\centering
	\includegraphics[width=0.6\textwidth]{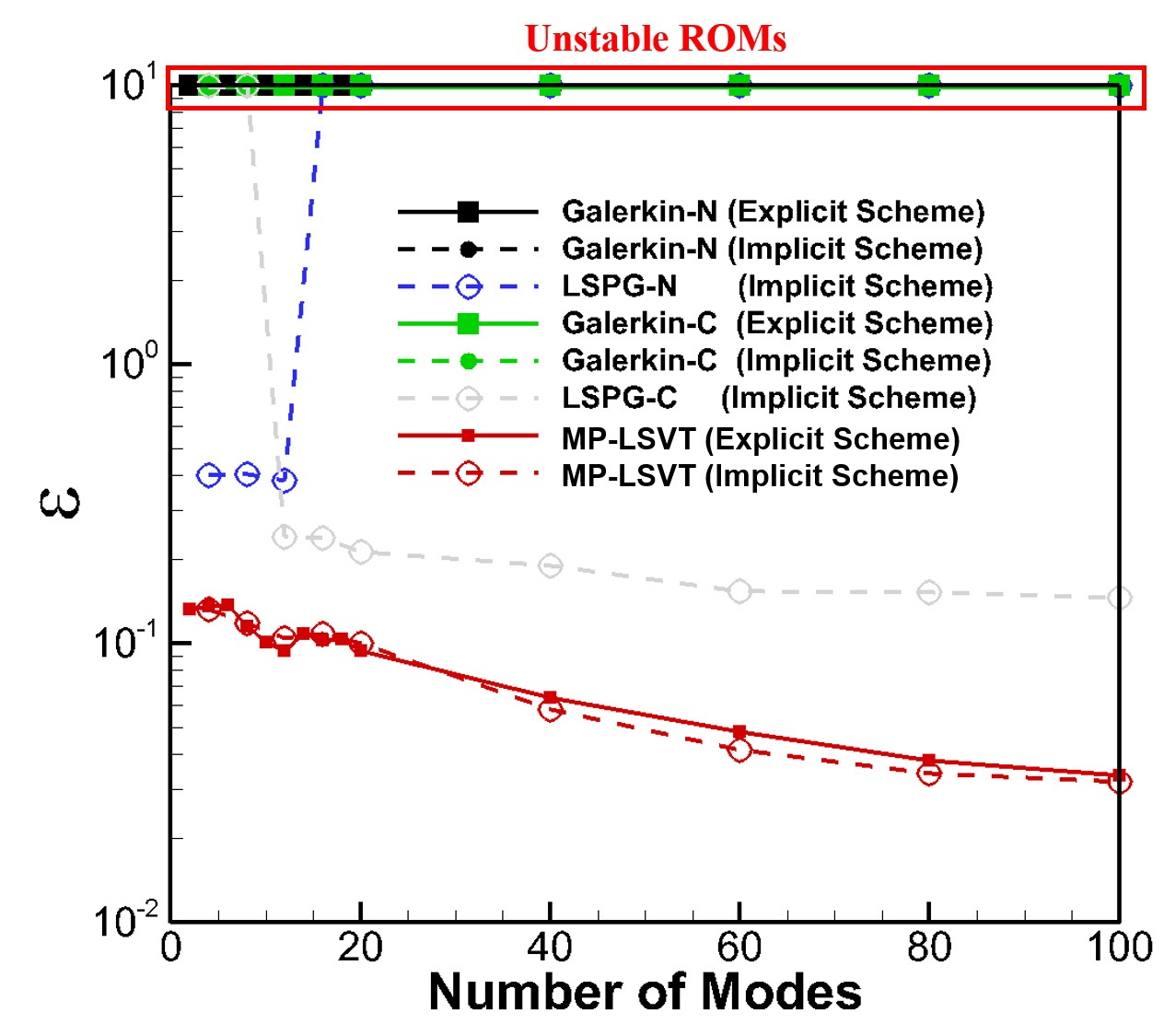}
	\caption{Global ROM reconstruction error comparisons between Galerkin, LSPG and {\color{black}MP-LSVT} ROM methods with different time integration schemes for the 2D reacting  injector simulation.}\label{2d:rom_recon_err} 
\end{figure}

Figure~\ref{2d:rom_recon_err} presents the ROM reconstruction error using Galerkin projection, LSPG projection, and the {\color{black}MP-LSVT} formulation with different time integration schemes and a varying number of modes retained in the trial basis. Results are shown for total mode numbers ranging from 4 to 100. To obtain a comprehensive assessment of different ROM methods, Galerkin and LSPG ROMs are developed based on both the conservative governing equations (Eq.~\ref{cfd:discretized}) and the non-conservative equations (Eq.~\ref{cfd:discretized_qp_ge}). ROMs using the former class of method are denoted as ``Galerkin-C'' and ``LSPG-C'', and those using the latter are denoted as as ``Galerkin-N'' and ``LSPG-N''. It is emphasized that Galerkin and LSPG projection are derived based on the premise that the governing equations and the solution variables are consistent (i.e. the conservative variables $\solConsFOM$ are chosen as the solution variables for Eq.~\ref{cfd:discretized} and the non-conservative variable $\solPrimFOM$ for Eq.~\ref{cfd:discretized_qp_ge}). On the other hand, {\color{black}MP-LSVT} ROMs are developed based on the conservative governing equations with transformed solution variables (Eq.~\ref{cfd:discretized_qp}). 

All the ROMs in Fig.~\ref{2d:rom_recon_err} are calculated with the temperature limiter introduced in Section~\ref{limiters}, with $T_{min} = 270$ K and $T_{max} = 2{,}850$ K, determined by the lowest temperature (the temperature of the $T_1$ stream, 300 K) and the highest temperature (the adiabatic flame temperature, 2,700 K) respectively. This promotes physically-realizable temperature fields during the calculations. Species limiters are not included for results presented in this section. ROMs solved via an explicit time integrator use an explicit 4-stage Runge--Kutta method. ROMs solved via an implicit time integrator use the second-order accurate backwards differentiation formula. It shall be pointed out that implicit time integration requires additional evaluations of Jacobian matrix $\jacobConsROM$, which are not needed for explicit time integration as summarized in Algorithm \ref{splsvt_algorithm}. Moreover, both implicit and explicit time integrators require an inversion of an $\numSolModes \times \numSolModes$ matrix during the ROM calculations while it is expected that the implicit time integrator requires more iterative steps for convergence. It can be readily seen that all the ROMs using Galerkin projection method are unstable. LSPG projection generally generates ROMs which are more stable than those using Galerkin projection, though most of the LSPG-N ROMs are still unstable with increased trial basis mode count. The LSPG-C ROMs are stabilized as more trial basis modes are included, but produce more than 10\% error on average. On the other hand, the {\color{black}MP-LSVT} ROMs are all stable (even with only 4 modes) and exhibit good error convergence with increased mode count.  

The comparisons between the Galerkin and LSPG ROMs are consistent with the observations made by Carlberg et al.~\cite{CarlbergConsvLSPG}. However, it should be pointed out that, unlike LSPG projection which requires an implicit time integrator, the {\color{black}MP-LSVT} ROM formulation is also applicable to explicit time integrators. Further, {\color{black}MP-LSVT} ROMs using an explicit time integrator perform similarly to those using an implicit time integrator, as can be seen from Fig.~\ref{2d:rom_recon_err}. This allows for more flexibility and more importantly, opens possibilities for further computational efficiency enhancements by using explicit schemes over implicit schemes.

\subsubsection{Linear Stability}
\label{2d:LinearStability}
Linear stability of the various ROMs is assessed by investigating the system matrix ($\linStabMat$) of the linearized ROM  defined in~\ref{appendix:ROMJacobMatrix}, which describes the update of the ROM state ($\solConsROMRed$)
\begin{equation}
    \solConsROMRed^{\iterIdx} = \linStabMat \solConsROMRed^{\iterIdx-1}.
    \label{rom:Cmatrix}
\end{equation}

\begin{figure}
     \centering
     \begin{subfigure}[b]{\textwidth}
         \centering
         \includegraphics[width=0.9\textwidth]{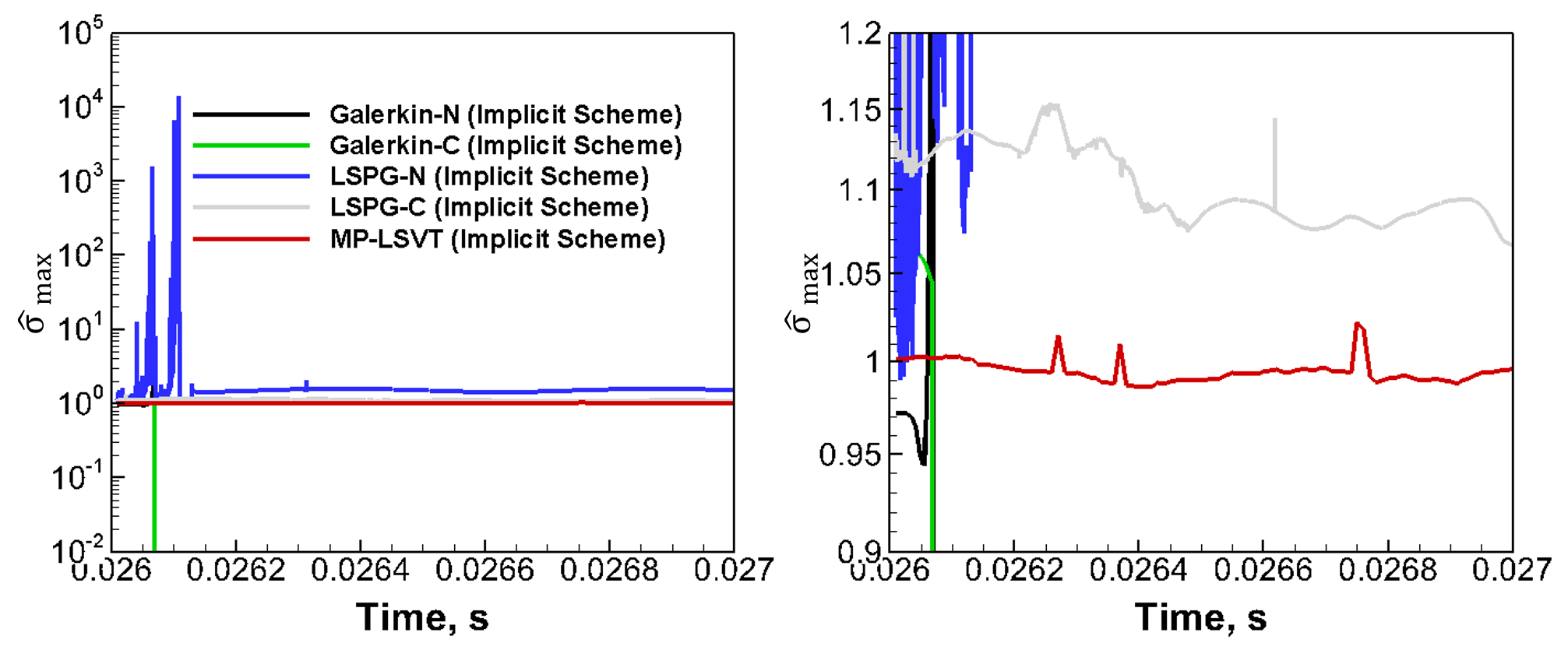}
         \caption{Implicit Scheme}
         \label{2d:rom_sigma_Imp}
     \end{subfigure}
     
     \begin{subfigure}[b]{\textwidth}
         \centering
         \includegraphics[width=0.9\textwidth]{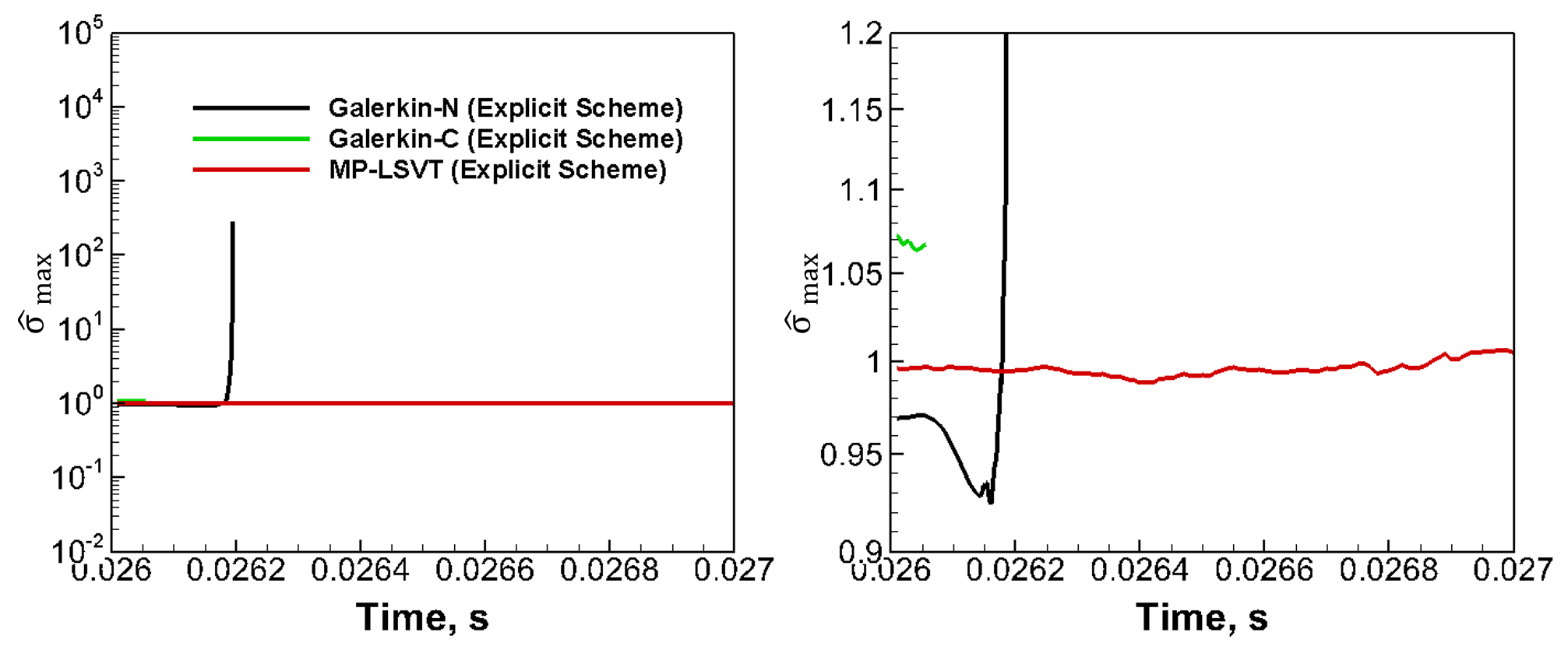}
         \caption{Explicit Scheme}
         \label{2d:rom_sigma_Exp}
     \end{subfigure}
     
     \caption{Linear stability comparisons based on the maximum singular values ($\sigmaLinStab_\text{max}$) of ROM Jacobian matrices between Galerkin, LSPG and {\color{black}MP-LSVT} ROMs with different time integration schemes for the 2D reacting  injector simulation.}\label{2d:rom_singular_value}
\end{figure}

The singular values ($\sigmaLinStab_{\genIdx}$) of the matrix $\linStabMat$ are computed at each time step of the ROM calculations. The time history of the maximum singular value ($\sigmaLinStab_\text{max}$) of each ROM method (Galerkin, LSPG and {\color{black}MP-LSVT}) and time integration scheme (explicit and implicit) are shown in in Fig.~\ref{2d:rom_singular_value}. If $\norm{\linStabMat} = \sigmaLinStab_{\text{max}}(\linStabMat) \leq 1$, the linearized ROM system is considered asymptotically stable; otherwise, it is linearly unstable. Due to the wide range of values spanned by $\sigmaLinStab_\text{max}$, Figs.~\ref{2d:rom_sigma_Imp} (left) and ~\ref{2d:rom_sigma_Exp} (left) show a wide range for consistent comparisons between all methods. Smaller ranges of y-axis centered around 1.0 are highlighted in Figs.~\ref{2d:rom_sigma_Imp} (right) and ~\ref{2d:rom_sigma_Exp} (right). 

As can be readily seen in Fig.~\ref{2d:rom_singular_value}, all the Galerkin ROMs (Galerkin-N and Galerkin-C) are unstable (indicated in Fig.~\ref{2d:rom_recon_err} as well) with the ROM calculations terminated before the end of the training time period (27 ms) with  $\sigmaLinStab_\text{max} > 1$ for a large portion of the calculations, which is a direct indicator of  instability. Though the LSPG-N method enables the ROM to complete the calculation spanning the entire 1.0 ms training time period, the accuracy of the resulting ROM is poor as seen in Fig.~\ref{2d:rom_recon_err}. This  can be attributed to the instability indicated by the high $\sigmaLinStab_\text{max} > O(10^3)$. The LSPG-C ROM shows more reasonable $\sigmaLinStab_\text{max}$ values and produces stable ROMs with acceptable accuracy, but $\sigmaLinStab_\text{max}$ remains higher than 1.0 during the entire 1.0 ms training time period. As a consequence, the  the accuracy of the resulting ROM is contaminated as seen in Fig.~\ref{2d:rom_recon_err}. On the other hand, both implicit and explicit {\color{black}MP-LSVT} ROMs exhibit significant improvement compared to Galerkin and LSPG ROMs with $\sigmaLinStab_\text{max} \approx 1.0$, and remaining below 1.0 for most of the training period. It is of course recognized that in these highly non-linear problems, stability based on linearized systems should be considered as one (and possibly imprecise) indicator.

Furthermore, the stiffness of the system matrix ($\linStabMat$) is investigated  by assessing the condition number ($\kappaLinStab$) defined as $\kappaLinStab \triangleq \sigmaLinStab_\text{max} / \sigmaLinStab_\text{min}$. The condition numbers $\kappaLinStab$ are calculated at each time step for different ROM methods. The highest ($\kappaLinStab_{max}$) and lowest ($\kappaLinStab_\text{min}$) condition numbers are compared between Galerkin, LSPG and {\color{black}MP-LSVT} and summarized in Table~\ref{table:2D_condition_number}. It can be readily seen that the {\color{black}MP-LSVT} ROM formulation results in nearly optimal conditioning. LSPG-C and Galerkin-C (Explicit Scheme) methods also produce well-conditioned systems. Unsurprisingly, the Galerkin-C ROM with an explicit scheme is unstable as indicated in Figs.~\ref{2d:rom_recon_err} and~\ref{2d:rom_sigma_Exp}. LSPG-N, Galerkin-N and Galerkin-C (Implicit Scheme) techniques are seen to generate extremely ill-conditioned matrices ($\kappaLinStab \approx O(10^{12})$).

\begin{table}
\centering
\begin{tabular}{lll} 
\toprule
ROM Method & $\kappaLinStab_\text{max}$ & $\kappaLinStab_\text{min}$ \\
\midrule
LSPG-N & $O(10^{4})$ & $1.28$ \\ 
Galerkin-N (Explicit Scheme) & $O(10^{6})$ & $4.17$ \\ 
Galerkin-N (Implicit Scheme) & $O(10^{10})$ & $1.28$ \\ 
LSPG-C & $2.0$ & $1.21$ \\ 
Galerkin-C (Explicit Scheme) & $1.8$ & $1.75$ \\ 
Galerkin-C (Implicit Scheme) & $O(10^{12})$ & $1.25$\\ 
{\color{black}MP-LSVT} (Explicit Scheme) & $1.3$ & $1.15$ \\
{\color{black}MP-LSVT} (Implicit Scheme) & $1.1$ & $1.05$ \\
\bottomrule
\end{tabular}
\caption{\label{table:2D_condition_number} Comparisons of highest condition numbers of the system matrix $\linStabMat$ with different time integration schemes for the 2D reacting injector.}
\end{table}

\begin{figure}
     \centering
     \begin{subfigure}[t]{0.45\textwidth}
         \centering
         \includegraphics[width=1.0\textwidth]{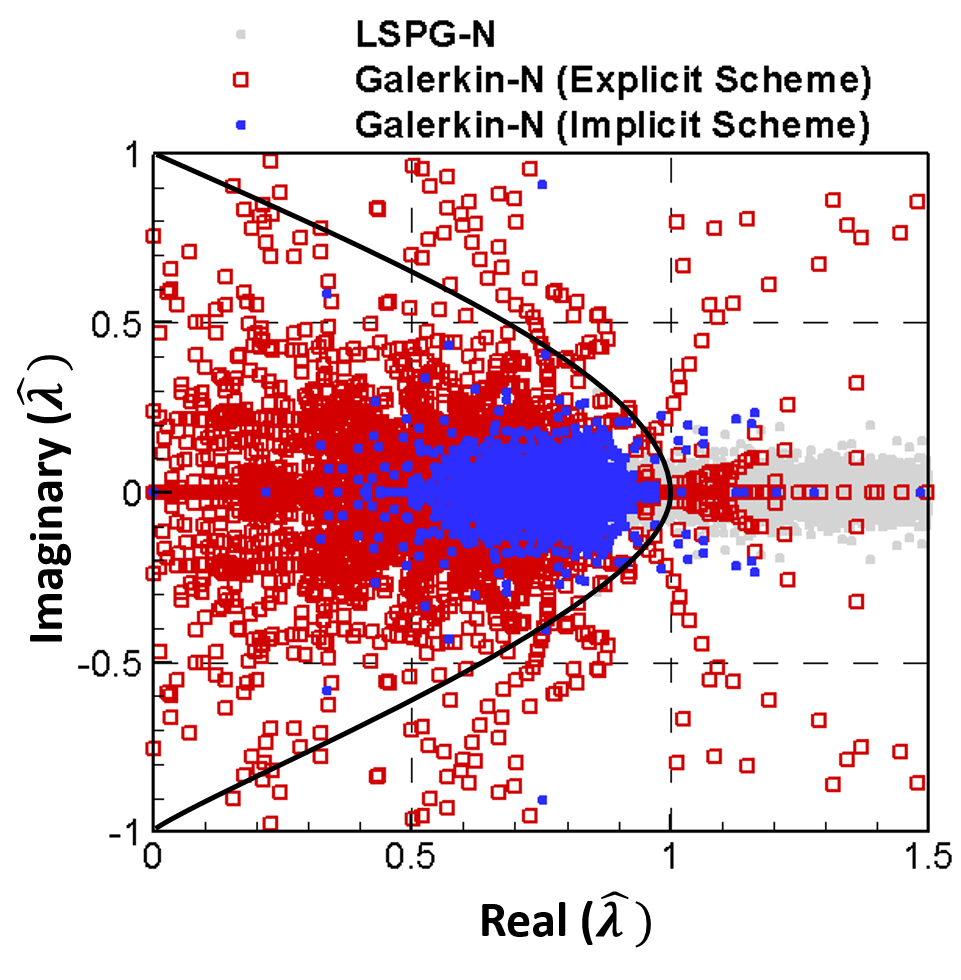}
         \caption{Galerkin-N and LSPG-N}
         \label{2d:rom_eigen_GvLSPG-N}
     \end{subfigure}
     \begin{subfigure}[t]{0.45\textwidth}
         \centering
         \includegraphics[width=1.0\textwidth]{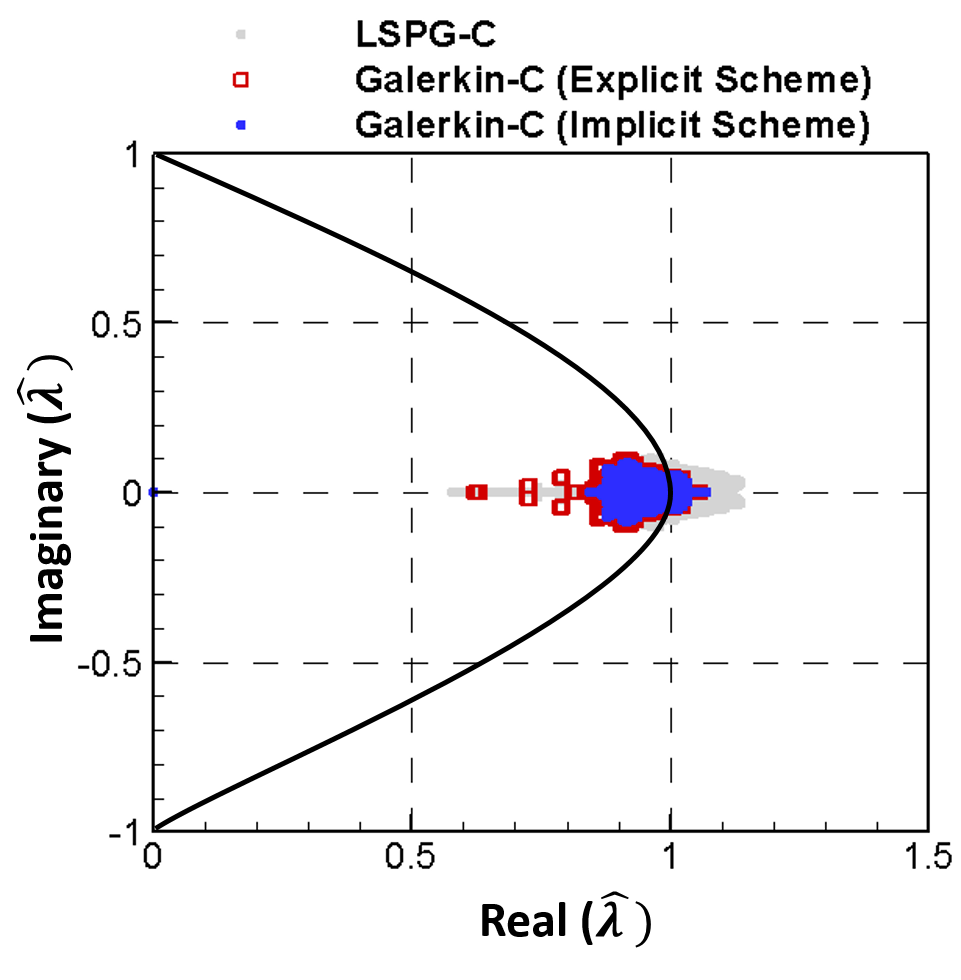}
         \caption{Galerkin-C and LSPG-C}
         \label{2d:rom_eigen_GvLSPG-C}
     \end{subfigure}
     \begin{subfigure}[t]{0.45\textwidth}
         \centering
         \includegraphics[width=1.0\textwidth]{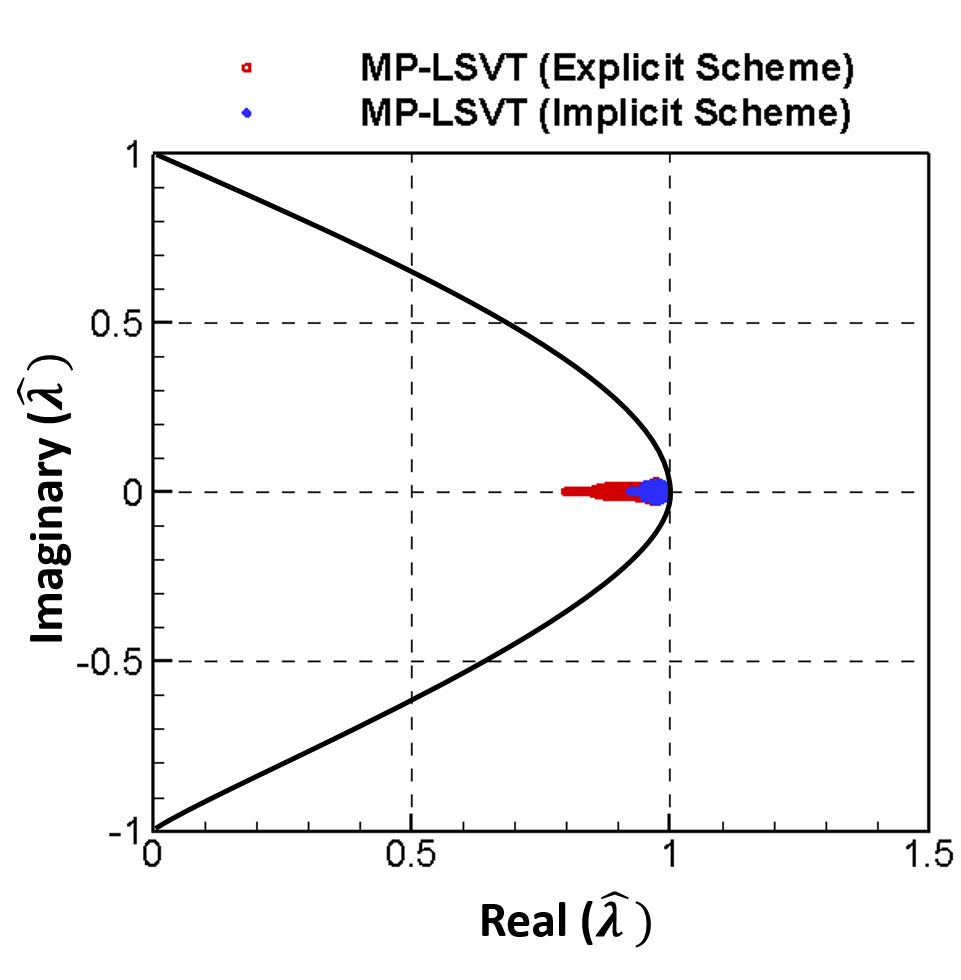}
         \caption{{\color{black}MP-LSVT}}
         \label{2d:rom_eigen_LSPT-VT}
     \end{subfigure}
     \caption{Linear stability comparisons based on eigenvalues ($\lambdaLinStab$) of ROM Jacobian matrices between Galerkin, LSPG and {\color{black}MP-LSVT} ROMs with different time integration schemes for the 2D reacting  injector simulation.}\label{2d:rom_eigen_value}
\end{figure}

As an additional indicator, the eigenvalues ($\lambdaLinStab$) of the matrix $\linStabMat$ are  displayed in Fig.~\ref{2d:rom_eigen_value}, with the solid black line depicting the unit circle. It can be seen in Figs.~\ref{2d:rom_eigen_GvLSPG-N} and~\ref{2d:rom_eigen_GvLSPG-C} that eigenvalues exist outside the unit circle for the both Galerkin and LSPG methods, although switching from non-conservative (Galerkin-N and LSPG-N) to conservative (Galerkin-C and LSPG-C) variables does improve the linear stability. On the other hand, {\color{black}MP-LSVT} ROMs indicate a stable behavior. Overall, the analysis based on singular values and eigenvalues show consistent linear stability and conditioning improvements using {\color{black}MP-LSVT} method over Galerkin and LSPG methods.

\subsubsection{Time Step Size Sensitivity}
\begin{figure}
	\centering
	\includegraphics[width=0.8\textwidth]{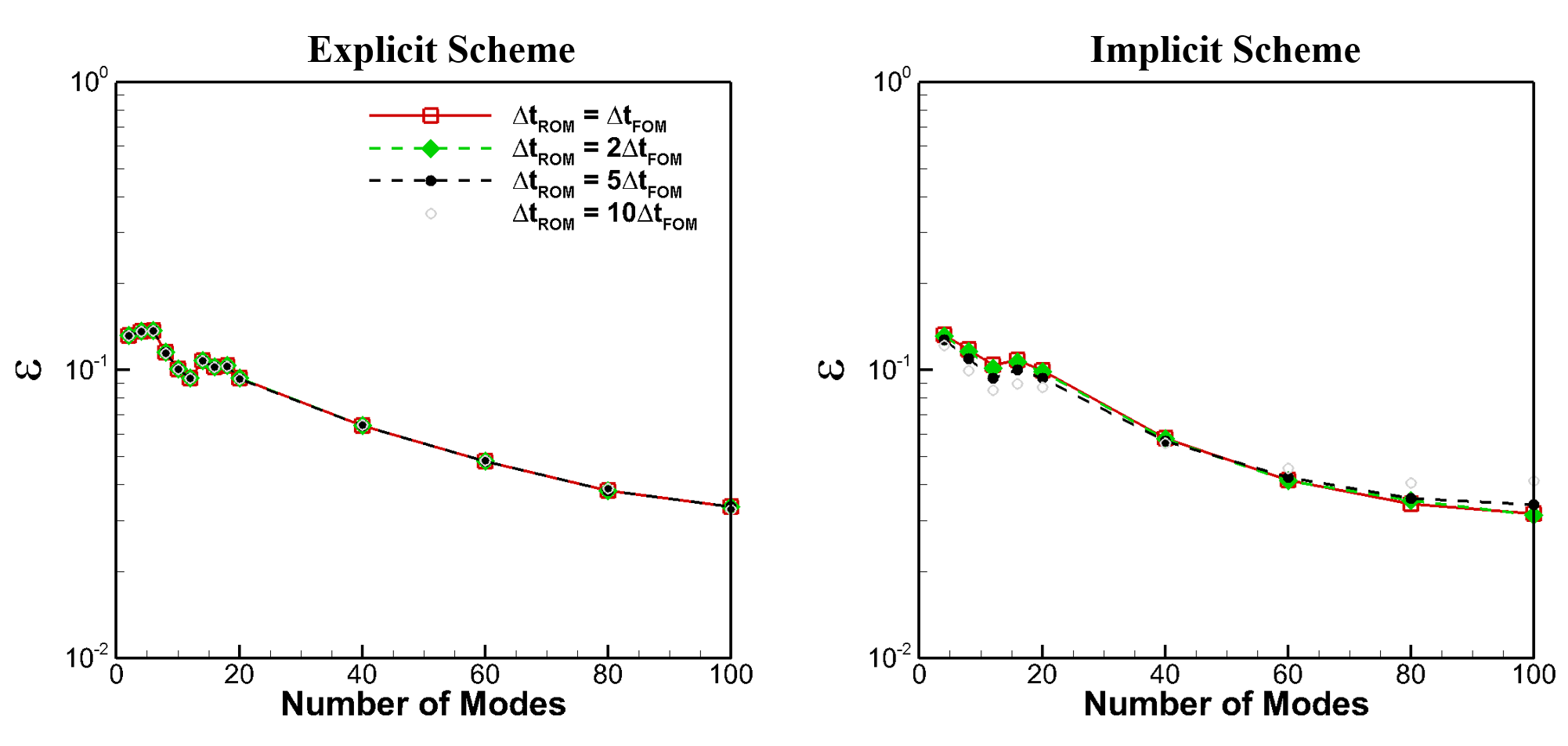}
	\caption{Sensitivity of {\color{black}MP-LSVT} ROMs to physical time step size for both explicit and implicit time integration schemes for the 2D reacting  injector simulation.}\label{2d:rom_recon_err_onDt} 
\end{figure}

\begin{figure}
	\centering
	\includegraphics[width=0.4\textwidth]{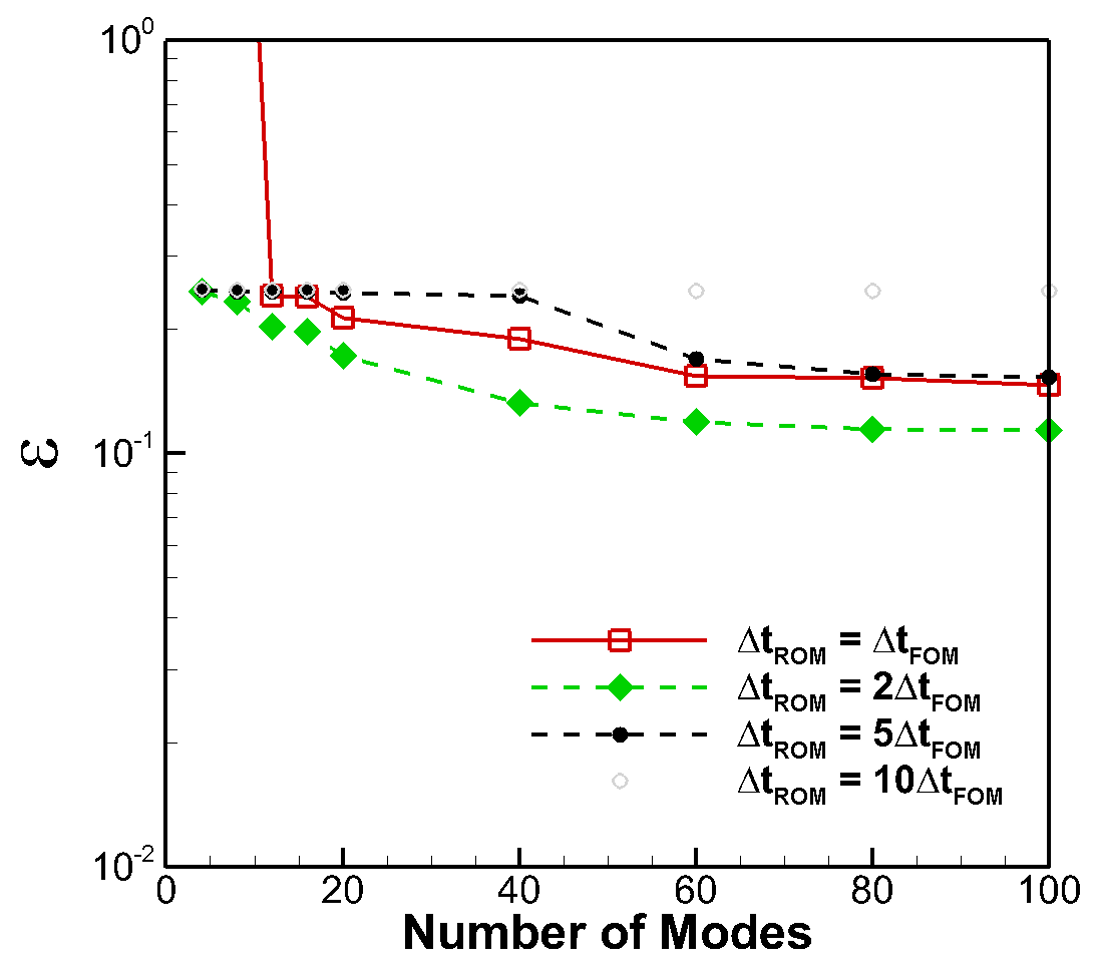}
	\caption{Sensitivity of LSPG-C ROMs with implicit time integration to physical time step size for the 2D reacting  injector simulation.}\label{2d:lspgC_rom_recon_err_onDt} 
\end{figure}

The sensitivity of the {\color{black}MP-LSVT} ROMs to the physical time step size is evaluated at four different time step sizes, $\dt_\text{ROM} \in \{ \dt_\text{FOM}, 2\dt_\text{FOM}$, $5\dt_\text{FOM}$, $10\dt_\text{FOM}\}$. The reconstruction error, as defined in Eq.~\ref{rom:reconstr_err}, is evaluated for both explicit and implicit time integration schemes. The results are shown in Fig.~\ref{2d:rom_recon_err_onDt}, displaying largely overlapping plots. Changing the time-step has a very minor impact on the solutions integrated with an implicit scheme and an even smaller effect on those integrated with an explicit scheme. Using an explicit time integration scheme, setting the physical time step to ten times that of the FOM ($\dt_\text{ROM} = 10\dt_\text{FOM}$) gives essentially identical reconstruction error to that of a ROM using the same physical time step as the FOM ($\dt_\text{ROM} = \dt_\text{FOM}$). With an implicit time integration scheme, minor differences show that the ROM solutions with a larger time step ($\dt_\text{ROM} = 5\dt_\text{FOM}$ and $10 \dt_\text{FOM}$) are marginally less accurate. The results in Fig.~\ref{2d:rom_recon_err_onDt} indicate that the {\color{black}MP-LSVT} ROMs are not as sensitive to the time step size as one might expect from the analysis by Carlberg et al.~\cite{Carlberg2017}, which showed that LSPG ROMs achieve optimal accuracy at intermediate time step sizes. Such a sensitivity to the time step size for LSPG ROMs is also observed in the current test problem, as shown in Fig.~\ref{2d:lspgC_rom_recon_err_onDt} for LSPG-C ROMs, which are identified to be the most stable method among the Galerkin and LSPG ROMs. Using a time step size of $\dt_\text{ROM} = 2\dt_\text{FOM}$ in the LSPG-C ROMs helps stabilize the unstable ROMs and improves their accuracy, but become less accurate as the time step increases to $5\dt_\text{FOM}$ and $10\dt_\text{FOM}$. This sensitivity to time step in the least-squares formulation does not appear to be a significant issue for the {\color{black}MP-LSVT} method, an aspect that requires further investigation and analysis. 
 
\subsubsection{{\color{black}MP-LSVT} ROM Enhancement with Species Limiters}
\label{sec:2d_limiters}
It has been demonstrated in Fig.~\ref{2d:rom_recon_err} that the {\color{black}MP-LSVT} ROMs are able to accurately reproduce the FOM solutions (i.e. pressure, velocities, temperature and species mass fraction fields) with approximately 3\% error using 100 trial basis modes. We now evaluate the ability of the ROM in representing the heat release rate, an important quantity of interest for reacting flow problems. 

\begin{figure}
	\centering
	\includegraphics[width=0.8\textwidth]{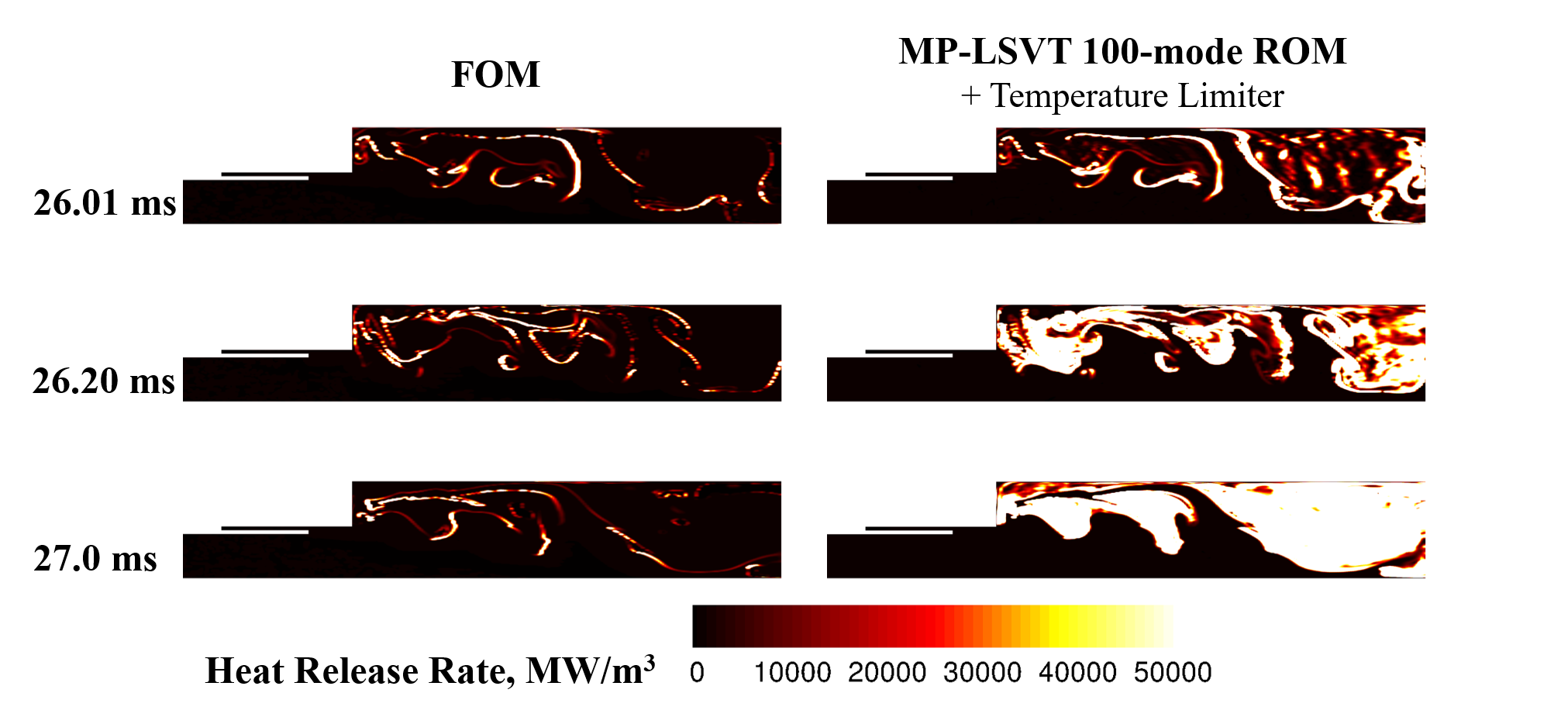}
	\caption{Comparisons of unsteady heat release rate field at representative time instances between FOM and {\color{black}MP-LSVT} 100-mode ROM with temperature limiter only.}\label{2d:q_cmp_FOMvsLSPGVTROM_noSpecLimiter} 
\end{figure}

\begin{figure}
	\centering
	\includegraphics[width=0.8\textwidth]{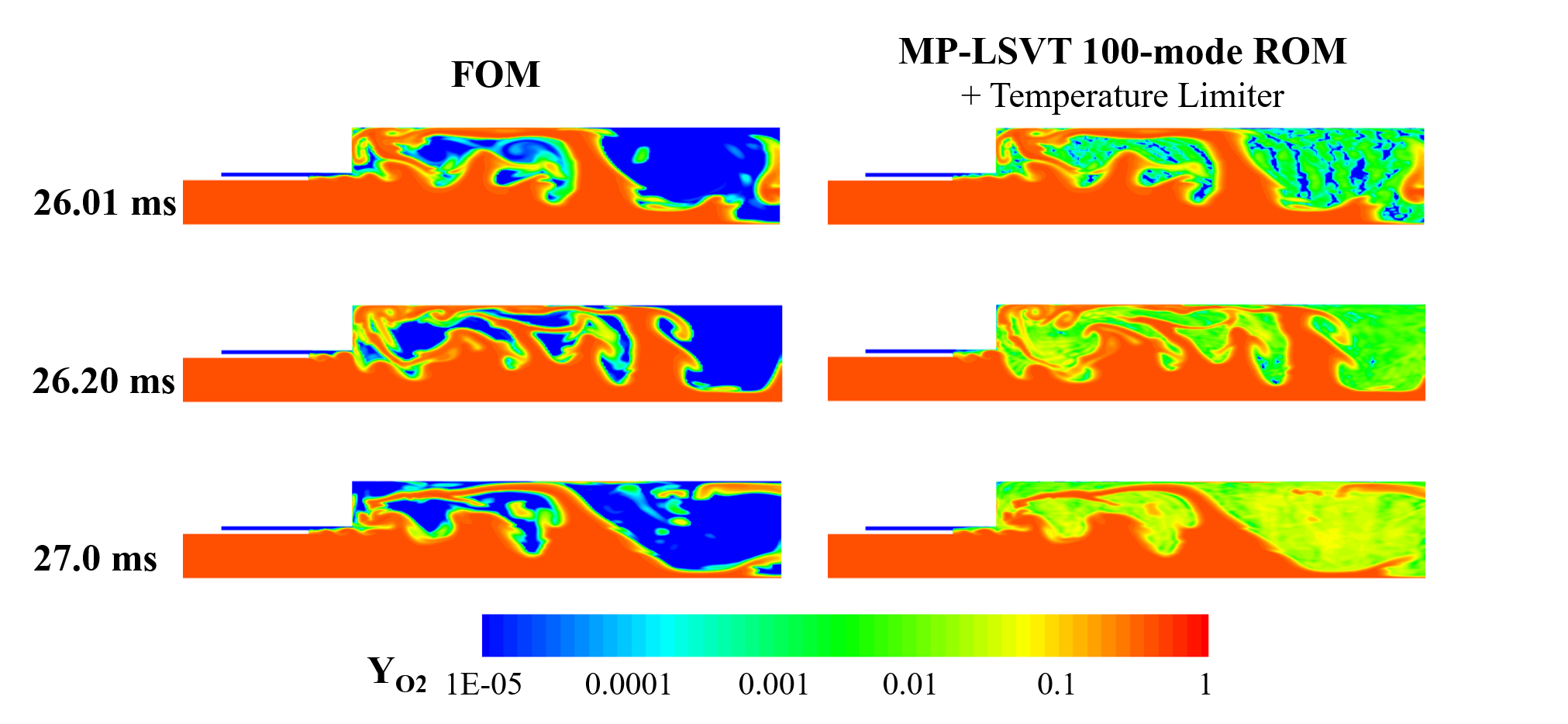}
	\caption{Comparisons of \oxygen\ mass fraction field at representative time instances between FOM and {\color{black}MP-LSVT} 100-mode ROM with temperature limiter only.}\label{2d:O2_cmp_FOMvsLSPGVTROM_noSpecLimiter} 
\end{figure}

Comparisons of the instantaneous heat release rate fields at three different time instances are shown in Fig.~\ref{2d:q_cmp_FOMvsLSPGVTROM_noSpecLimiter}. The {\color{black}MP-LSVT} ROM uses 100 trial basis modes, solved using the explicit time integration scheme, and applies the temperature limiter with bounds $T_{min} = 270$ K and $T_{max} = 2{,}850$ K. The species limiter introduced in Section~\ref{limiters} is not yet applied for this simulation. It is seen that the reconstruction of the heat release rate  is very inaccurate with  excessive burning occurring within the recirculation region. Such errors can usually be attributed to the spurious oscillations in species mass fractions near sharp gradients in the species mass fraction field. Figure~\ref{2d:O2_cmp_FOMvsLSPGVTROM_noSpecLimiter} compares the \oxygen\ mass fraction field between the FOM and ROM at the same time instances in Fig.~\ref{2d:q_cmp_FOMvsLSPGVTROM_noSpecLimiter}. The contour levels of the \oxygen\ mass fraction field are selected to be in log scale to highlight the presence of these small magnitude ($<$ 0.1\%) spurious oscillations in the ROM. These are concentrated within the recirculation region and coincide with the excessive heat release rate (Fig.~\ref{2d:q_cmp_FOMvsLSPGVTROM_noSpecLimiter}). 

\begin{figure}
	\centering
	\includegraphics[width=0.8\textwidth]{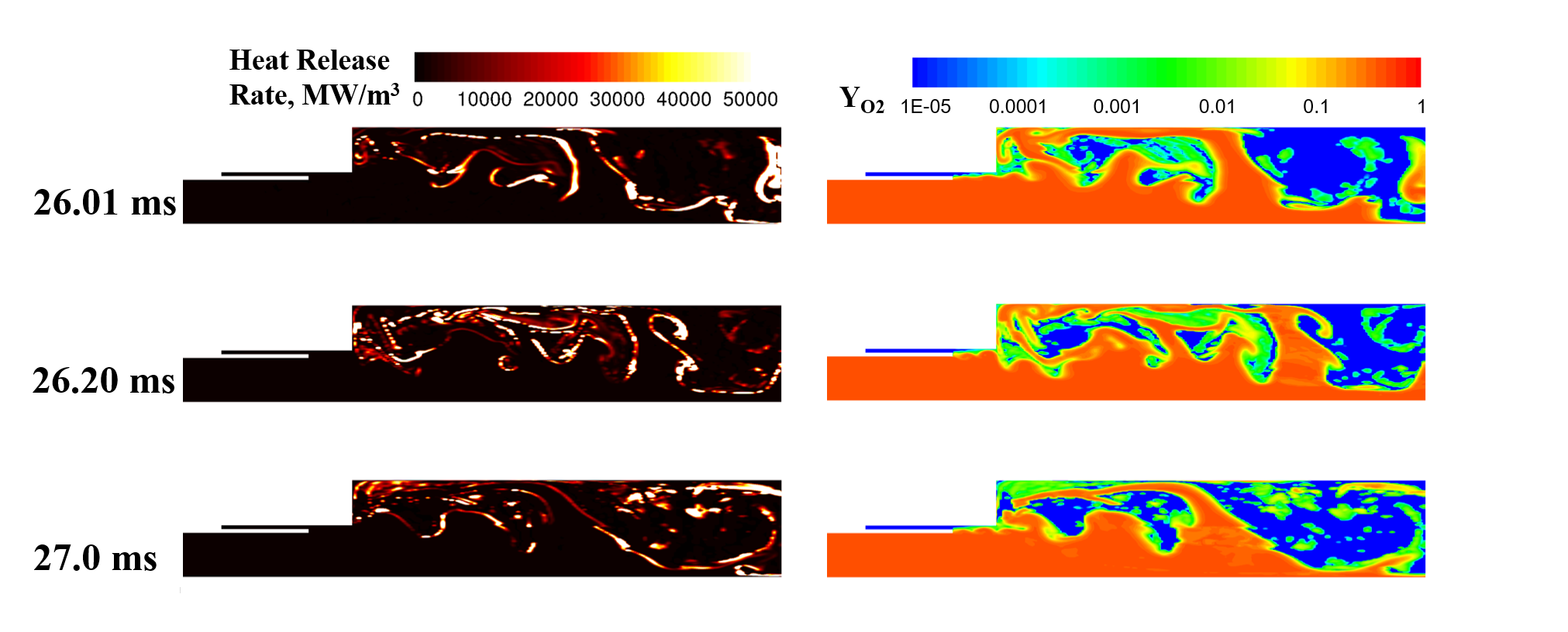}
	\caption{Unsteady heat release rate and $\text{O}_2$ mass fraction fields at representative time instances for {\color{black}MP-LSVT} 100-mode ROM with temperature and species limiters.}\label{2d:qO2_cmp_FOMvsLSPGVTROM_SpecLimiter} 
\end{figure}

Interestingly, the spurious oscillations in the \oxygen\ mass fraction field and the resulting excessive burning did not make the ROM unstable in this case, and the ROM can still provide a reasonable representation of the solution variables (pressure, velocities, temperature and species mass fractions). However, the ability of the ROM to represent the heat release rate is  lacking.  Therefore, the species limiter introduced in Eq.~\ref{rom:species_limiter_nonpremixed} is applied to the {\color{black}MP-LSVT} 100-mode ROM here with $T_{th} = 2{,}200$ K (80\% of the adiabatic flame temperature, $T_{ad}$) and $\delta = 1 \times 10^{-5}$. The resulting flow field snapshots, shown in Fig.~\ref{2d:qO2_cmp_FOMvsLSPGVTROM_SpecLimiter}, display significant improvements in representing the heat release rate. It can be readily seen that most of the spurious oscillations in Fig.~\ref{2d:O2_cmp_FOMvsLSPGVTROM_noSpecLimiter} have been eliminated by the species limiter, which yields a significantly more accurate representation of the heat release rate compared to Fig.~\ref{2d:q_cmp_FOMvsLSPGVTROM_noSpecLimiter}. The species limiter proves to be effective and necessary in enabling the ROM to provide an accurate representation of important features in reacting flow problems (e.g. heat release rate and flame propagation speed) as also illustrated in~\ref{appendix:species_oscillations}. 

\subsection{3D Reacting  Injector}
\label{3d:injector}
\begin{figure}
	\centering
	\includegraphics[width=0.8\textwidth]{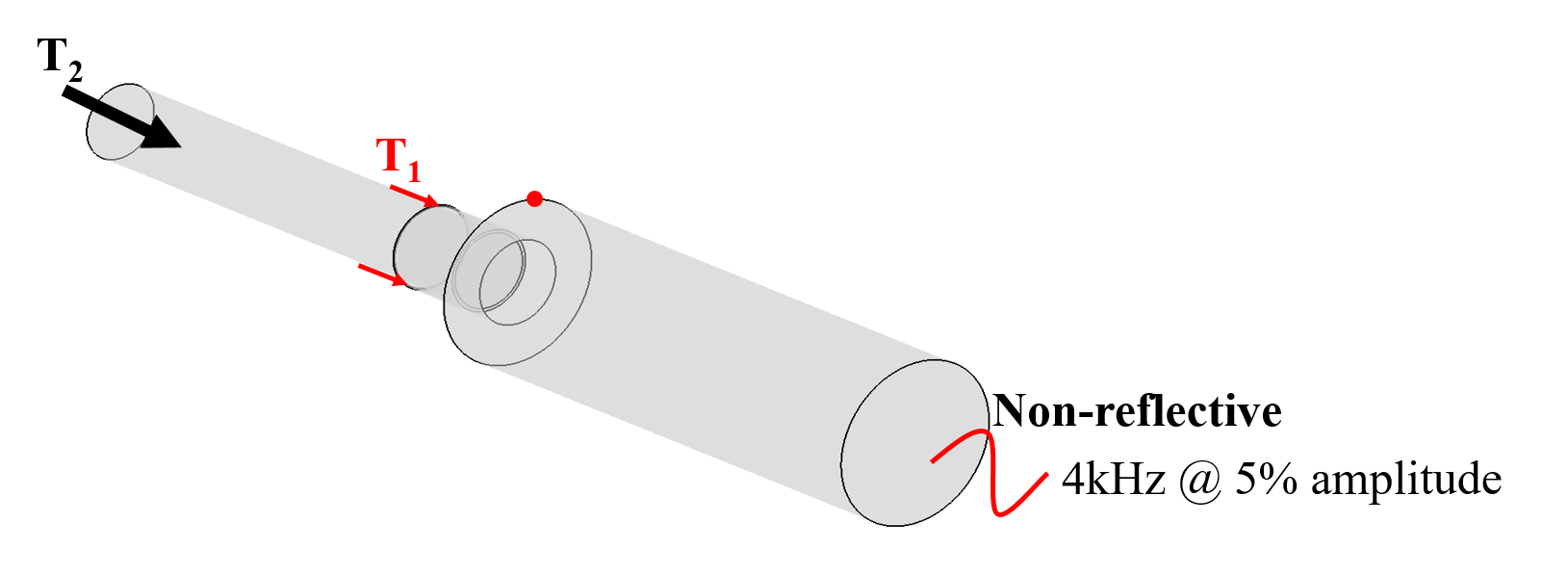}
	\caption{Computational configuration of 3D reacting  injector.}\label{3d:geometry} 
\end{figure}

Next, we extend the investigations to a  3D representation of the generic laboratory-scale single injector combustor~\cite{YuJPP}. The configuration is shown in Fig.~\ref{3d:geometry}. Similar to the 2D case in Section~\ref{2d:injector}, the 3D problem consists of a shear coaxial injector with an outer passage, $T_1$, that introduces fuel (100\% \methane\ at 300 K) near the downstream end of the coaxial center passage, $T_2$, that feeds oxidizer (42\% gaseous \oxygen\ by mass and 58\% gaseous \water\ by mass at 660 K) to the combustion chamber. 

The operating conditions for this 3D reacting flow simulation are maintained similar to the conditions in the laboratory combustor~\cite{YuJPP,YuPhD}, with an adiabatic flame temperature of approximately 2,600 K and a mean chamber pressure of 1.1 MPa. The $T_1$ and $T_2$ streams are both fed with constant mass flow rates: 0.027 kg/s and 0.32 kg/s, respectively. As in the 2D case, a non-reflective boundary condition is enforced at the downstream outflow boundary. A sinusoidal pressure perturbation (with amplitude 5\% of the mean pressure) at 4,000 Hz is imposed at this boundary. Different from the finite rate global reaction model used for the 2D problem, combustion is represented by the flamelet progress variable (FPV) model~\cite{Pierce2004_FPV} with more detailed chemical kinetics, GRI-1.2~\cite{GRI_12}, which consists of 32 species and 177 chemical reactions. The chemical species are treated as thermally perfect gases. Note that although 32 chemical species are modeled, the FPV model only solves transport equations for three scalar quantities: the mean mixture fraction (${Z}_{mean}$), the mixture fraction variance (${Z''^2}$), and the reaction progress variable (${C}_{mean}$) as described in~\ref{appendix:fom_eq}. Individual chemical species mass fractions are looked up from pre-computed flamelet manifolds~\cite{FlameletJointPDFModel}.

\begin{figure}
	\centering
	\includegraphics[width=0.8\textwidth]{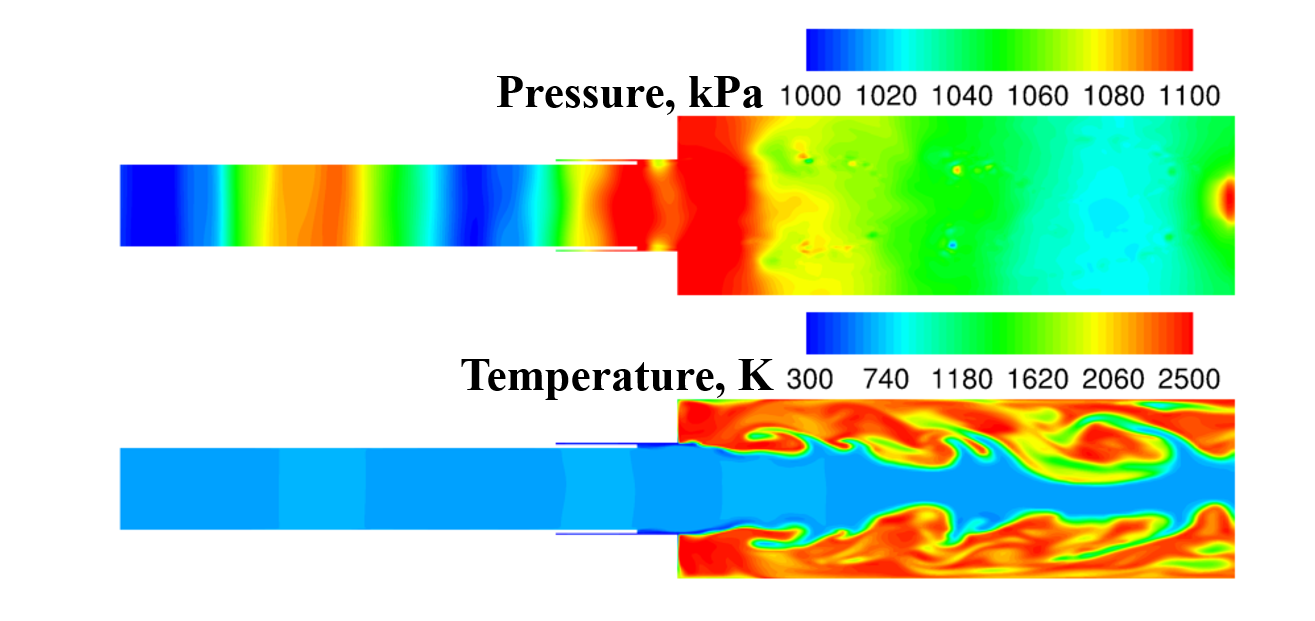}
	\caption{Representative instantaneous snapshots of pressure (top) and temperature (bottom) fields from FOM simulation of the 3D reacting  injector.}\label{3d:fom_snapshots} 
\end{figure}

The 3D FOM is solved using the second-order accurate backwards differentiation formula and dual time-stepping, with a constant physical time step size of 1.0 $\mu$s. The computational mesh is composed of a total of 589,395 finite volume cells, resulting in a total of 4,715,160 degrees of freedom. A representative instantaneous snapshot is shown in Fig.~\ref{3d:fom_snapshots}. Similar to the 2D solutions in Fig.~\ref{2d:fom_snapshots}, the pressure exhibits global dynamics while the combustion dynamics, characterized by temperature, show more local convection-dominated features. However, due to the improved modeling of mixing in 3D simulations, the combustion dynamics exhibit more coherent structures than in the 2D case where the dynamics are more intermittent. Solution snapshots are stored at every physical time step over a time duration of 1.0 ms, corresponding to a total of 1,000 snapshots. All snapshots are used to generate POD bases for ROM construction.

\subsubsection{POD and ROM Characteristics}
\label{3d:results-pod_rom}

\begin{figure}
	\centering
	\includegraphics[width=0.6\textwidth]{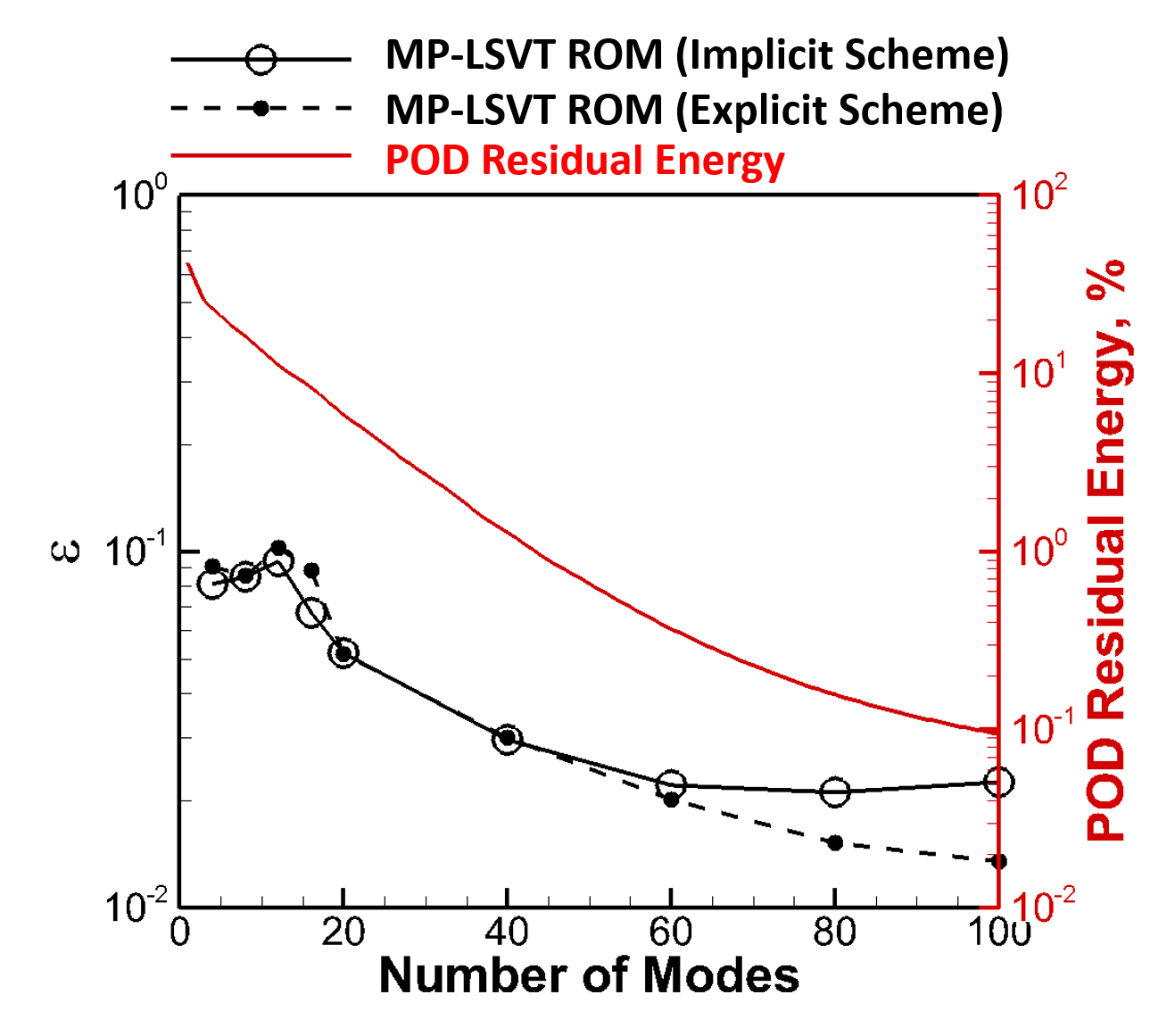}
	\caption{Global ROM reconstruction error for {\color{black}MP-LSVT} method with different time integration schemes and POD residual energy distribution for the 3D reacting  injector simulation.}\label{3d:romErr_podEnergy} 
\end{figure}

The characteristics of the POD bases and ROMs are evaluated based on the POD residual energy (Eq.~\ref{pod:res_energy}) and ROM reconstruction error (Eq.~\ref{rom:reconstr_err}), shown in Fig.~\ref{3d:romErr_podEnergy}. The POD residual energy still shows a relatively slow decay similar to the 2D case (Fig.~\ref{2d:pod_res_energy}). More than 43 POD modes are required to retrieve 99\% of the energy, and 100 POD modes are required to retrieve 99.9\%. ROMs are constructed using the {\color{black}MP-LSVT} formulation with either the explicit four-stage Runge--Kutta method or the implicit second-order accurate backwards differentiation formula.  The temperature limiter ($T_{min} = 270$ K and $T_{max} = 2{,}800$ K), determined by the temperature of the $T_1$ stream (300 K), and the adiabatic flame temperature ($2,600$ K). In addition, species limiters ($T_{th} = 2{,}500$ K, $C_\text{ref} = 0.85$, $Z_\text{st} = 0.096$, $\delta = 1\times 10^{-5}$) are applied during the {\color{black}MP-LSVT} ROM calculations to ensure reasonable modeling of the heat release rate. 
It can be readily seen in Fig.~\ref{3d:romErr_podEnergy} that all {\color{black}MP-LSVT} ROMs are stable using both explicit and implicit time integration schemes.  Less than 3\% $L^2$ error can be reached by using more than 40 POD trial basis modes.

\subsubsection{Hyper-reduction}\label{3d:injector_hyperReduction}
\begin{figure}
	\centering
	\includegraphics[width=0.7\textwidth]{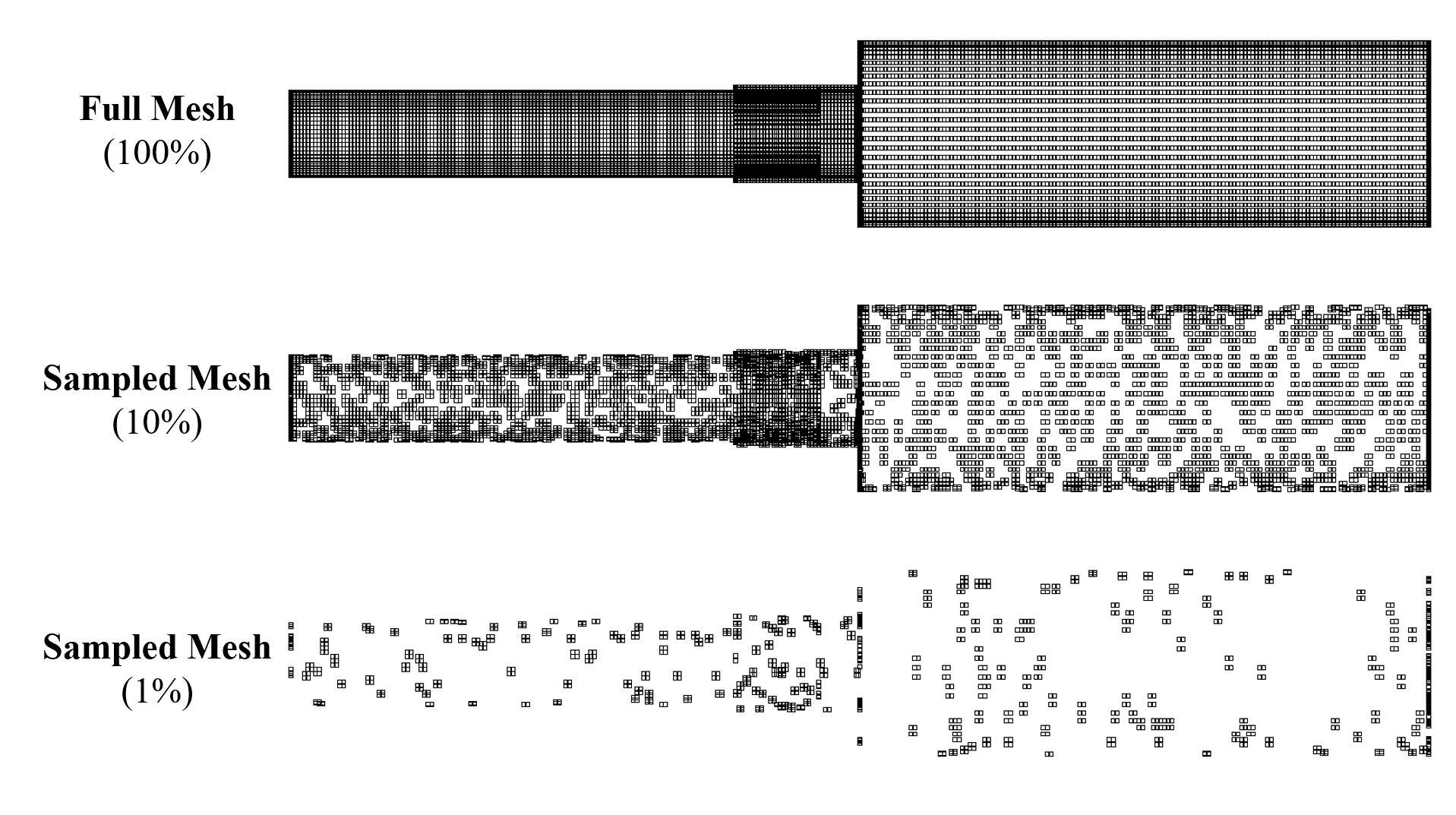}
	\caption{Comparisons of sampled meshes for the 3D reacting  injector simulation.}\label{3d:rom_sampling} 
\end{figure}

In this section, we extend the investigations to {\color{black}MP-LSVT} ROMs with hyper-reduction, as introduced in Section~\ref{hyper_reduction}, to achieve enhancement in computational efficiency. As mentioned previously, the rank-revealing QR factorization and randomized oversampling~\cite{PeherstorferODEIM} are used to determine the selection of the mesh points. Figure~\ref{3d:rom_sampling} displays sampled cells of the two sparse sampled meshes investigated here, alongside the fully-sampled mesh.

The hyper-reduced {\color{black}MP-LSVT} ROMs are evaluated based on their accuracy (reconstruction error) and computational efficiency. Computational efficiency is measured by both wall clock time (real-world time to complete the simulation) and CPU time (wall clock time multiplied by the number of computational cores used). The FOM computations were performed on 4 compute nodes (two Haswell CPUs @ 2.60GHz per node, 20 cores per node) and 128 GB of RAM. The training data generation required approximately  710 core-hours, to simulate 1 ms (1,000 snapshots). Results using the explicit time integrator and implicit time integrator are displayed in Table~\ref{table:3D_hyperreduction_exp} and Table~\ref{table:3D_hyperreduction_imp}, respectively, for three different physical time step sizes. Computational efficiency results are reported as the ratio of the FOM computational time to the ROM computational time.

It can be readily seen that the reconstruction errors for the hyper-reduced ROMs are similar to that of the unsampled ROM. All {\color{black}MP-LSVT} ROMs achieve a reconstruction error below 2.5\% using the explicit time integration scheme, and the hyper-reduced ROMs lead to greatly improved efficiency with fewer sampled degrees of freedom and larger physical time steps. This does not seem to affect the accuracy of the explicit {\color{black}MP-LSVT} ROMs, consistent with the observations for the 2D case in Fig.~\ref{2d:rom_recon_err_onDt}. However, the reconstruction errors from implicit {\color{black}MP-LSVT} ROMs are sensitive to the physical time steps, with error increasing with the time step. A possible contributing factor to the sensitivity of implicit {\color{black}MP-LSVT} ROMs to the physical time steps is that implicit time integration schemes require the calculation of Jacobians from the ROM-reconstructed solutions. It is possible that accuracy may be impacted not only by the truncation of trial basis modes (i.e. loss of small-scale dynamics) but also by errors in the computed Jacobians. In general, Jacobian calculations are more challenging for extremely stiff reacting flow simulations, and these errors may escalate in regions where the solution deviates from the original calculation when a larger physical time step is used.

It should be pointed out that even without hyper-reduction, a larger time step and explicit time integration scheme can already achieve $> O(20)$ acceleration in wall clock and CPU times. With hyper-reduction, more than a factor of $O(300)$ improvement in wall clock time and a factor of $O(600)$ improvement in CPU time can be achieved with an explicit time integration scheme and a physical time step 10 times that of the FOM, producing less than 2\% reconstruction error.

\begin{table}
\centering
\begin{tabular}{lllll} 
\toprule
Time Step Ratio & Mesh & $\errROMVsFOM$ & Wall Clock Time Ratio & CPU Time Ratio \\
\midrule
\multirow{3}{1em}{1.0} & Full Mesh (100\%) & $2.0 \times 10^{-2}$ & 2.6 & 2.6 \\
& Sampled Mesh (10\%) & $1.3 \times 10^{-2}$ & 7.3 & 7.3 \\
& Sampled Mesh (1\%) & $1.5 \times 10^{-2}$ & 32.1 & 64.3 \\
\midrule
\multirow{3}{1em}{5.0} & Full Mesh (100\%) & $1.9 \times 10^{-2}$ & 13.1 & 13.1 \\
& Sampled Mesh (10\%) & $1.8 \times 10^{-2}$ & 37.1 & 37.1 \\
& Sampled Mesh (1\%) & $2.3 \times 10^{-2}$ & 161.8 & 323.6 \\
\midrule
\multirow{3}{1em}{10} & Full Mesh (100\%) & $1.6 \times 10^{-2}$ & 26.2 & 26.2 \\
& Sampled Mesh (10\%) & $1.4 \times 10^{-2}$ & 71.7 & 71.7 \\
& Sampled Mesh (1\%) & $1.6 \times 10^{-2}$ & 319.0 & 638.0 \\
\bottomrule
\end{tabular}
\caption{\label{table:3D_hyperreduction_exp} Summary of global ROM reconstruction error, wall clock time and CPU time ratios of the hyper-reduced {\color{black}MP-LSVT} ROMs using explicit scheme with different number of sampling points and time steps sizes.}
\end{table}

\begin{table}
\centering
\begin{tabular}{lllll} 
\toprule
Time Step Ratio & Mesh & $\errROMVsFOM$ & Wall Clock Time Ratio & CPU Time Ratio \\
\midrule
\multirow{3}{1em}{1.0} & Full Mesh (100\%) & $2.2 \times 10^{-2}$ & 1.8 & 1.8 \\
& Sampled Mesh (10\%) & $1.5 \times 10^{-2}$ & 5.8 & 5.8 \\
& Sampled Mesh (1\%) & $1.6 \times 10^{-2}$ & 24.7 & 49.5 \\
\midrule
\multirow{3}{1em}{5.0} & Full Mesh (100\%) & $6.2 \times 10^{-2}$ & 4.5 & 4.5 \\
& Sampled Mesh (10\%) & $4.2 \times 10^{-2}$ & 27.6 & 27.6 \\
& Sampled Mesh (1\%) & $3.8 \times 10^{-2}$ & 124.5 & 249.0 \\
\midrule
\multirow{3}{1em}{10} & Full Mesh (100\%) & $8.0 \times 10^{-2}$ & 9.0 & 9.0 \\
& Sampled Mesh (10\%) & $7.0 \times 10^{-2}$ & 56.7 & 56.7 \\
& Sampled Mesh (1\%) & $6.6 \times 10^{-2}$ & 249.3 & 498.7 \\
\bottomrule
\end{tabular}
\caption{\label{table:3D_hyperreduction_imp} Summary of global ROM reconstruction error, wall clock time and CPU time ratios of the hyper-reduced {\color{black}MP-LSVT} ROMs using implicit scheme with different number of sampling points and time steps sizes.}
\end{table}

\subsubsection{Evaluations of Future-state Prediction Capabilities}

\begin{figure}
     \centering
     \includegraphics[width=0.6\textwidth]{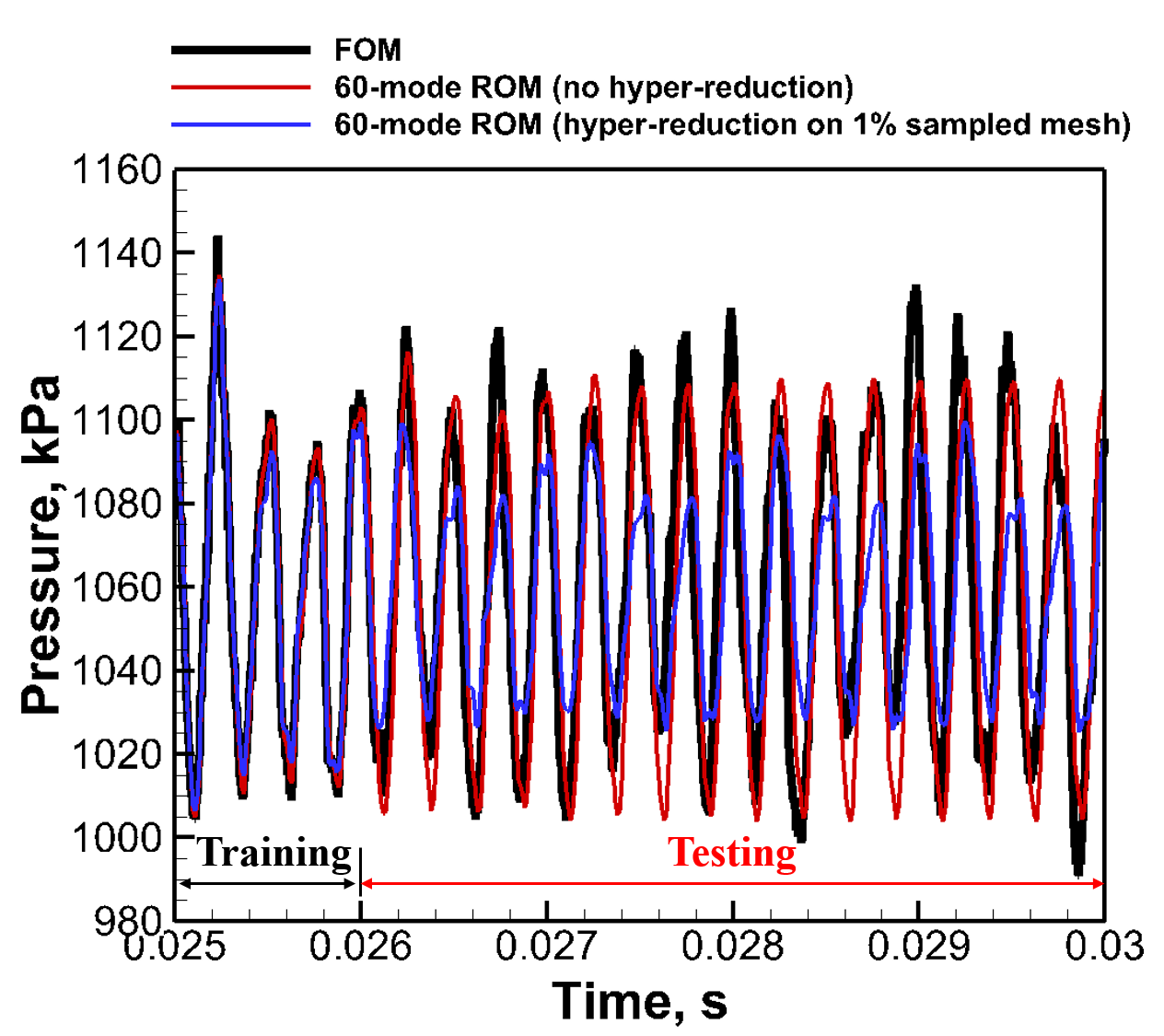}
     \caption{Comparisons of the local pressure time histories between FOM and ROM (with and without hyper-reduction)}\label{3d:pressure_cmp}
\end{figure}

The final evaluation is performed on the ability of the {\color{black}MP-LSVT} ROMs to provide future-state predictions beyond the 1.0 ms training duration used to compute the POD trial bases. The evaluation is performed for {\color{black}MP-LSVT} ROMs using 60 trial basis modes and an explicit time integration scheme. ROMs are constructed with and without hyper-reduction, where the hyper-reduced ROM uses the 1\% sampled mesh shown in Fig.~\ref{3d:rom_sampling}. All ROMs are computed with a time step ten times that of the FOM time step. The local pressure time histories, obtained at the highlighted location in Fig.~\ref{3d:geometry}, are compared for qualitative assessment as shown in Fig.~\ref{3d:pressure_cmp} within the training period ($\timeVar = [25, 26]$ ms) and beyond -- labeled as the testing period ($\timeVar = (26,30]$ ms). Within the training period, both ROMs with and without hyper-reduction are able to accurately capture the characteristics of the pressure oscillations in amplitude and phase. As expected, some discrepancies between FOM and ROMs arise beyond the training period but the ROMs are able to accurately predict the phase of the pressure oscillations. The ROM without hyper-reduction is able to predict the amplitude reasonably well (slight over-prediction) while the ROM with hyper-reduction seems to under-predict the  amplitude. 

The accuracy is further quantified and evaluated based on the ROM error in representing the $\varIdx^\text{th}$ physical quantity (e.g. $p, \mathbf{u}, T, Y_j$), estimated at each time step, $n$, defined in two levels. First, the ROM error is evaluated based on the projected FOM solutions as follows:
 \begin{equation}
    \errROMVsProj^{\iterIdx}_{\varIdx} = \frac{\norm{\solPrimROMFullVar^{\iterIdx} - \solPrimROMProjVar^{\iterIdx}}}{\norm{\solPrimROMProjVar^{\iterIdx}}},
    \label{rom:proj_err_onProjFOM}
\end{equation}
where $\solPrimROMFullVar^{\iterIdx}$ represents the $\varIdx^\text{th}$ solution variable of the state vector, $\solPrimROMFull^{\iterIdx}$, at time step $\iterIdx$ from ROM simulations following Eq.~\ref{pod:expansion_qp}. We define $\solPrimROMProjVar^{\iterIdx}$ to represent the $\iterIdx^\text{th}$ solution variable of the state vector, $\solPrimROMProj^{\iterIdx}$, at time step $\iterIdx$, evaluated as $\solPrimROMProj^{\iterIdx}= \solPrimFOMRef + \scaleMatPrim^{-1} \trialBasisPrim \trialBasisPrim^T \solPrimFOM^{\iterIdx}$. similar to Eq.~\ref{pod:expansion_qp}. This is referred to as the projected FOM solution.

\begin{figure}
	\centering
    \begin{subfigure}[b]{\textwidth}
         \centering
         \includegraphics[width=0.7\textwidth]{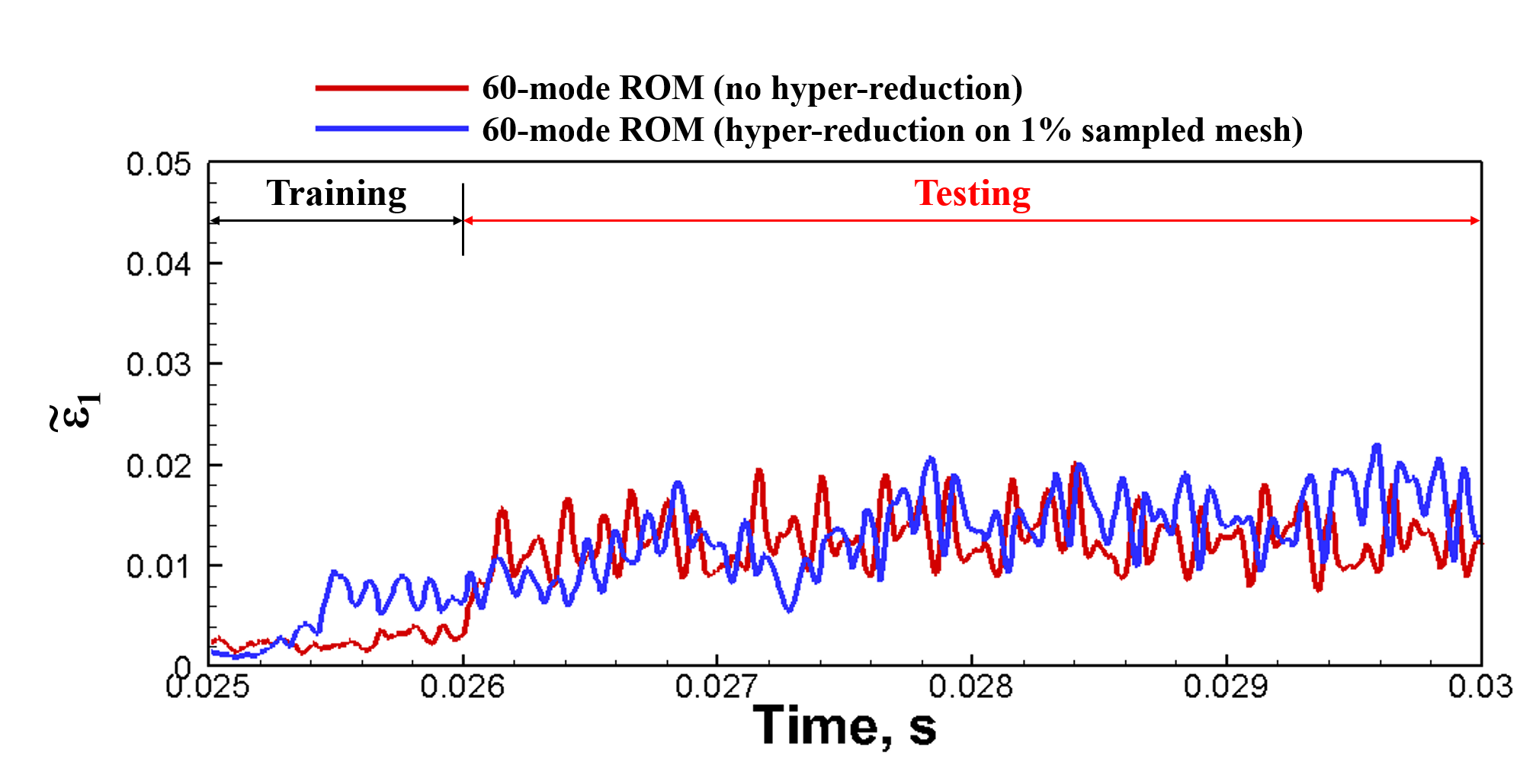}
         \caption{Pressure error time history.}
          \label{3d:SP-LSVT-errPonPOD}
    \end{subfigure}
    \centering
    \begin{subfigure}[b]{\textwidth}
         \centering
         \includegraphics[width=0.7\textwidth]{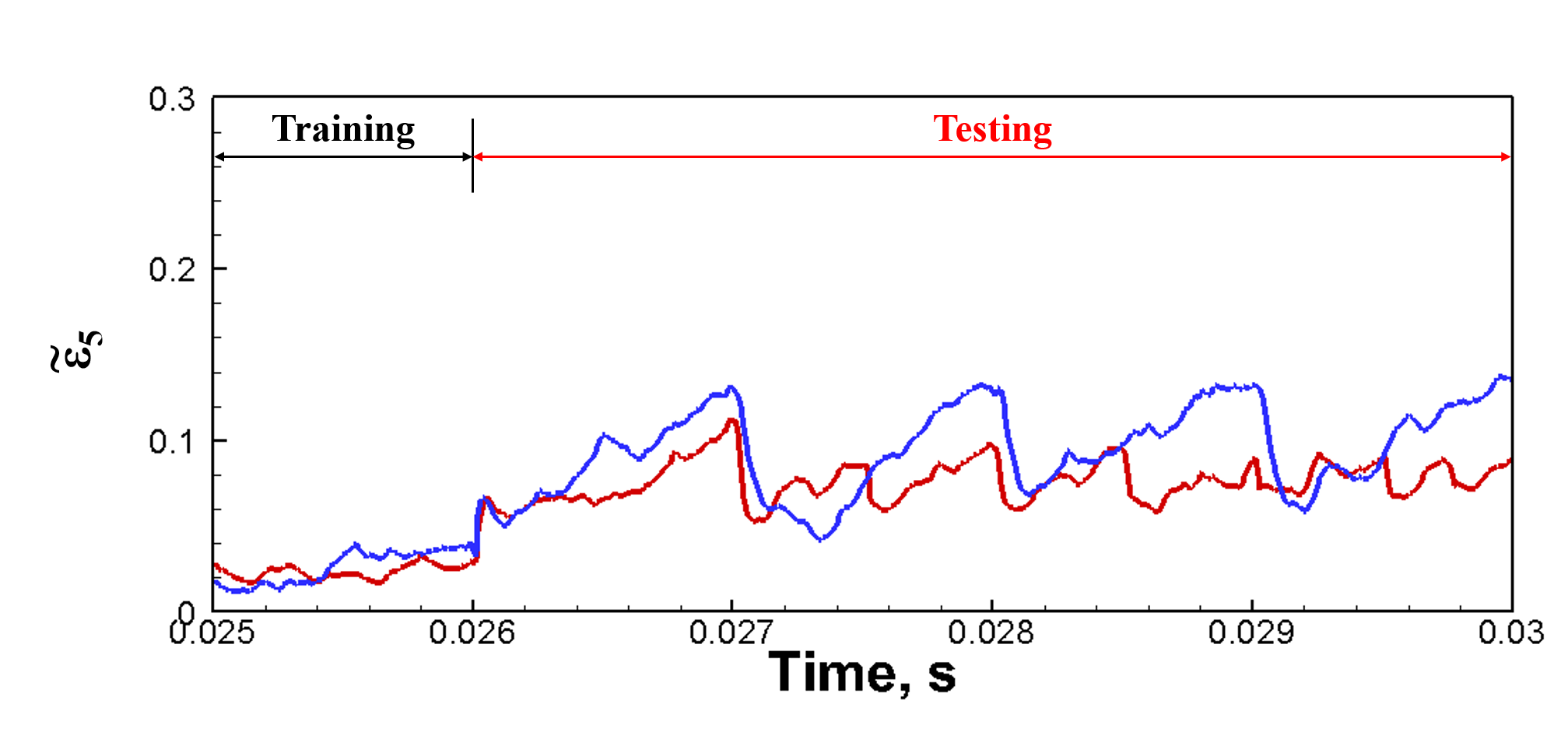}
         \caption{Temperature error time history.}
          \label{3d:SP-LSVT-errTonPOD}
    \end{subfigure}
	\caption{Comparisons of time evolution of errors with respect to the projected FOM solutions.}\label{3d:rom_future_prjErronPOD} 
\end{figure}

The ROM error (with respect to the projected FOM solutions) in the pressure ($\varIdx=1$ in Eq.~\ref{rom:proj_err_onProjFOM}) and temperature ($\varIdx=5$ in Eq.~\ref{rom:proj_err_onProjFOM}) fields are compared in Fig.~\ref{3d:rom_future_prjErronPOD}. Within the training period, the ROMs are able to track the projected FOM solutions for both pressure and temperature accurately (well below 1\% error for pressure and well below 4\% error for temperature). Beyond the training period, however, the error start to increase noticeably. In the testing period, error in the pressure field is still well below 2\% error, which indicates good accuracy of the ROMs in matching the projected pressure field. On the other hand, the time evolution of error in the temperature field exhibits error as high as 12\%, with large-amplitude oscillations between 4\% and 12\%. This is a particular issue for the ROM with hyper-reduction. These oscillations result in approximately 8\% error on average, which can still be considered marginally accurate given the fact that the temperature field is characterized with sharp gradients. Such sharp gradients can be challenging features to match based on $L^2$-norm error, as measured by Eq.~\ref{rom:proj_err_onProjFOM}.

\subsubsection{Limitations of Linear Static Bases}

Though the {\color{black}MP-LSVT} ROMs perform well in comparison to  the projected FOM solutions in both training and testing periods, it is necessary to evaluate the ROM predictions based on the truth (i.e. the FOM solutions). Therefore, the second-level evaluation bases the ROM error on the FOM solutions

\begin{equation}
    \errROMVsFOM^{\iterIdx}_{\varIdx} = \frac{\norm{\solPrimROMFullVar^{\iterIdx} - \solPrimFOMVar^{\iterIdx}}}{\norm{\solPrimFOMVar^{\iterIdx}}},
    \label{rom:proj_err}
\end{equation}
where $\solPrimFOMVar^{\iterIdx}$ is directly obtained from the FOM solutions. Since the error of POD-based ROM is bounded by the projection error of the POD basis, the POD projection error is quantified as a baseline
\begin{equation}
    \errProjVsFOM^{\iterIdx}_{\varIdx} = \frac{\norm{\solPrimROMProjVar^{\iterIdx} - \solPrimFOMVar^{\iterIdx}}}{\norm{\solPrimFOMVar^{\iterIdx}}}.
    \label{rom:proj_err_pod}
\end{equation}

\begin{figure}
	\centering
    \begin{subfigure}[b]{\textwidth}
         \centering
         \includegraphics[width=0.7\textwidth]{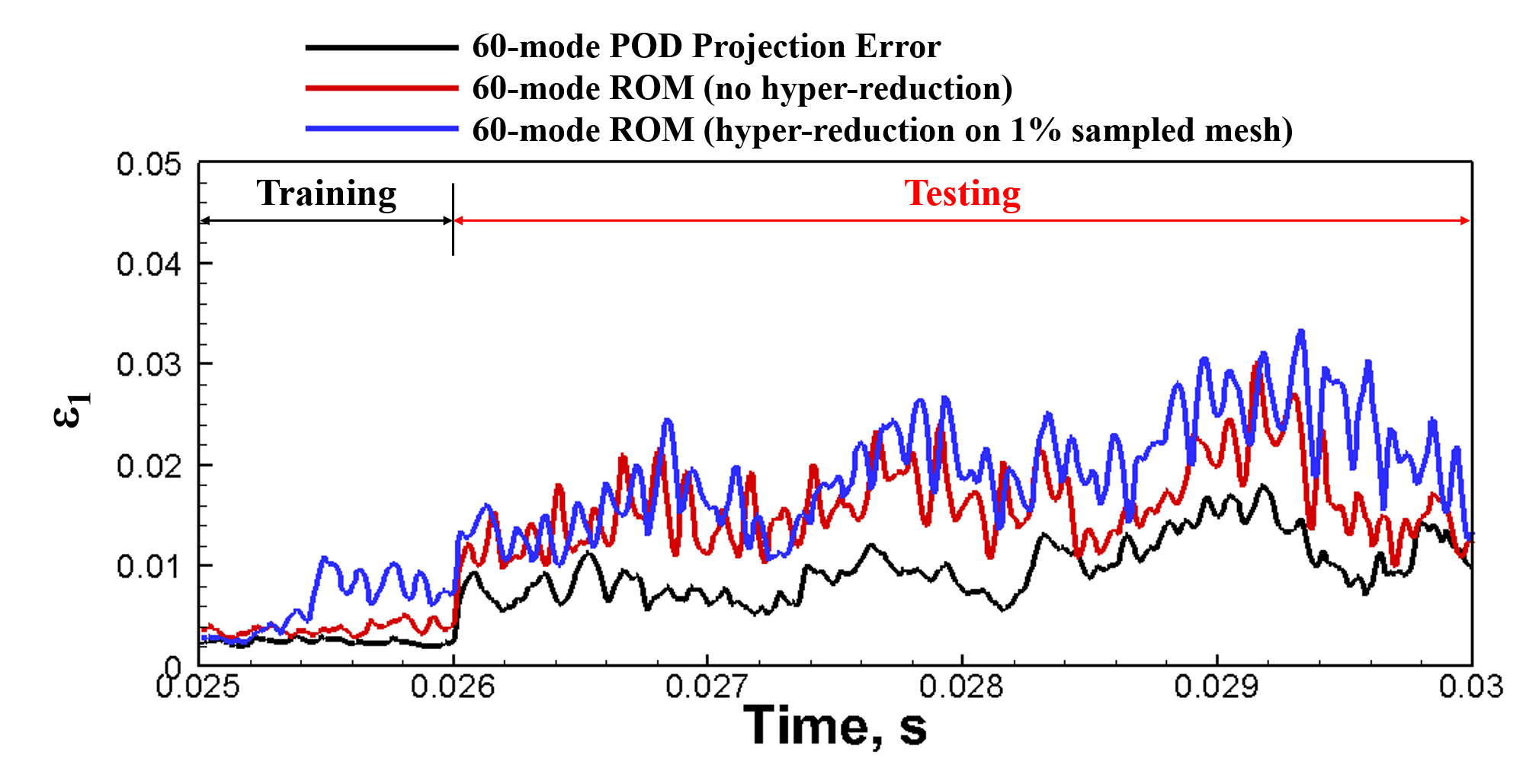}
         \caption{Pressure error time history.}
          \label{3d:SP-LSVT-errP}
    \end{subfigure}
    \centering
    \begin{subfigure}[b]{\textwidth}
         \centering
         \includegraphics[width=0.7\textwidth]{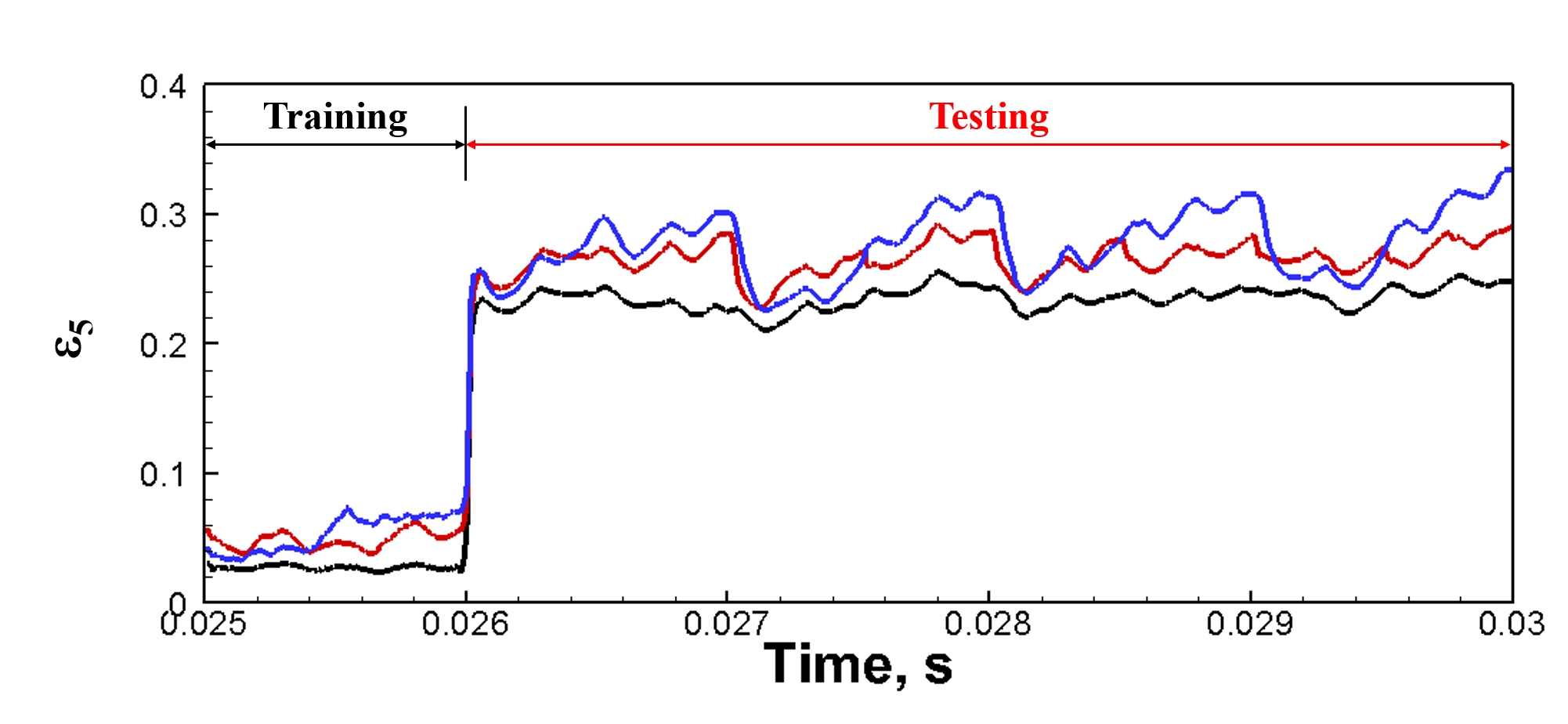}
         \caption{Temperature error time history.}
          \label{3d:SP-LSVT-errT}
    \end{subfigure}
	\caption{Comparisons of time evolution of errors with respect to the FOM solution.}\label{3d:rom_future_prjErr} 
\end{figure}

Similarly, the ROM errors (with respect to the FOM solutions in the pressure ($\varIdx = 1$ in Eqs.~\ref{rom:proj_err} and~\ref{rom:proj_err_pod}) and temperature ($\varIdx = 5$ in Eqs.~\ref{rom:proj_err} and~\ref{rom:proj_err_pod}) fields are compared in Fig.~\ref{3d:rom_future_prjErr}. Within the training period in which the dynamics are well-represented, as indicated by low POD projection error  ($<$ 0.25\% error for pressure and $<$ 2\% error for temperature), the ROMs are able to represent both pressure and temperature accurately ($<$ 1\% error for pressure and $<$ 5\% error for temperature), consistent with Fig.~\ref{3d:rom_future_prjErronPOD}. However, challenges arise during future-state predictions in the testing period. Though the POD trial basis and the {\color{black}MP-LSVT} ROMs are able to reasonably represent the future-state pressure dynamics (well below 3\% error), they are unable to accurately represent the dynamics of the temperature field after the end of the training period ($\timeVar = 26$ ms). In contrast to the small error increase (from 4\% to 6\%) at $\timeVar = 26$ ms in Fig.~\ref{3d:rom_future_prjErronPOD}, there is a significant increase in error at this point, from less than 5\% to more than 20\%.  We emphasize  that this significant error increase is mainly due to the POD projection error (and not the ROM formulation), and thus reflects the insufficiency of the POD trial basis in accurately representing the system dynamics beyond the training period. 

The observed challenges in future-state projection error can be largely attributed to the chaotic nature of the dynamics as shown in Fig~\ref{3d:fom_snapshots}. The pressure field exhibits organized dynamics due to the strong forcing introduced downstream, which allows the ROMs to provide reasonable predictions in the future state as the POD trial basis generated within the training period are able to easily represent such organized dynamics. However, in turbulent reacting flows (characterized by transport of strong temperature gradients), chaotic and non-stationary features present a major  challenge. Although the basis can represent convection features within the training period, it is unable to represent features beyond the training period, and therefore produce significant errors in the ROMs. This is not a flaw in the {\color{black}MP-LSVT} technique, but rather a limitation of using a linear and static basis set. Nevertheless, the {\color{black}MP-LSVT} ROMs are able to provide accurate predictions of pressure dynamics, which is an important quantity of interest for reacting flow problems, especially in practical engineering applications.

\section{Conclusion}\label{sec:conclusion}
A comprehensive projection-based reduced-order model formulation is presented for multi-scale and multi-physics problems. A {\color{black}model-form preserving} least-squares with variable transformation ({\color{black}MP-LSVT}) formulation is derived to provide global stability enhancement. The new formulation builds on recent developments in least-squares-based ROMs~\cite{Carlberg2017,GrimbergPROM2020}, and relies on least-squares minimization of the discrete equation residuals to guarantee symmetrization and discrete consistency with the full-order model at both the physical-time and the sub-iteration levels. It allows for the selection of arbitrary (but complete) state solution variables while preserving the structure of the governing equations so that the consistency between FOM and ROM is achieved at the fully discrete level. It is applicable to both implicit and explicit time integrators (in contrast to LSPG projection~\cite{Carlberg2017,GrimbergPROM2020}, which requires implicit time integrators) and provides more flexibility in constructing and computing the ROMs. Detailed evaluations of the {\color{black}MP-LSVT} formulation are performed based on a representative 2D reacting flow problem with a stiff chemical kinetic model that can often result in challenging numerical robustness issues even for full-order models. The {\color{black}MP-LSVT} formulation is shown to produce ROMs with significantly improved stability and accuracy over the standard projection-based ROM techniques. The {\color{black}MP-LSVT} ROMs remain stable and exhibit largely monotonic error decay with respect to the number of modes included in the trial basis. To address a gap in the literature, we present a proof of linear stability of least-squares-based ROMs for linear time invariant systems. This is followed by a numerical analysis of linearized version of the reacting flow equations, further confirming the linear stability and improved conditioning offered by the {\color{black}MP-LSVT} procedure.


For local stabilization, physical realizability in reacting flow ROMs is improved by enforcing limiters on the temperature and species mass fraction fields to mitigate the production of local spurious oscillations near sharp gradients. 
The importance of a species mass fraction limiter is demonstrated for both 1D and 2D reacting flow problems with stiff chemical kinetics. It is shown to be especially effective in improving the accuracy of heat release rate and flame  speed predictions,  both of which are important quantities of interest in reacting flows. 

Enhancement of computational efficiency is demonstrated in a 3D reacting single injector problem via hyper-reduction using gappy POD.  A factor of 319 speedup in wall clock time and 638 in CPU time can be achieved with {\color{black}MP-LSVT} ROM on a 1\% sampled mesh, calculated using an explicit time integrator and a physical time step 10 times that of the FOM.

Finally, the capability of the {\color{black}MP-LSVT} ROM in predicting the  dynamics beyond the training window is assessed in the 3D case. Though both the {\color{black}MP-LSVT} ROMs  are able to provide reasonably accurate predictions of the pressure field,  future-state predictions of the temperature field remain challenging. This is, however, not a limitation of the {\color{black}MP-LSVT} procedure, as the projection error is itself high at the future state -- a consequence of a static linear basis approximation. To improve on this, adaptive bases~\cite{peherstorfer2015online,PeherstorferADEIM} or non-linear manifold ROMs~\cite{LeeNonlinearManifold2020} can be leveraged with the {\color{black}MP-LSVT} formulation. Future work will also include explorations of explicit constraint preservation (e.g. enforcing conservation~\cite{CarlbergConsvLSPG}).

\section*{Acknowledgments}
{The authors acknowledge  support from the Air Force under the Center of Excellence grant FA9550-17-1-0195, titled Multi-Fidelity Modeling of Rocket Combustor Dynamics (Program Managers: Dr. Mitat Birkan and Dr. Fariba Fahroo).
Computing resources were provided by the NSF via grant 1531752 MRI: Acquisition of Conflux, A Novel Platform for Data-Driven Computational Physics (Program Manager: Dr. Stefan Robila).
}
\appendix

\section{Linear Algebra Theorems for Proofs in Section~\ref{proof}}

\subsection{Poincar\'{e} Separation Theorem}\label{appendix:poincare}

 Let $\mb{A} \in \mathbb{R}^{\numDOF \times \numSolModes}$ be a symmetric matrix and $\mb{F} \in \mathbb{R}^{\numDOF \times \numSolModes}$ be a semi-orthogonal matrix such that $\mb{F}^T \mb{F} = \mb{I}_{\numSolModes}$. Then Ref.~\cite{bellman1997introduction} gives
 
 \begin{equation}
 \lambda_{\numDOF - \numSolModes + i}(\mb{A})  \leq  \lambda_{i}(\mb{F}^T \mb{A} \mb{F}) \leq   \lambda_i(\mb{A}) \ \ \ \  1 \leq i \leq \numSolModes,
\end{equation}
where the eigenvalues $\lambda_i$ are arranged in descending order. 

\subsection{Singular Value Transformation Identity}\label{appendix:singValIdent}

Adapting Lemma 3.3.1 in Ref.~\cite{horn1994topics}, let $\mb{A} \in \mathbb{R}^{\numDOF \times \numDOF}$ and $\mb{F} \in \mathbb{R}^{\numDOF \times \numSolModes}$ be a semi-orthogonal matrix such that $\mb{F}^T\mb{F} = \mb{I}_{\numSolModes}$. Then

\begin{equation}
\sigma_{i}(\mb{F}^T \mb{A} \mb{F}) \leq \sigma_i(\mb{A})   \ \ \ \ 1 \leq i \leq \numSolModes.
\end{equation}

\subsection{Relationship Between Singular Values of a Matrix and the Eigenvalues of its Hermitian}\label{appendix:evalsHermitian}

Lemma 3.1.5 in Ref.~\cite{horn1994topics} shows that, for $\mb{A} \in \mathbb{R}^{\numDOF \times \numDOF}$,
\begin{equation}
\sigma_i(\mb{A}) \geq \lambda_i \left( \frac{\mb{A}+\mb{A}^T}{2}   \right) \ \ 1 \leq i \leq \numDOF. 
\end{equation}

\section{ Governing Equations for the Full Order Model}
\label{appendix:fom_eq}

The full order model computations are carried out with an in-house CFD code, the General Equations and Mesh Solver (GEMS), the capabilities of which have been successfully demonstrated in modeling rocket combustion instabilities~\cite{HarvazinskiPoF}. GEMS solves the conservation equations for mass, momentum, energy and species transport in a coupled fashion:

\begin{equation}
    \pde{Q}{\timeVar} + \pde{F_i}{x_i} - \pde{F_{v,i}}{x_i} = H,
    \label{fom:governing}
\end{equation}
Here, $Q$ is the vector of conserved variables defined as, $Q=\left(\begin{array}{cccccc}
        \rho & \rho{u_i} & \rho{h^0-p} & \rho{Y_l}\\
    \end{array}\right)^T$,
with $\rho$ representing density, $u_i$ representing the velocity in the $i^{th}$ spatial dimension, and $p$ representing static pressure. The total enthalpy $h^0$ is defined as, 
\begin{equation}
    h^0 = h + \frac{1}{2} u^2_i = \sum_l h_l Y_l + \frac{1}{2} u^2_i. 
\end{equation}
$Y_l$ represents the $l^{th}$ species mass fraction if a multi-species model is used to describe the chemical reaction, as is the case for results presented in Section~\ref{2d:injector}. If a flamelet progress variable (FPV) model is used, $Y_l$ represents the $l^{th}$ transported scalar of the model. In that case, $Y_1$ is the mean mixture fraction (${Z}_{mean}$), $Y_2$ is the mean mixture fraction variance (${Z''^2}$), and $Y_3$ is the mean progress variable (${C}_{mean}$). Such an FPV model is used to generate the results presented in Section~\ref{3d:injector}.

The fluxes have been separated into inviscid ($F_i$) and viscous ($F_{v,i}$) terms. The inviscid fluxes are given by
\begin{equation}
    F_i = \left(\begin{array}{c}
        \rho{u_i} \\
        \rho{u_i u_j} + p \delta_{ij} \\
        \rho{u_i h^0} \\
        \rho{u_i Y_l} \\
    \end{array} \right), 
    \label{fom:inviscid_fluxes}
\end{equation}
The viscous fluxes are
\begin{equation}
    F_{v,i} = \left(\begin{array}{c}
        0 \\
        \tau_{ij} \\
        u\tau_{ii}+v\tau_{ji}+w\tau_{ki}-q_i \\
        \rho{D_{l}}\left( {\partial{Y_l}}/{\partial{x_i}} \right) \\
    \end{array} \right),
    \label{fom:viscous_fluxes}
\end{equation}
where $D_l$ is the mass diffusivity of the $l^{th}$ species. This is an approximation used to model the multicomponent diffusion as the binary diffusion of each species into a mixture.

The heat flux in the $i^{th}$ direction, $q_i$, is defined as
\begin{equation}
    q_i = -K \pde{T}{x_i} + \rho \sum^{\numSpec}_{l=1} D_l \pde{Y_l}{x_i} h_l + \mathbf{\dot{Q}}_{\text{source}}.
    \label{fom:heat_flux}
\end{equation}
The three terms in the heat flux represent the heat transfer due to conduction, species diffusion, and heat generation from a volumetric source (e.g. radiation or an external source) respectively.

The  stress tensor $\tau$ is defined as in terms of the molecular viscosity and strain-rate 
\begin{equation}
    \tau_{ij} = \mu \lp \pde{u_i}{x_j} + \pde{u_j}{x_i} - \frac{2}{3} \pde{u_m}{x_m} \delta_{ij} \rp.
    \label{fom:shear_stress}
\end{equation}

The source term $H$ includes a single entry for each of the scalar transport equations signifying the production or destruction of the $l^{th}$ species, $\dot{\omega}_l$, which is determined by the chemical kinetics
\begin{equation}
    H=\left(\begin{array}{cccccc}
        0 & 0 & 0 & 0 & 0 & \dot{\omega}_l\\
    \end{array}\right)^T.
    \label{cfd:source_term}
\end{equation}

\section{Test Basis for {\color{black}MP-LSVT} Formulation in Physical Time}
\label{appendix:SP-LSVTTestFunctionPhysicalTime}
We derive the test basis $\testBasisPrim^{\iterIdx}$ for {\color{black}MP-LSVT} in physical time, based on its definition in Eq.~\ref{rom:lspg_w} 
\begin{equation}
    \testBasisPrim^{\iterIdx} = \pde{\scaleMatCons \resFunc{\solPrimROMFull^{\iterIdx}}}{\solPrimROMRed^{\iterIdx}}.
\end{equation}
With the equation residual, $\res$, for a linear multi-step method defined in Eq.~\ref{EqRes:LinearMultiStep_qp}, substitute the approximate solution, $\solPrimROMFull^{\iterIdx} = \solPrimFOMRef + \scaleMatPrim^{-1} \trialBasisPrim \solPrimROMRed^{\iterIdx}$ and $\solPrimROMFull^{\iterIdx-\genIdx} = \solPrimFOMRef + \scaleMatPrim^{-1} \trialBasisPrim \solPrimROMRed^{\iterIdx black-\genIdx}$,
\begin{equation}
    \resFunc{\solPrimROMFull^{\iterIdx}} = \solConsFOMFunc{\solPrimROMFull^{\iterIdx}} + \sum^{\itermaxLinMS}_{\sumIdx=1} \alpha_\sumIdx \solConsFOMFunc{\solPrimROMFull^{\iterIdx-\sumIdx}} - \dt \beta_0 \rhsFunc{\solPrimROMFull^{\iterIdx}, \timeVar^{\iterIdx}} - \dt \sum^{\itermaxLinMS}_{\sumIdx=1} \beta_\sumIdx \rhsFunc{\solPrimROMFull^{\iterIdx-\sumIdx}, \timeVar^{\iterIdx-\sumIdx}}.
\end{equation}
Therefore
\begin{equation}
    \pde{\resFunc{\solPrimROMFull^{\iterIdx}}}{\solPrimROMRed^{\iterIdx}} = \pde{\solConsFOMFunc{\solPrimROMFull^{\iterIdx}}}{\solPrimROMRed^{\iterIdx}} - \dt \beta_0 \pde{\rhsFunc{\solPrimROMFull^{\iterIdx}}}{\solPrimROMRed^{\iterIdx}},
\end{equation}
where
\begin{equation}
    \pde{\solConsFOMFunc{\solPrimROMFull^{\iterIdx}}}{\solPrimROMRed^{\iterIdx}} = \pde{\solConsFOMFunc{\solPrimROMFull^{\iterIdx}}}{\solPrimROMFull^{\iterIdx}} \pde{\solPrimROMFull^{\iterIdx}}{\solPrimROMRed^{\iterIdx}} = \gmROM^{\iterIdx} \scaleMatPrim^{-1} \trialBasisPrim,
\end{equation}
and
\begin{equation}
    \pde{\rhsFunc{\solPrimROMFull^{\iterIdx}}}{\solPrimROMRed^{\iterIdx}} = \pde{\rhsFunc{\solPrimROMFull^{\iterIdx}}}{\solPrimROMFull^{\iterIdx}} \pde{\solPrimROMFull^{\iterIdx}}{\solPrimROMRed^{\iterIdx}} = \jacobPrimROM^{\iterIdx} \scaleMatPrim^{-1} \trialBasisPrim,
\end{equation}
The Jacobian $\jacobPrimROM^{\iterIdx}$ is defined as
\begin{align}
\begin{split}
\jacobPrimROM^{\iterIdx} &= \left[ \pde{\rhs}{\solPrimFOM} \right]^{\iterIdx}_{\solPrimFOM = \solPrimROMFull} \\ 
&= \left[ \pde{\rhs}{\solConsFOM} \right]^{\iterIdx}_{\solPrimFOM = \solPrimROMFull} \left[ \pde{\solConsFOM}{\solPrimFOM} \right]^{\iterIdx}_{\solPrimFOM = \solPrimROMFull} \\
&= \jacobConsROM^{\iterIdx} \gmROM^{\iterIdx}.
\end{split}
\end{align}
Hence, the test basis $\testBasisPrim^{\iterIdx}$ for a linear multi-step method becomes
\begin{equation}
    \testBasisPrim^{\iterIdx} = \scaleMatCons \lp \gmROM^{\iterIdx} - \dt \beta_0 \jacobConsROM^{\iterIdx} \gmROM^{\iterIdx} \rp \scaleMatPrim^{-1} \trialBasisPrim.
\end{equation}

The process is similar for a Runge--Kutta method, with the equation residual $\res$ defined in Eq.~\ref{EqRes:RK_qp}. We again substitute the approximated solution, $\solPrimROMFull^{\iterIdx}$ and $\solPrimROMFull{\iterIdx-j}$, in Eq.~\ref{pod:expansion_qp} to evaluate $\res$
\begin{equation}
    \resFunc{\solPrimROMFull^{\iterIdx}} = \solConsFOMFunc{\solPrimROMFull^{\iterIdx}} - \solConsFOMFunc{\solPrimROMFull^{\iterIdx-1}} - \dt \sum^{\itermaxRK}_{\sumIdx=1} b_\sumIdx \rkSol_\sumIdx.
\end{equation}
where $\rkSol_1 = \rhsFunc{\solPrimROMFull^{\iterIdx-1}, \timeVar^{\iterIdx-1}}$ and
\begin{equation}
    \rkSol_{\sumIdx} = \rhsFunc{ \solPrimROMFull^{\iterIdx-1} + \dt \sum^{\sumIdx-1}_{\sumIdxTwo=1} a_{\sumIdx \sumIdxTwo} \gm_{\sumIdxTwo}^{-1} \rkSol_{\sumIdxTwo}, \timeVar^{\iterIdx-1} + c_\sumIdx \dt},
\end{equation}
with $\gm_1 = \gm \lp \solPrimROMFull^{\iterIdx-1},\timeVar^{\iterIdx-1} \rp$ and
\begin{equation}
    \gm_{\sumIdx} = \gm \lp \solPrimROMFull^{\iterIdx-1} +\dt \sum^{\sumIdx-1}_{\sumIdxTwo=1} a_{\sumIdx \sumIdxTwo} \gm_{\sumIdxTwo}^{-1} \rkSol_{\sumIdxTwo}, \timeVar^{\iterIdx-1} + c_{\sumIdx} \dt \rp.
\end{equation}
Therefore
\begin{equation}
    \pde{\resFunc{\solPrimROMFull^{\iterIdx}}}{\solPrimROMRed^{\iterIdx}} = \pde{\solConsFOMFunc{\solPrimROMFull^{\iterIdx}}}{\solPrimROMRed^{\iterIdx}} = \pde{\solConsFOMFunc{\solPrimROMFull^{\iterIdx}}}{\solPrimROMFull^{\iterIdx}} \pde{\solPrimROMFull^{\iterIdx}}{\solPrimROMRed^{\iterIdx}} = \gmROM^{\iterIdx} \scaleMatPrim^{-1} \trialBasisPrim.
\end{equation}
Hence, the test basis $\testBasisPrim^\iterIdx$ for a Runge--Kutta method becomes
\begin{equation}
    \testBasisPrim^\iterIdx = \scaleMatCons \gmROM^{\iterIdx} \scaleMatPrim^{-1} \trialBasisPrim.
\end{equation}

\section{Derivations of Jacobian Matrices }
\label{appendix:ROMJacobMatrix}

We derive expressions for the linearized matrices $\linStabMat$ for different ROM methods (Galerkin, LSPG and {\color{black}MP-LSVT}) used for linear stability analysis in section~\ref{2d:LinearStability}. For simplicity, We choose the 1-step version of linear multi-step methods in Eq.~\ref{EqRes:LinearMultiStep} to illustrate our derivation. We start with the FOM residual
\begin{equation}
    \resFunc{\solConsFOM^\iterIdx} = \solConsFOM^{\iterIdx}-\solConsFOM^{\iterIdx - 1} - \dt \beta_0 \rhsFunc{\solConsFOM^\iterIdx, \timeVar^\iterIdx} - \dt (1 - \beta_0) \rhsFunc{\solConsFOM^{\iterIdx - 1}, \timeVar^{\iterIdx - 1}}.
    \label{EqRes:OneStep}
\end{equation}
If $\beta_0 = 0$ , the method is explicit; otherwise, the method is implicit.

\subsection{Galerkin ROM}

Starting with Eq.~\ref{rom:galerkin_ode}, the Galerkin technique (Section~\ref{galerkin_rom}) leads to 
\begin{equation}
    \solConsROMRed^{\iterIdx} - \solConsROMRed^{\iterIdx - 1} - \dt \beta_0 \trialBasisCons^T \scaleMatCons \rhsFunc{\solConsROMFull^\iterIdx, \timeVar^\iterIdx} - \dt (1 - \beta_0) \trialBasisCons^T \scaleMatCons \rhsFunc{\solConsROMFull^{\iterIdx - 1}, \timeVar^{\iterIdx - 1}} = 0,
    \label{rom:galerkin_eq}
\end{equation}
where $\solConsROMFull = \solConsFOMRef + \scaleMatCons^{-1}\trialBasisCons \solConsROMRed$ following Eq.~\ref{pod:expansion_consv}. With rearrangement, the linearized version of Eq.~\ref{rom:galerkin_eq}  is
\begin{equation}
    \left[ \mathbf{I} - \dt \beta_0 \trialBasisCons^T \scaleMatCons \jacobConsROM^\iterIdx \scaleMatCons^{-1} \trialBasisCons \right] \solConsROMRed^{\iterIdx} = \left[ \mathbf{I} + \dt (1 - \beta_0) \trialBasisCons^T \scaleMatCons \jacobConsROM^{\iterIdx-1} \scaleMatCons^{-1}\trialBasisCons \right] \solConsROMRed^{\iterIdx - 1},
    \label{rom:galerkin_eq_linear}
\end{equation}
where $\jacobConsROM^\iterIdx=\left[{\partial \rhs} / {\partial \solConsFOM} \right]^{\iterIdx}_{\solConsFOM = \solConsROMFull}$. Then the final form of Eq.~\ref{rom:galerkin_eq_linear} for linear stability analysis is
\begin{equation}
    \solConsROMRed^{\iterIdx} = \linStabMat_\text{Galerkin} \solConsROMRed^{\iterIdx-1},
    \label{rom:galerkin_eq_linear_final}
\end{equation}
where $\linStabMat_\text{Galerkin}$ is defined as the Jacobian matrix of the Galerkin ROM as follows
\begin{equation}
    \linStabMat_\text{Galerkin} \triangleq \left[ \mathbf{I} - \dt \beta_0 \trialBasisCons^T \scaleMatCons \jacobConsROM^\iterIdx \scaleMatCons^{-1}\trialBasisCons \right]^{-1} \left[ \mathbf{I} + \dt (1 - \beta_0) \trialBasisCons^T \scaleMatCons \jacobConsROM^{\iterIdx - 1} \scaleMatCons^{-1} \trialBasisCons \right].
    \label{rom:Cmatrix_galerkin}
\end{equation}

\subsection{LSPG ROM}

The standard LSPG method (Section~\ref{slspg_rom})  results in a fully discrete ROM system given by Eq.~\ref{rom:slspg_proj}.  
Since LSPG requires an implicit scheme ($\beta_0 = 1$) in Eq.~\ref{rom:galerkin_eq},  Eq.~\ref{rom:slspg_proj} becomes
\begin{equation}
    \lp \testBasisCons^\iterIdx \rp^T \scaleMatCons \left[ \solConsROMFull^{\iterIdx} - \solConsROMFull^{\iterIdx - 1} - \dt \rhsFunc{\solConsROMFull^\iterIdx, \timeVar^\iterIdx} \right] = 0.
    \label{rom:slspg_proj_appendix}
\end{equation}

Therefore, the linearized version of Eq.~\ref{rom:slspg_proj_appendix} with $\solConsROMFull = \solConsFOMRef + \scaleMatCons^{-1} \trialBasisCons \solConsROMRed$ is
\begin{equation}
    \lp \testBasisCons^\iterIdx \rp^T \left[ \trialBasisCons - \dt \scaleMatCons \jacobConsROM^\iterIdx \scaleMatCons^{-1} \trialBasisCons \right] \solConsROMRed^{\iterIdx} = \lp \testBasisCons^\iterIdx \rp^T \trialBasisCons \solConsROMRed^{\iterIdx - 1}.
    \label{rom:slspg_eq_linear}
\end{equation}
Then the final form of Eq.~\ref{rom:slspg_eq_linear} for linear stability analysis is
\begin{equation}
    \solConsROMRed^{\iterIdx} = \linStabMat_\text{LSPG} \solConsROMRed^{\iterIdx-1},
    \label{rom:slspg_eq_linear_final}
\end{equation}
where $\linStabMat_\text{LSPG}$ is defined as the Jacobian matrix of the LSPG ROM as follows
\begin{equation}
    \linStabMat_\text{LSPG} \triangleq \left[ \lp \testBasisCons^\iterIdx \rp^T \testBasisCons^\iterIdx \right]^{-1}  \lp \testBasisCons^\iterIdx \rp^T \trialBasisCons,
    \label{rom:Cmatrix_slspg}
\end{equation}
where $\testBasisCons^\iterIdx \triangleq \trialBasisCons - \dt \scaleMatCons \jacobConsROM^\iterIdx \scaleMatCons^{-1} \trialBasisCons$ as defined in Eq.~\ref{rom:slspg_w_LinearMultistep}.

\subsection{{\color{black}MP-LSVT} ROM}

With the FOM equation residual defined in Eq.~\ref{EqRes:OneStep}, Eq.~\ref{rom:lspg_proj} becomes
\begin{equation}
    \lp \testBasisPrim^\iterIdx \rp^T \scaleMatCons \left[ \solConsFOMFunc{\solPrimROMFull^\iterIdx} - \solConsFOMFunc{\solPrimROMFull^{\iterIdx-1}} - \dt \beta_0 \rhsFunc{\solPrimROMFull^\iterIdx, \timeVar^\iterIdx} - \dt (1 - \beta_0) \rhsFunc{\solPrimROMFull^{\iterIdx-1}, \timeVar^{\iterIdx-1}} \right] = 0.
    \label{rom:lspg_proj_appendix}
\end{equation}
Then we linearize Eq.~\ref{rom:lspg_proj_appendix} with $\solPrimROMFull = \solPrimFOMRef + \scaleMatPrim^{-1} \trialBasisPrim \solPrimROMRed$ and rearrange terms, which gives
\begin{equation}
    \lp \testBasisPrim^\iterIdx \rp^T \scaleMatCons \lp \gmROM^\iterIdx - \dt  \beta_0 \jacobConsROM^\iterIdx \gmROM^\iterIdx \rp \scaleMatPrim^{-1} \trialBasisPrim \solPrimROMRed^{\iterIdx} = \lp \testBasisPrim^\iterIdx \rp^T \scaleMatCons \left[ \gmROM^{\iterIdx - 1} + \dt (1 - \beta_0) \jacobConsROM^{\iterIdx - 1} \gmROM^{\iterIdx - 1} \right] \scaleMatPrim^{-1} \trialBasisPrim\solPrimROMRed^{\iterIdx - 1}.
    \label{rom:lspg_eq_linear}
\end{equation}
Then the final form of Eq.~\ref{rom:lspg_eq_linear} for linear stability analysis is
\begin{equation}
    \solPrimROMRed^{\iterIdx} = \linStabMat_\text{{\color{black}MP-LSVT}} \solPrimROMRed^{\iterIdx-1},
    \label{rom:lspg_eq_linear_final}
\end{equation}
where $\linStabMat_\text{{\color{black}MP-LSVT}}$ is defined as the Jacobian matrix of the {\color{black}MP-LSVT} ROM as follows
\begin{equation}
    \linStabMat_\text{{\color{black}MP-LSVT}} \triangleq \left[ \lp \testBasisPrim^\iterIdx \rp^T \lp \testBasisPrim^\iterIdx \rp \right]^{-1} \lp \testBasisPrim^\iterIdx \rp^T \scaleMatCons \left[ \gmROM^{\iterIdx-1} + \dt (1-\beta_0) \jacobConsROM^{\iterIdx-1} \gmROM^{\iterIdx-1} \right] \scaleMatPrim^{-1} \trialBasisPrim
    \label{rom:Cmatrix_lspg}
\end{equation}
where $\testBasisPrim^\iterIdx \triangleq \scaleMatCons (\gmROM^\iterIdx - \dt \beta_0 \jacobConsROM^\iterIdx \gmROM^\iterIdx) \scaleMatPrim^{-1} \trialBasisPrim$ as defined in Eq.~\ref{rom:lspg_w_LinearMultistep}.

For brevity, the above derivations for {\color{black}MP-LSVT} are based on the  physical time formulation in section~\ref{lspg-vt}. Extension to the pseudo-time in section~\ref{lspg-vt-pseudo} is straightforward. 

The aforementioned Jacobian matrices ($\linStabMat_\text{Galerkin}$, $\linStabMat_\text{LSPG}$ and $\linStabMat_\text{SP-LSPG}$) are used in linear stability analysis of the Galerkin, LSPG and {\color{black}MP-LSVT} ROM methods in section~\ref{2d:LinearStability}.

\section{Effects of Spurious Oscillations in Species Mass Fractions}
\label{appendix:species_oscillations}
We consider a one-dimensional, freely propagating, premixed laminar flame for detailed investigations and diagnosis of the ROM robustness issues from spurious oscillations in the species mass fractions. The 1D problem is calculated using the governing equation in Eq.~\ref{cfd:discretized} with simplified single-step, two-species reaction in which both the reactant and product species are treated as calorically perfect gases with identical molecular weights. Pertinent physical properties, which are summarized in Table~\ref{table:1DProblem}, have been chosen to model the reactant and product mixtures in the 2D reacting injector in Section~\ref{2d:injector}. The computational domain has a length of 10 mm, discretized with 200 uniform finite volume cells. This has been confirmed to be sufficient to resolve the flame thickness ($\sim$1 mm). The FOM solution is computed using the second-order accurate backwards differentiation formula with dual time-stepping, and a constant physical time step size of $\dt = 0.05 \; \mu$s.

\begin{table}
\centering
\begin{tabular}{ lllllll } 
\toprule
Species & MW (g/mol) & $c_p$ (kJ/kg/K) & Pr & Sc & $\mu_{ref}$ (kg/m/s) & $h_{ref}$ (kJ/kg) \\
\midrule
Reactant & 21.32 & 1.538 & 0.713 & 0.62 & 7.35$\times 10^{-4}$ &	-7,432 \\
Product	& 21.32 & 1.538 & 0.713 & 0.62 & 7.35$\times 10^{-4}$ & -10,800 \\
\bottomrule
\end{tabular}
\caption{\label{table:1DProblem}Properties of species reactant and product.}
\end{table}

The expression for the chemical reaction source term for the reactant follows the Arrhenius form 
\begin{equation}
    \dot{\omega}_\text{Reactant} = -MW_\text{Reactant} \cdot A \text{exp} \left( \frac{-E_A/R_u}{T} \right) \left[ \frac{\rho Y_\text{Reactant}}{MW_\text{Reactant}} \right]^a ,
    \label{1Dsource_term}
\end{equation}
with the pre-exponential factor $A = 2 \times 10^{10}$, the activation energy $E_A/R_u = 24{,}358$ K, and the concentration exponent $a = 1.0$. The authors acknowledge that labeling such a two-species problem a ``premixed'' flame may be confusing, but the above reaction model is indeed designed to model the conversion of perfectly-mixed reactants to completely-burned products via a one-dimensional flame.

\begin{figure}
	\centering
	\includegraphics[width=0.6\textwidth]{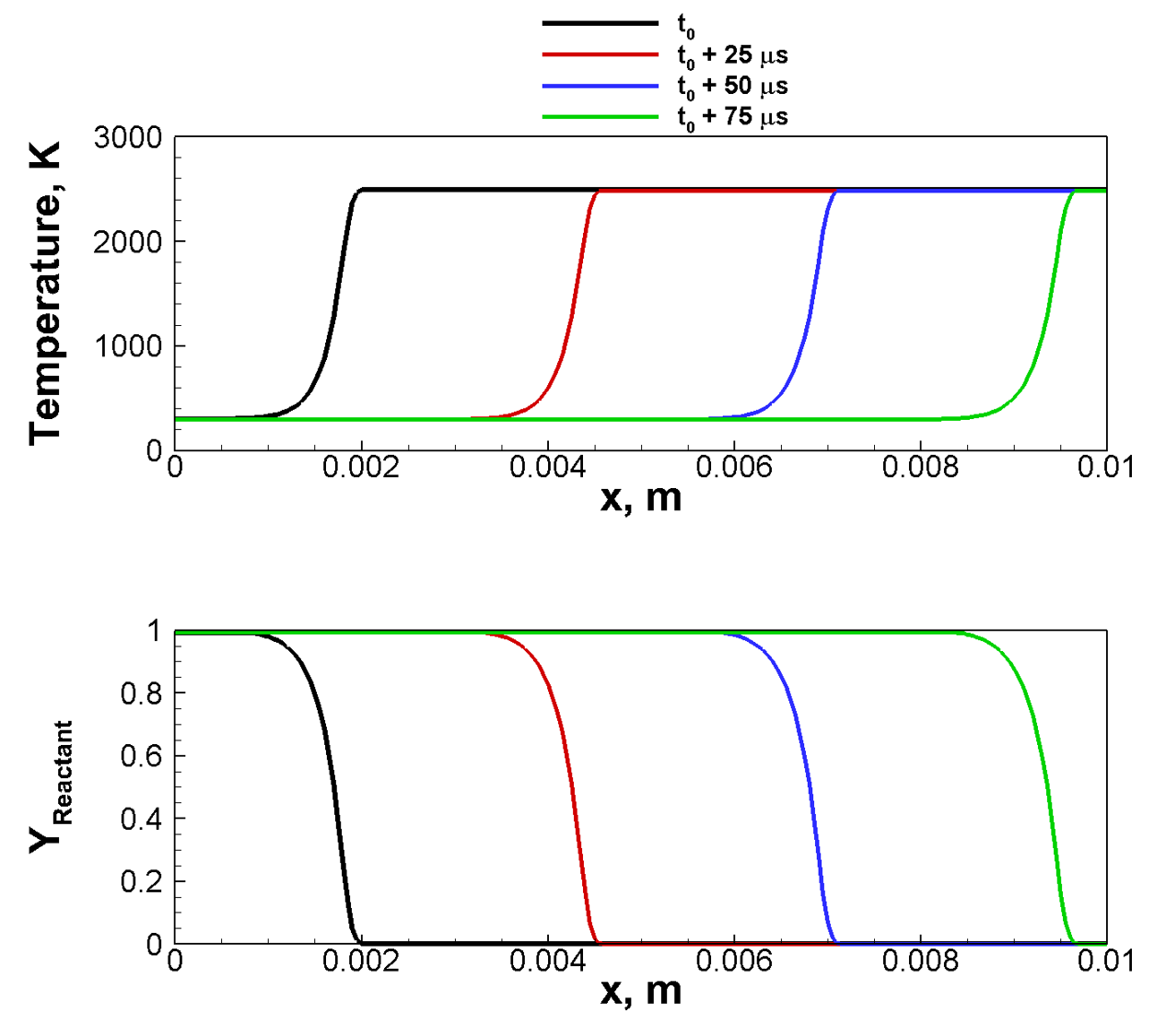}
	\caption{Representative FOM solutions from 1D premixed laminar flame.}\label{1d:fom} 
\end{figure}

The unsteady solution is advanced from $t_0$ to $t_0 + 75 \; \mu$s. Every tenth time step is saved, resulting in a total of 150 snapshots to generate the POD trial basis for the ROMs. Representative FOM temperature and reactant mass fraction fields are shown in Fig.~\ref{1d:fom} at different time instants. The dynamics is dominated by the convection of the sharp flame front accompanied by a temperature rise from 300 K to 2,500 K and a drop in reactant mass fraction from 1 to 0.

\begin{figure}
	\centering
	\includegraphics[width=0.6\textwidth]{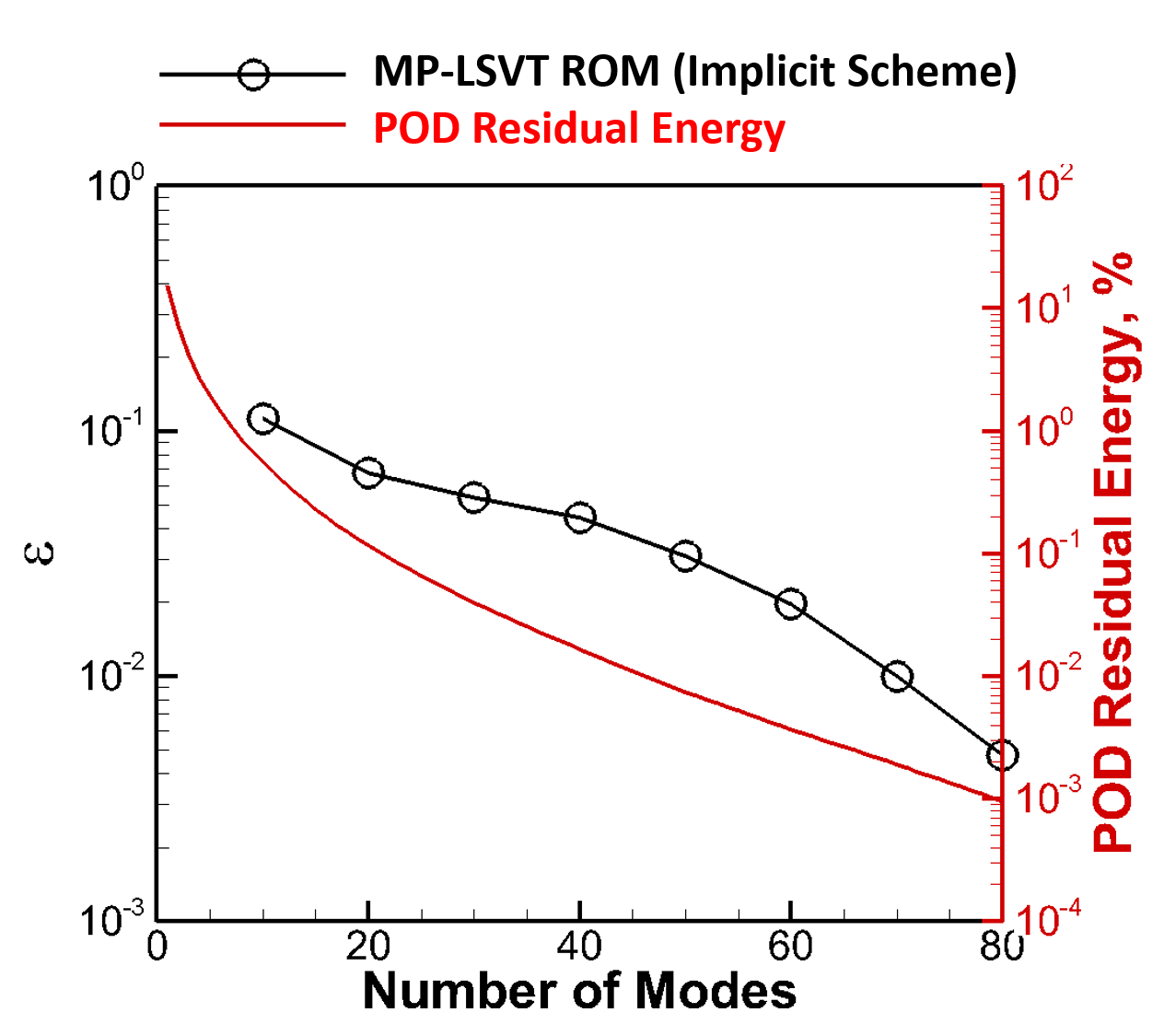}
	\caption{{\color{black}MP-LSVT} ROM error convergence and POD residual energy decay for 1D premixed laminar flame.}\label{1d:rom} 
\end{figure}

The {\color{black}MP-LSVT} ROM is computed using the physical time formulation (Section~\ref{lspg-vt}), with the implicit second-order accurate backwards differentiation formula. A temperature limiter with $T_{min} = 270$ K and $T_{max} = 2{,}750$ K is imposed during ROM calculations. The reconstruction error (Eq.~\ref{rom:reconstr_err}) and the POD residual energy (Eq.~\ref{pod:res_energy}) are shown in Fig.~\ref{1d:rom} and exhibit a monotonic decay with the number of POD trial basis modes. {\color{black}MP-LSVT} ROMs using an explicit time integrator have also been investigated and shown to produce very similar results. 

Even for this apparently simple 1D problem, however, the ROM reconstruction error shows a slow decay with the number of POD trial basis modes, and at least 70 modes are needed to reach approximately 1\% error. This reflects major challenges in ROM development for convection-dominated problems~\cite{HuangAIAAJ2019}, and is consistent with the challenges experienced in the 2D and 3D reacting flow simulations introduced in this manuscript.

\begin{figure}
	\centering
	\includegraphics[width=0.6\textwidth]{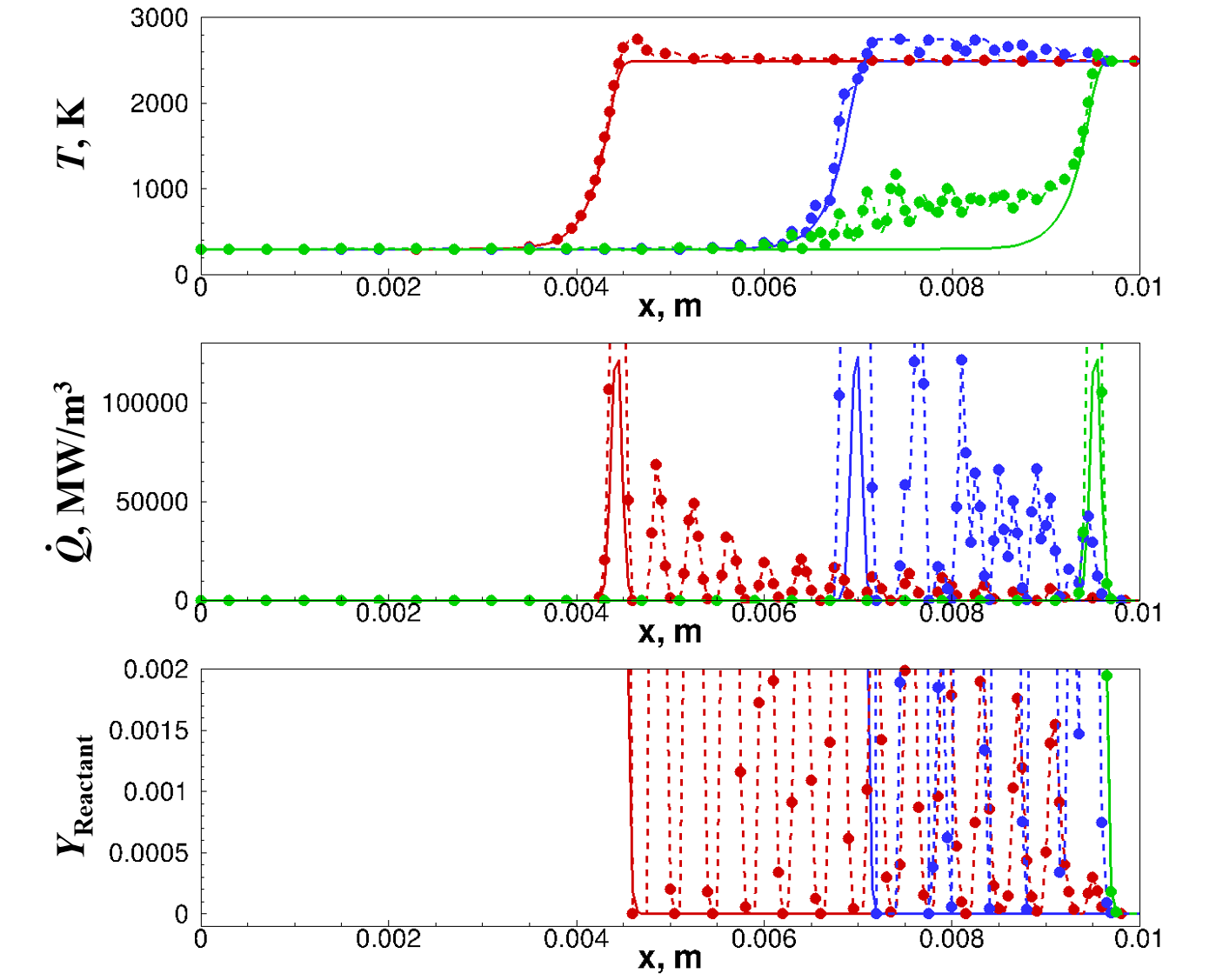}
	\caption{Solution comparisons between FOM (solid line) and 50-mode {\color{black}MP-LSVT} ROM (dashed line with symbol) at representative time instants for 1D premixed laminar flame.}\label{1d:romVSfomNS} 
\end{figure}

The temperature ($T$), heat release rate ($\dot{Q}$) and reactant mass fraction ($Y_\text{Reactant}$) fields of the 50-mode ROM solutions are compared with FOM solutions at three representative time instants in Fig.~\ref{1d:romVSfomNS}. Note that the three fields exhibit highly disparate scales throughout the spatial domain. Intervals for $T$ (top) and $\dot{Q}$ (middle)  cover the full range from unburned to burned temperatures and from zero to the peak heat release, while the ordinate for $Y_\text{Reactant}$ (bottom) is restricted to the very narrow range between 0 and 0.002. This small range for the $Y_\text{Reactant}$ field plot is chosen to highlight the presence of spurious fluctuations in the ROM reactant mass fraction on the burned side of the flame. The FOM solution gives a reactant mass fraction of zero downstream of the flame ($Y_\text{Reactant}$ steps from unity to zero across the flame over a few grid points), but the ROM solution predicts reactant mass fractions of $O(0.001)$. Although this 0.1\% error appears to be insignificant, the presence of tiny concentrations of reactant in the high temperature region downstream of the flame leads to excessive reaction, which increases the peak heat release rate in the ROM calculation. This increased heating in the post-flame region increases the temperature downstream of the flame which, in turn, leads to even higher peak heat release rates as a consequence of the Arrhenius term in Eq.~\ref{1Dsource_term}. Since the flame speed is inversely proportional to the reaction rate, the flame propagation in the ROM solution is decreased by the elevated heat release rate. The mechanism behind the ROM robustness issues can thus be tracked back to the spurious fluctuations in temperature and species mass fractions near the sharp flame front. While temperature oscillations can be alleviated by imposing a temperature limiter, it is also necessary to identify a limiter for the species mass fractions to constrain the level of spurious oscillations. As a final observation, we note that similar fluctuations in $Y_\text{Reactant}$ are also seen on the cold side of the flame. Unlike in the high-temperature side, however, the low temperature upstream of the flame precludes any reaction and the resulting sensitive coupling between small errors in $Y_\text{Reactant}$ and the heat release rate. The Arrhenius term allows much more substantial $Y_\text{Reactant}$ errors to be tolerated on the cold side of the flame.

\begin{figure}
	\centering
	\includegraphics[width=0.6\textwidth]{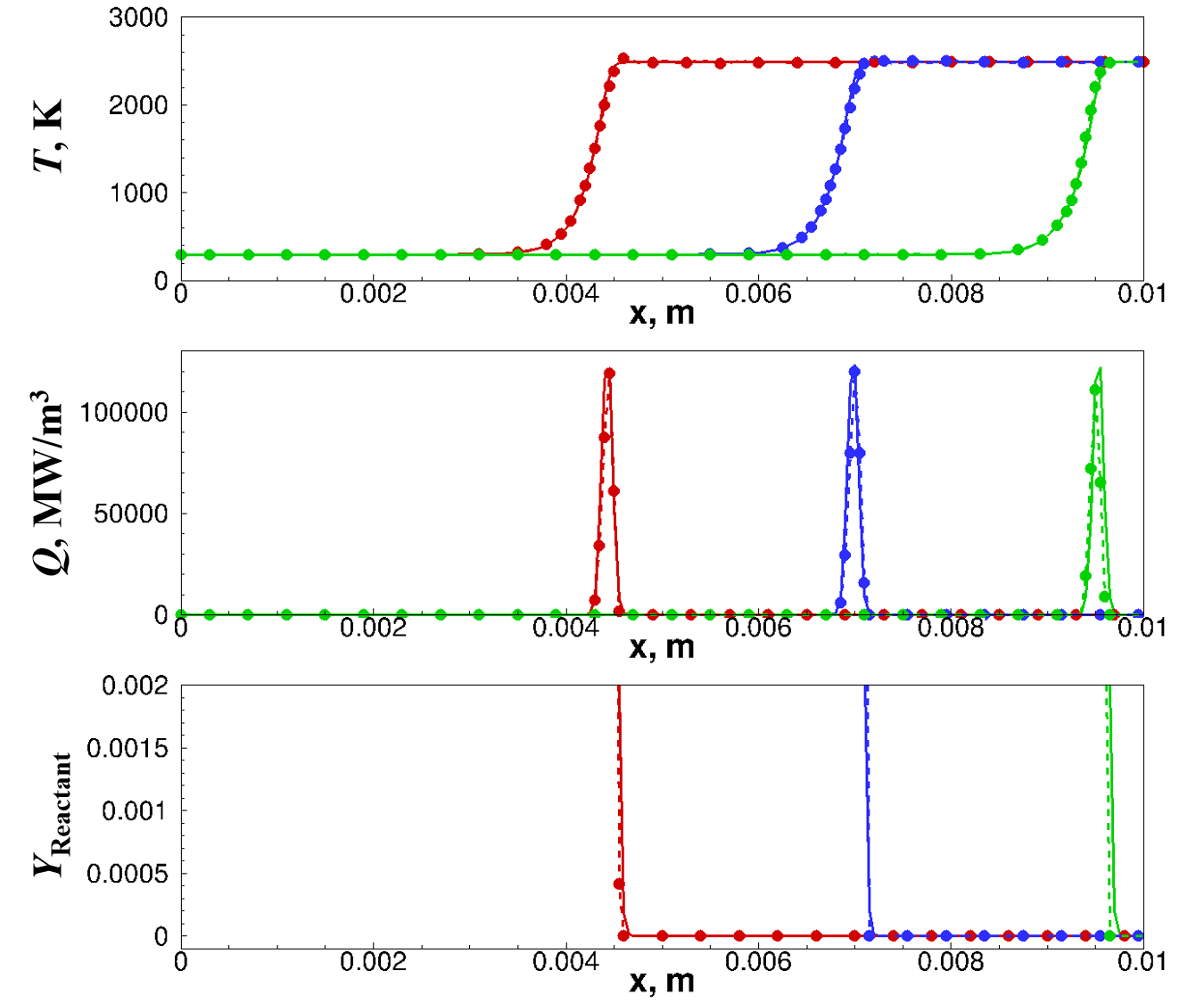}
	\caption{Solution comparisons FOM (solid line) and 50-mode {\color{black}MP-LSVT} ROM (dashed line with symbol) \emph{with species limiter} at representative time instants for 1D freely propagating laminar flame.}\label{1d:romVSfomS} 
\end{figure}

\begin{figure}
	\centering
	\includegraphics[width=0.6\textwidth]{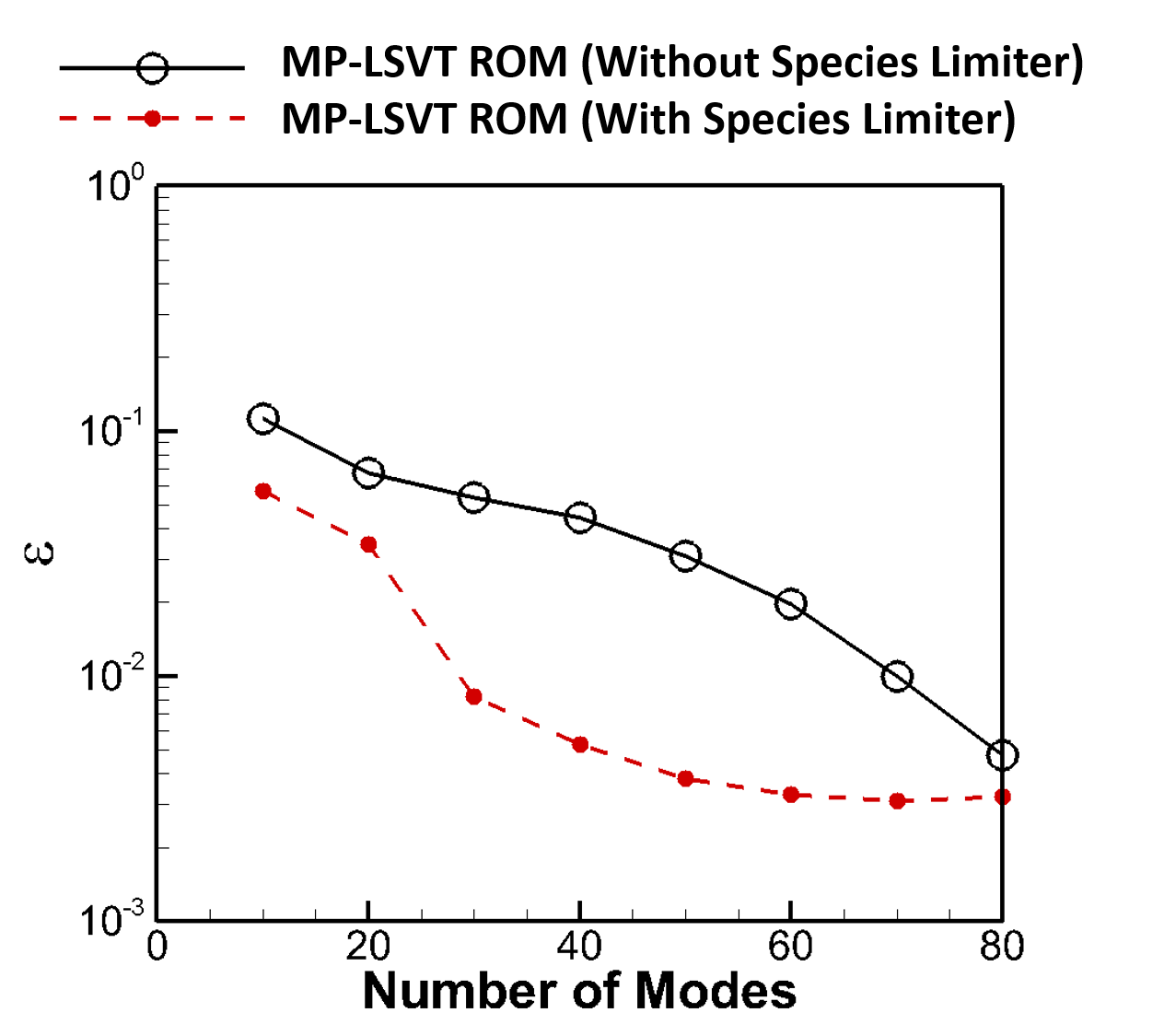}
	\caption{ROM error convergence comparisons between {\color{black}MP-LSVT} ROM with and without species limiter for 1D premixed laminar flame.}\label{1d:romYlimiter} 
\end{figure}

The species limiter with $T_{th} = 2{,}450$ K and $\delta = 0$ is now introduced in the ROM calculations. The 50-mode ROM with the species limiter included are compared with the FOM solution in Fig.~\ref{1d:romVSfomS}. The results show significant improvement over Fig.~\ref{1d:romVSfomS} with diminished fluctuations and accurate ROM reproduction of flame propagation and heat release rate. The ROM reconstruction error with the species limiter employed is compared with the reconstruction error without limiting as a function of the number of trial basis modes in Fig.~\ref{1d:romYlimiter}. The limiter provides significant and consistent improvement in ROM accuracy, about one order of magnitude for lower trial basis resolutions. 

\section{Hyper-reduction Implementation Details}
\label{appendix:hyperreductionDetails}

In this section, we provide further details of scalable implementation of hyper-reduction.

\subsection{Sampling Coupled Systems of Equations}
As mentioned briefly in Section~\ref{hyper_reduction}, for systems of coupled equations $\mathbf{S}^T \resFunc{\solPrimROMFull} \neq \resFunc{\mathbf{S}^T \solPrimROMFull}$.  For example, in the Navier-Stokes equations, computing the inviscid density flux requires access to not only the density, but also the velocity field. Furthermore, for a cell-centered finite volume spatial discretization, the state at adjacent cells are required to correctly compute face fluxes, reconstruct the state at cell vertices, and calculate gradients for higher-order flux schemes.

\begin{figure}[H]
    \begin{minipage}{0.49\linewidth}
    \centering
    \includegraphics[width=0.54\linewidth]{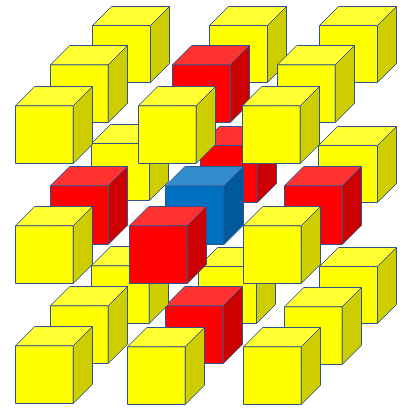}
    \end{minipage}
    \begin{minipage}{0.49\linewidth}
    \centering
    \includegraphics[width=0.75\linewidth]{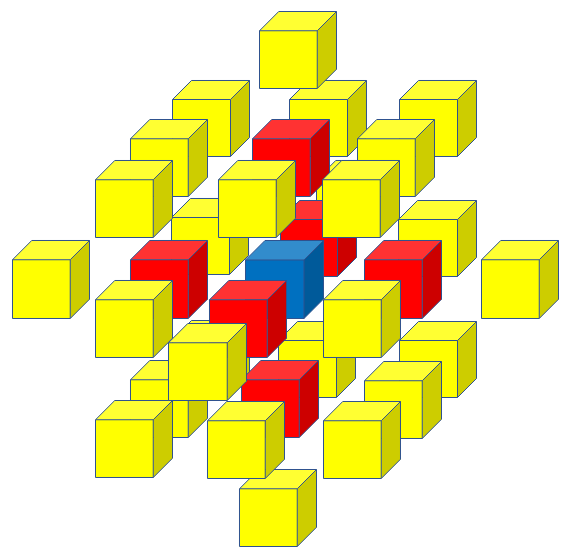}
    \end{minipage}
    \caption{Sampling for first-order (left) and second-order (right) flux schemes in 3D. Blue cells are directly sampled, red are required for face flux calculations, and yellow are required for gradient/vertex state calculations.}
    \label{fig:sampling3d}
\end{figure}

To optimize calculations and minimize memory requirements, special designations to cells must be assigned:
\begin{enumerate}
    \item Directly sampled: cells for which at least one of its associated degrees of freedom is sampled and interpolated. All state variables at this cell are included.
    \item Flux cells: cells which share a face with a directly sampled cell. All state variables at this cell are included.
    \item Gradient/vertex cells: cells which share a vertex with a directly sampled cell, or in the case of second-order accurate fluxes, share a face with a flux cell. All state variables at this cell are included.
    \item Unsampled: cells which do not fall into any of the above designations. Memory does not need to be allocated for these cells, and they do not contribute to the calculations except to visualize the flow field.
\end{enumerate}
The above designations are limited to second-order accurate flux schemes, but can be extended as needed. For a 3D mesh composed of hexahedral elements, these designations are visualized in Fig.~\ref{fig:sampling3d}.

\begin{figure}[H]
    \centering
    \includegraphics[width=0.7\linewidth]{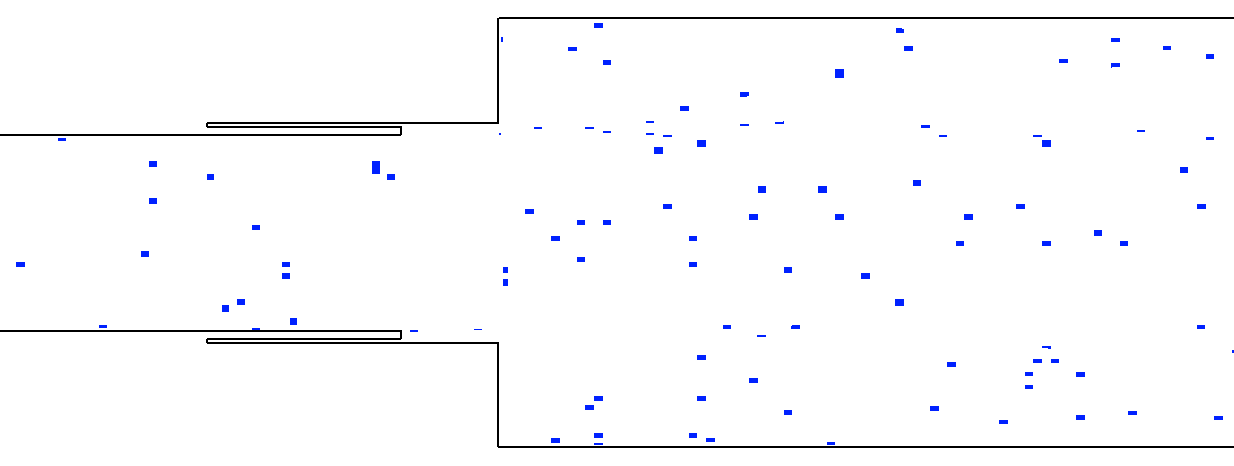}
    \includegraphics[width=0.7\linewidth]{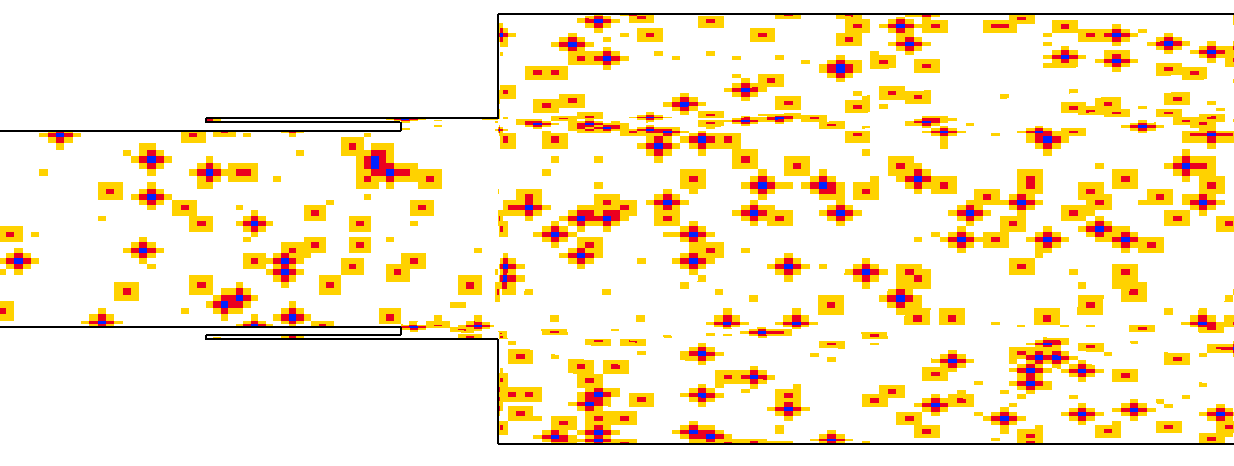}
    \caption{Contrast between directly sampled cells only (top) and all required cells (bottom) in a zoomed-in slice of the 3D combustor, 1\% sampling mesh.}
    \label{fig:3d_samplingComparison}
\end{figure}

These cell designations hint at a challenging reality in hyper-reduction: to sampled a certain number of degrees of freedom in computing $\mathbf{S}^T \mathbf{r}$, many additional degrees of freedom will be required. This is visualized in Fig.~\ref{fig:3d_samplingComparison}. Both images show the same zoomed-in slice from the 1\% sampled mesh evaluated in Section~\ref{3d:injector_hyperReduction}. The top image displays only the directly sampled cells, showing the $\sim$1\% sampling rate expected. The bottom image displays all cells required for computing the sampled non-linear functions. The  number of  cells in each designation as a percentage of the entire mesh, is shown in Table~\ref{tab:sampling}.

\begin{table}
\centering
\begin{tabular}{rrcrrcrr}
\toprule
 \multicolumn{2}{c}{Directly sampled} & \phantom{abc} & \multicolumn{2}{c}{Flux} & \phantom{abc} & \multicolumn{2}{c}{Gradient/vertex} \\
\cmidrule{1-2} \cmidrule{4-5} \cmidrule{7-8}
Count & \% && Count & \% && Count & \% \\ \midrule
5,893 & 1.0 && 34,958 & 5.9 && 124,631 & 21.1 \\
58,939 & 10.0 && 252,187 & 42.8 && 259,724 & 44.1 \\
589,395 & 100.0 && 0 & 0.0 && 0 & 0.0 \\
\bottomrule
\end{tabular}
\caption{\label{tab:sampling} Cell designation counts and percentage of $\numElements$ for 1\% sampled, 10\% sampled, and fully-sampled meshes.}
\end{table}

While this may negatively impact expected memory requirements, significant speedup can still be achieved as demonstrated in the 3D injector case, with speedup results shown in Tables~\ref{table:3D_hyperreduction_exp} and~\ref{table:3D_hyperreduction_imp}. Proper care must be taken to avoid any calculations which are not strictly necessary for computing $\mathbf{S}^T \mathbf{r}$. However, due to the above requirements, it should not be expected that the scaling between the number of directly sampled cells and speedup be one-to-one, or even linear (e.g. a 1\% sampled mesh will not result in a 100-fold speedup).

\subsection{Load Balancing}

If ROM simulations are to be applied to many-query problems such as parametric design, uncertainty quantification, or real-time future state prediction, they must be amenable for execution on memory-restricted computers such as desktop workstations or embedded systems. In the current context, computational tasks to be load balanced can be roughly divided into three categories: mathematical operations (usually floating-point operations for computing physical or numerical quantities), communications (data transfers between processing units in a distributed-memory framework), and reading from/writing to high-capacity data storage (I/O). The elimination of I/O tasks in a projection-based ROM is straightforward. Only the trial basis coefficients ($\solPrimROMRed$) need to be written to disk, and the full system state can be easily reconstructed ($\solPrimROMFull = \solPrimFOMRef + \scaleMatPrim^{-1} \trialBasisPrim \solPrimROMRed$) using very few computational resources. Proper load balancing of mathematical calculations and communications requires more care.

We employ METIS~\cite{metisOriginal} to extract optimal partitions based on the sparsely-sampled mesh (rather than the original mesh used by the FOM) while minimizing communication between the partitions. The finite volume mesh is described as an unstructured graph, in which each cell is a vertex of the graph, and two cells which interact with each other (e.g. through fluxes, gradient calculations) are connected by an edge of the graph. The sparsely-sampled meshes (such as the one displayed in Fig.~\ref{fig:3d_samplingComparison}) can further be considered as a collection of disjoint sub-graphs. Flux cells and gradient/vertex cells only share edges with the directly sampled cell from which they originate, and with any flux/gradient/vertex cells similarly associated.

\begin{figure}
\centering
    \begin{minipage}{0.49\textwidth}
        \includegraphics[width=\linewidth]{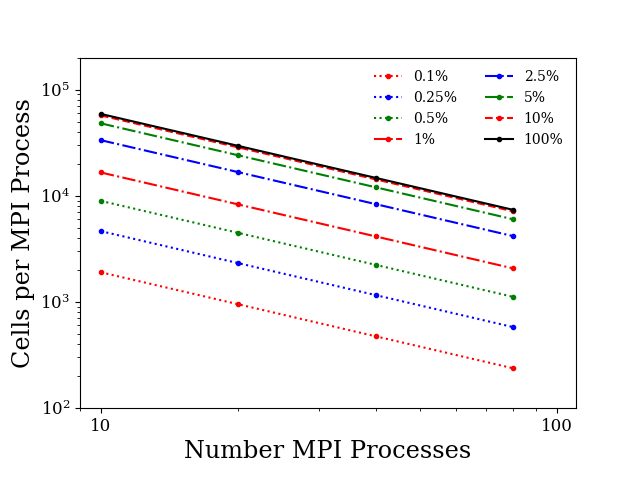}
        \caption{Average number of cells allocated to each MPI process for sparsely sampled meshes, 3D  injector.}
        \label{fig:cellsPerProcCount}
    \end{minipage}
    \hfill
    \centering
    \begin{minipage}{0.49\textwidth}
        \includegraphics[width=\linewidth]{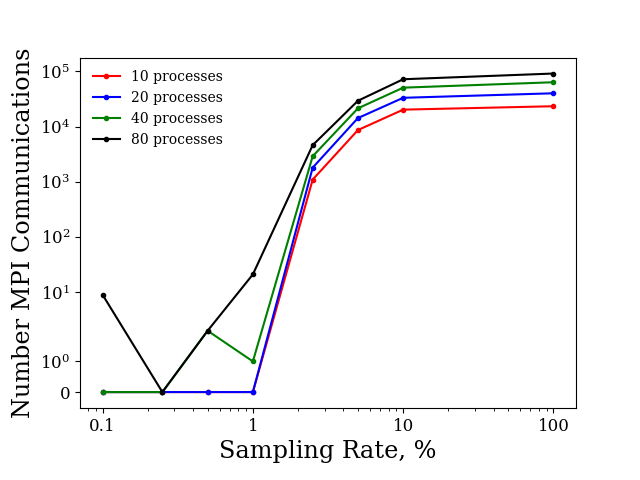}
        \caption{Total number of MPI communications for sparsely sampled meshes, 3D injector.}
        \label{fig:mpiCommCount}
    \end{minipage}
\end{figure}

We provide results for the sparsely-sampled computations of the 3D injector introduced in Section~\ref{3d:injector}. As in the hyper-reduction experiments, the first 60 sampled cells are selected by the rank-revealing QR factorization, while the remainder are sampled from a uniform random distribution. Figure~\ref{fig:cellsPerProcCount} shows that, given a sparsely sampled mesh, halving the MPI process count results in, on average, a doubling in the number of cells assigned to each process. The standard deviations on this data are not shown as they are extremely small, no more than 2\% of the average. The behavior of MPI communications in the partitioned mesh (represented by METIS as edge-cuts in the graph) is quite revealing. Figure~\ref{fig:mpiCommCount} shows that there is a precipitous drop-off in the number of MPI communications when the mesh is sampled below 2.5\%, regardless of the number of MPI processes to which the partitions are assigned. We believe this may be linked to a greater level of disjointedness between sub-graphs, allowing for fewer constraints due to load balancing. This motivates the question as to whether different sampling techniques (e.g. deterministic oversampling) might lead to lower-quality partitions; and is a topic for future work. Unsurprisingly, decreasing the number of processes generally leads to a decrease in the number of MPI communications. 

\section{Algorithm for the {\color{black}Model-form Preserving} Least-Squares with Variable Transformation ({\color{black}MP-LSVT}) Formulation}
\label{appendix:splsvt_algorithm}
\begin{algorithm}
	\caption{{\color{black}Model-form Preserving} Least-Squares with Variable Transformation ({\color{black}MP-LSVT})} 
	\label{splsvt_algorithm}
	\begin{algorithmic}
	    \State \textbf{Input}: The number of physical time steps, $M$; the number of pseudo time steps, $K$, within each physical time step; the error tolerance for convergence, $\epsilon_\text{tol}$
	    \State Solve the full order model (Eq.~\ref{cfd:discretized_qp}) to collect the full-state data $\mathbf{q}_p$
	    \State Compute POD bases $\trialBasisPrim$ from $\mathbf{q}_p$ using Eq.~\ref{pod:expansion_qp}
	    \State Initialize $\solPrimROMRed^{0} = \scaleMatPrim (\solPrimROMFull^0 - \solPrimFOMRef)$
		\For {$\iterIdx = 1,\ldots,M$}
    		\For {$k = 1,\ldots,K$}
    		    \State Approximate solution variables, $\solPrimROMFull^{\subIdx-1} = \solPrimFOMRef + \scaleMatPrim^{-1}\trialBasisPrim \solPrimROMRed^{\subIdx-1}$
    		    
    			\If{minimize the fully-discrete FOM equation residual at the physical-time level, $\resFunc{\solPrimROMFull^{\iterIdx}}$}
    			
    			\If{\emph{Linear multi-step methods}}
    			\State Set $\testBasisPrim^{\subIdx-1} = \scaleMatCons \lp \gmROM^{\subIdx-1} - \dt \beta_0 \hat{\mathbf{J}}^{\subIdx-1} \gmROM^{\subIdx-1} \rp \scaleMatPrim^{-1}\trialBasisPrim$ based on Eq.~\ref{rom:lspg_w_LinearMultistep}  
    			
    			\ElsIf{\emph{Explicit and diagonally implicit Runge–Kutta methods}}
    			\State Set $\testBasisPrim^\iterIdx = \scaleMatCons \gmROM^\iterIdx \scaleMatPrim^{-1} \trialBasisPrim$ based on Eq.~\ref{rom:lspg_w_RK}
    			
    			\ElsIf{\emph{Implicit Runge–Kutta methods}}
    			\State Set $\testBasisPrim^\iterIdx = \scaleMatCons \lp \gmROM^\iterIdx - \dt b_s \jacobConsROM^\iterIdx \gmROM^\iterIdx \rp \scaleMatPrim^{-1} \trialBasisPrim$ based on Eq.~\ref{rom:lspg_w_RK_imp}
    			
    			\EndIf
    			
    			\If{\emph{Newton's method}}
    			\State $\solPrimROMRed^{\subIdx} = \solPrimROMRed^{\subIdx-1} - \left[ \lp \testBasisPrim^{\subIdx-1} \rp^T \testBasisPrim^{\subIdx-1} \right]^{-1} \lp \testBasisPrim^{\subIdx-1} \rp^T \scaleMatCons \resFunc{\solPrimROMFull^{\subIdx-1}}$
    			
    			\ElsIf{\emph{Pseudo-time stepping introduced to the fully-discrete FOM equation}}
    			\State $\solPrimROMRed^{\subIdx} = \solPrimROMRed^{\subIdx-1} - \left[ \lp \testBasisPrim^{\subIdx-1} \rp^T \lp \frac{\dt}{\dtau} \gmROM^{\subIdx-1} + \testBasisPrim^{\subIdx - 1} \rp \right]^{-1} \lp \testBasisPrim^{\subIdx-1} \rp^T \scaleMatCons \resFunc{\solPrimROMFull^{\subIdx-1}}$
    			
    			\ElsIf{\emph{Pseudo-time stepping introduced to the fully-discrete ROM equation}}
    			\State $\solPrimROMRed^{\subIdx} = \solPrimROMRed^{\subIdx-1} - \lp 1 + \frac{\dt}{\dtau} \rp^{-1} \left[ \lp \testBasisPrim^{\subIdx-1} \rp^T \testBasisPrim^{\subIdx-1} \right]^{-1} \lp \testBasisPrim^{\subIdx-1} \rp^T \scaleMatCons \resFunc{\solPrimROMFull^{\subIdx-1}}$
    			
    			\EndIf
    			
    			\ElsIf{minimize the fully-discrete FOM equation residual at the sub-iteration level, $\resPrimFunc{\solPrimROMFull^{\subIdx}}$ for linear multi-step methods}
    			\State Set $\testBasisPrimPrim^{\subIdx} = \scaleMatCons \left[ \lp \frac{\dt}{\dtau} + 1 \rp \gmROM^{\subIdx-1} - \dt \beta_0 \jacobConsROM^{\subIdx-1} \gmROM^{\subIdx-1} \right] \scaleMatPrim^{-1} \trialBasisPrim$ based on Eq.~\ref{rom:lspg_w_qp_DualTime}
    			
    			\State then $\solPrimROMRed^{\subIdx} = \solPrimROMRed^{\subIdx-1} - \left[ \lp \testBasisPrimPrim^{\subIdx} \rp^T \testBasisPrimPrim^{\subIdx} \right]^{-1} \lp \testBasisPrimPrim^{\subIdx} \rp^T \scaleMatCons \resFunc{\solPrimROMFull^{\subIdx-1}}$
    			\EndIf
    			
    			\If{$\norm{\solPrimROMFull^{\subIdx} - \solPrimROMFull^{\subIdx-1}} / \norm{\solPrimFOMRef} < \epsilon_\text{tol}$}
    			\State exit the loop
    			\EndIf
    			
    		\EndFor
    		\State Update $\solPrimROMFull^{\iterIdx} = \solPrimFOMRef + \scaleMatPrim^{-1} \trialBasisPrim \solPrimROMRed^{\subIdx}$
		\EndFor
	\end{algorithmic} 
\end{algorithm}
\bibliographystyle{elsarticle-num}
\bibliography{ref.bib}

\end{document}